\documentclass[a4paper,oneside,11pt]{article}
\usepackage{amsmath}
\usepackage{amssymb}
\usepackage{amsthm}
\usepackage{MnSymbol}
\usepackage{enumerate}
\usepackage{bm}
\usepackage{graphicx}
\usepackage[margin = 3cm]{geometry}
\usepackage{color}
\usepackage{hyperref,url}
\usepackage{enumitem}
\usepackage{placeins} 

\allowdisplaybreaks[3]
\newcommand{\N}{\mathbb{N}}

\newcommand{\Z}{\mathbb{Z}}
\newcommand{\R}{\mathbb{R}}

\newcommand{\EE}{\mathcal{E}}
\newcommand{\OO}{\mathcal{O}}

\newcommand{\dHaus}{\, d \Gamma \,}
\newcommand{\dd}{\, d \,}
\newcommand{\dz}{\, dz \,}
\newcommand{\dx}{\, dx \,}
\newcommand{\ds}{\, ds \,}
\newcommand{\dt}{\, dt \,}

\newcommand{\pd}{\partial}

\newcommand{\eps}{\varepsilon}
\newcommand{\der}{ \mathrm{D} }

\newcommand{\abs}[1]{\left| #1 \right|}
\newcommand{\norm}[1]{\| #1 \|}
\newcommand{\bignorm}[1]{\left\| #1 \right\|}
\newcommand{\inner}[2]{\langle #1 , #2 \rangle}

\newcommand{\Laplace}{\Delta}
\newcommand{\surf}{\nabla_{\Sigma}}

\newcommand{\md}{\pd_{t}^{\bullet}}
\newcommand{\pdnu}{\pd_{\bm{\nu}}}

\newcommand{\para}{\alpha}
\newcommand{\jump}[1]{ \left [ #1 \right ]_{1}^{2}}
\newcommand{\velo}{\mathcal{V}}

\newcommand{\mean}[1]{\overline{#1}}
\renewcommand{\div}{\, \mathrm{div}\,}

\newtheorem{thm}{Theorem}[section]

\newtheorem{remark}{Remark}[section]

\newtheorem{assump}{Assumption}[section]
\numberwithin{equation}{section}

\begin{document}
\title{A Hele--Shaw--Cahn--Hilliard model for incompressible two-phase flows with different densities}

\author{Luca Ded\`{e} \footnotemark[1] \and Harald Garcke \footnotemark[2] \and Kei Fong Lam \footnotemark[2]}

\date{ }

\maketitle

\renewcommand{\thefootnote}{\fnsymbol{footnote}}
\footnotetext[1]{CMCS – Chair of Modeling and Scientific Computing, MATHICSE – Mathematics Institute of Computational Science and Engineering, EPFL – \'{E}cole Polytechnique F\'{e}d\'{e}rale de Lausanne, Station 8, 1015 Lausanne, Switzerland ({\tt luca.dede@epfl.ch}).}
\footnotetext[2]{Fakult\"at f\"ur Mathematik, Universit\"at Regensburg, 93040 Regensburg, Germany
({\tt \{Harald.Garcke, Kei-Fong.Lam\}@mathematik.uni-regensburg.de}).}

\begin{abstract}
Topology changes in multi-phase fluid flows are difficult to model within a traditional sharp interface theory.  Diffuse interface models turn out to be an attractive alternative to model two-phase flows.  Based on a Cahn--Hilliard--Navier--Stokes model introduced by Abels, Garcke and Gr\"{u}n (Math. Models Methods Appl. Sci. 2012), which uses a volume averaged velocity, we derive a diffuse interface model in a Hele--Shaw geometry, which in the case of non-matched densities, simplifies an earlier model of Lee, Lowengrub and Goodman (Phys. Fluids 2002).  We recover the classical Hele--Shaw model as a sharp interface limit of the diffuse interface model.  Furthermore, we show the existence of weak solutions and present several numerical computations including situations with rising bubbles and fingering instabilities.
\end{abstract}



\section{Introduction}
Interfaces in fluid flow play an important role in many applications.  A mathematical description of such phenomena typically involves highly nonlinear equations due to the unknown interfaces.  Two-phase flow in the special case of a Hele--Shaw cell which involves the slow flow of a fluid between two parallel flat plates which are fixed at a small distance apart still contains many ingredients of more complicated systems.  Especially interesting instabilities like the Saffman--Taylor fingering instability \cite{Saffman} can occur and this instability has important applications in technology.  We refer for example to the analogy of the Hele--Shaw cell to instabilities that appear when one tries to extract residual oil from a porous rock, see \cite{Todd,Tryggvason}.  Water is pumped into the porous rocks to direct the oil to the producing wells, but it was observed that a lot of oil remained in the ground when the water appeared at the wells.  One explanation of this phenomenon was attributed to the instabilities of the oil-water interface, which allowed the water to flow through the porous rocks without displacing much of the oil.

In a sharp interface description, the Hele--Shaw model is given as follows.  The overall domain $\Omega$ is occupied by two fluids, modeled as time-dependent disjoint regions $\Omega_{1}$ and $\Omega_{2}$, which are separated by a time-dependent hypersurface $\Sigma$.  Introducing the fluid velocity $\bm{v}$, the viscosities $\eta_{i}$ and densities $\overline{\rho}_{i}$, $i = 1,2$, (which can be different in the two phases), the pressure $p$, the gravity vector $\bm{g} = g \hat{\bm{g}}$ with modulus $g$ and unit vector $\hat{\bm{g}}$, the unit normal $\bm{\nu}$ on $\Sigma$ pointing into $\Omega_{2}$, one has to study the following set of equations:
\begin{subequations}
\begin{alignat}{3}
\div \bm{v} & = 0 && \text{ in } \Omega_{1} \cup \Omega_{2}, \\
12 \eta_{1} \bm{v} = & - \nabla p + \overline{\rho}_{1} \bm{g} && \text{ in } \Omega_{1}, \\
12 \eta_{2} \bm{v} = & - \nabla p + \overline{\rho}_{2} \bm{g} && \text{ in } \Omega_{2}, \\
\jump{\bm{v}} \cdot \bm{\nu} & = 0 && \text{ on } \Sigma, \\
\jump{p} & = \sigma \kappa && \text{ on } \Sigma, \\
\velo & = \bm{v} \cdot \bm{\nu} && \text{ on } \Sigma.
\end{alignat}
\end{subequations}
Here, $\jump{\cdot}$ denotes the jump across the interface, $\sigma$ is the surface tension, $\kappa$ is the mean curvature and $\velo$ is the normal velocity.  These equations can be derived from more complete models involving the (Navier--)Stokes equations in situation where the flow is slow (small Reynolds number) and is confined between two parallel plates at a small distance apart, see for example \cite{Ockendon}.

Such a sharp interface description breaks down when the topology of the interface changes.  As a remedy, various diffuse interface models have been introduced to describe incompressible two-phase flows.  A first model restricted to equal densities was introduced by Hohenberg and Halperin \cite{HH}, while a first diffuse interface model for two-phase flow allowing for a density contrast was introduced by Lowengrub and Truskinowsky \cite{Lowengrub}.  However, the model in \cite{Lowengrub} leads to a velocity field which is not divergence-free (solenoidal) although both individual fluids are.  Let us remark that Lowengrub and Truskinowsky used a mass averaged velocity field to define their diffuse interface model.  More recently, Abels, Garcke and Gr\"{u}n \cite{AGG} introduced a diffuse interface model with a divergence-free velocity field which also allows for different densities.

We base our derivation of a diffuse interface model for a Hele--Shaw cell on the Cahn--Hilliard--Navier--Stokes model of \cite{AGG}, which in nondimensional form reads as
\begin{subequations}\label{AGGNS}
\begin{align}
\div \bm{v} & = 0, \label{AGGNS:div} \\
\pd_{t}(\rho_{*}(\varphi) \bm{v}) + \div (\rho_{*}(\varphi) \bm{v} \otimes \bm{v}) - \nabla p  & = \frac{1}{\mathrm{Re}} \div  (2 \eta_{*}(\varphi) \der  \bm{v}) + \bm{G} \label{AGGNS:velo} \\
\notag &  - \frac{\eps}{\mathrm{Ca}} \div (\nabla \varphi \otimes \nabla \varphi) + \tfrac{\overline{\rho}_{2} - \overline{\rho}_{1}}{2 \overline{\rho}_{2}} \div (m(\varphi) \bm{v} \otimes \nabla \mu),  \\
\md \varphi & = \div (m(\varphi) \nabla \mu), \label{AGGNS:varphi} \\
\mu & = \frac{1}{\eps} \Psi'(\varphi) - \eps \Laplace \varphi, \label{AGGNS:mu}
\end{align}
\end{subequations}
with suitable boundary and initial conditions.  Here, $\varphi$ is an order parameter which represents the difference in the volume fractions, such that $\{\varphi = -1\}$ represents fluid 1 and $\{ \varphi = 1\}$ represents fluid 2.  The function $\rho_{*}(\varphi) = \frac{\overline{\rho}_{2} - \overline{\rho}_{1}}{2 \overline{\rho}_{2}} \varphi + \frac{\overline{\rho}_{2} + \overline{\rho}_{1}}{2 \overline{\rho}_{2}}$ is the nondimensionalized density of the fluid mixture, $\der \bm{v} = \frac{1}{2} (\nabla \bm{v} + (\nabla \bm{v})^{\top})$ is the symmetric gradient for the volume-averaged velocity $\bm{v}$, $p$ denotes the pressure, $\eta_{*}(\varphi) = \frac{\eta_{2} - \eta_{1}}{2 \eta_{2}} \varphi + \frac{\eta_{2} + \eta_{1}}{2 \eta_{2}}$ is the nondimensionalized viscosity of the mixture, $\mathrm{Re}$ denotes the Reynolds number, $\mathrm{Ca}$ denotes the capillary number, $\eps > 0$ is a (small) parameter related to the thickness of the interfacial regions, $\Psi'$ is the derivative of a potential $\Psi$ which has equal minima at $\pm 1$, $\mu$ is the chemical potential, $m(\varphi)$ is a non-negative mobility which, in the case of a constant mobility $m(\varphi) = m$, can be seen as the reciprocal of the P\'{e}lect number $\mathrm{Pe}$, $\md \varphi = \pd_{t}\varphi + \nabla \varphi \cdot \bm{v}$ is the material derivative of $\varphi$, and $\bm{G}$ denotes an external body force. 

We will show, via a formal asymptotic analysis, that for slow flow in a Hele--Shaw cell geometry the above model leads to a Hele--Shaw--Cahn--Hilliard model
\begin{subequations}\label{Intro:DGL}
\begin{align}
\div \bm{v} & = 0, \\
12 \eta_{*}(\varphi) \bm{v} & = - \nabla p + \bm{G} - \frac{\eps}{\mathrm{Ca}} \div ( \nabla \varphi \otimes \nabla \varphi), \\
\md \varphi  & = \div (m(\varphi) \nabla \mu), \\
\mu & = \frac{1}{\eps} \Psi'(\varphi) - \eps \Laplace \varphi,
\end{align}
\end{subequations}
which inherits a divergence-free velocity field from the Cahn--Hilliard--Navier--Stokes model \eqref{AGGNS}.  In this paper, we will study the model \eqref{Intro:DGL} in detail both from an analytical and also from a numerical point of view.

An earlier Hele--Shaw--Cahn--Hilliard model was introduced by Lee, Lowengrub and Goodman \cite{LLG01, LLG02}.  However, they used the Cahn--Hilliard--Navier--Stokes model of Lowengrub and Truskinovsky \cite{Lowengrub} as a basis and obtained
\begin{subequations}\label{Intro:LLG}
\begin{alignat}{3}
\div \bm{v} - \frac{\alpha}{\mathrm{Pe}} \Laplace \mu & = 0, \label{Intro:LLG:div} \\
\rho(c) (\pd_{t}c + \nabla c \cdot \bm{v}) - \frac{1}{\mathrm{Pe}} \Laplace \mu & = 0, \label{Intro:LLG:phase} \\
\bm{v} + \frac{1}{12 \eta(c)} \left ( \nabla p + \frac{\mathrm{Ch}}{\mathrm{Ma}} \div(\rho(c) \nabla c \otimes \nabla c)  - \rho(c) \hat{\bm{g}} \right ) & = \bm{0}, \label{Intro:LLG:velo} \\
\mu - f'_{0}(c) +  \frac{\mathrm{Ch}}{\rho(c)} \div (\rho(c) \nabla c) -  \mathrm{Ma} \,  \alpha p & = 0, \label{Intro:LLG:chem}
\end{alignat}
\end{subequations}
where $\mathrm{Pe}$ is the P\'{e}lect number, $\mathrm{Ch}$ is the Cahn number, $\mathrm{Ma}$ is the Mach number, $c$ is the mass concentration of fluid 1, so that $ \{ c = 1 \}$ represents fluid 1 and $\{c = 0 \}$ represents fluid 2, $\rho(c)$ is the total density, $\alpha$ is the difference between the reciprocals of the actual mass densities of the fluid, $f_{0}(c) = c^{2}(1-c^{2})$ is a potential with two minima at $c = 0$ and $c = 1$, $\eta(c) = \eta_{1}c + \eta_{2}(1-c)$ is the interpolation of the two viscosities, and $\hat{\bm{g}}$ is the unit vector of gravity.  We refer the reader to Section \ref{sec:LLGModel} for more details.

It is important to note that the velocity $\bm{v}$ in \eqref{Intro:LLG} is the mass-averaged velocity, which is in contrast to the volume-averaged velocities in \eqref{AGGNS} and \eqref{Intro:DGL}.  One observes that the mass-averaged velocity is not divergence-free and that the pressure $p$ enters the equation for the chemical potential \eqref{Intro:LLG:chem}.  These facts make the analysis and the numerical approximation of this model quite involved.  We remark that Lee, Lowengrub and Goodman derived \eqref{Intro:DGL} from \eqref{Intro:LLG} in the case where a Boussinesq approximation is valid, i.e., the deviation of $\rho$ from its spatial average needs to be small which basically means that the densities of the two fluids are very close.  Our derivation however is valid for any density contrast among the fluids.

We spatially approximate the Hele--Shaw--Cahn--Hilliard equations by means of NURBS-based Isogeometric Analysis \cite{Cottrell09,Hughes05} as it allows a straightforward construction of the finite dimensional function spaces for high order problems \cite{Gomez08,Tagliabue}.  Indeed, in this paper, we formulate the Hele--Shaw--Cahn--Hilliard model \eqref{Intro:DGL} in terms of the pressure $p$ and order parameter $\varphi$, thus yielding a fourth order problem in the latter variable.  In this respect, our finite dimensional function spaces are built out of globally $C^{1}$-continuous B-spline basis functions of degree $2$ \cite{Piegl97}.  For the time discretization, we use Backward Differentiation Formulas (BDF) of order $2$ \cite{QSS07} with equal order extrapolation of the unknowns to obtain a semi-implicit formulation of the full discrete problem as e.g. in \cite{FortiDede}.

Finally, we propose and discuss numerical results for two benchmark problems: the rising bubble and viscous fingering tests \cite{Hysing,LLG01}.

The outline of this paper is as follows: In Section \ref{sec:derivation} we derive \eqref{Intro:DGL} from \eqref{AGGNS} by means of a formal asymptotic analysis.  In Section \ref{sec:analysis}, we derive the sharp interface limit of \eqref{Intro:DGL} and prove the existence of weak solutions to \eqref{Intro:DGL}.  In Section \ref{sec:approx} we present the numerical scheme for \eqref{Intro:DGL} reformulated in terms of the pressure $p$ and the order parameter $\varphi$, and in Section \ref{sec:reso} we present and discuss the numerical results. 

\section{Derivation of the Hele--Shaw--Cahn--Hilliard model}\label{sec:derivation}
\subsection{A Navier--Stokes--Cahn--Hilliard model for incompressible two-phase flows}\label{sec:AGG}
We start from the volume-averaged velocity model introduced by Abels, Garcke and Gr\"{u}n in \cite{AGG}:  For fluid $i$, $i = 1,2$, let $\rho_{i}$ denote the actual mass density, $\overline{\rho}_{i}$ the density of a pure component, $u_{i} := \frac{\rho_{i}}{\overline{\rho}_{i}}$ the volume fraction, $\bm{v}_{i}$ the individual velocity, and $\eta_{i}$ the viscosity.  The volume-averaged velocity for the fluid mixture is defined as
\begin{align*}
\bm{v} = u_{1} \bm{v}_{1} + u_{2} \bm{v}_{2}.
\end{align*}
We define the order parameter $\varphi$ as the difference in the volume fractions, i.e., $\varphi = u_{2} - u_{1}$, then we obtain the following system of equations:
\begin{subequations}\label{GravNS}
\begin{align}
\div \bm{v} & = 0, \label{GravNS:div} \\
\pd_{t}(\rho(\varphi) \bm{v}) + \div (\rho(\varphi) \bm{v} \otimes \bm{v}) & = \div (2 \eta(\varphi) \der  \bm{v}) - \nabla p + \bm{G} \label{GravNS:velo} \\
\notag &  - \sigma \eps \div (\nabla \varphi \otimes \nabla \varphi) + \tfrac{\overline{\rho}_{2} - \overline{\rho}_{1}}{2} \div (m(\varphi) \bm{v} \otimes \nabla \mu),  \\
\md \varphi & = \div (m(\varphi) \nabla \mu), \label{GravNS:varphi} \\
\mu & = \frac{\sigma}{\eps} \Psi'(\varphi) - \sigma \eps \Laplace \varphi. \label{GravNS:mu}
\end{align}
\end{subequations}
Here, $\rho(\varphi) = \frac{\overline{\rho}_{2} - \overline{\rho}_{1}}{2} \varphi + \frac{\overline{\rho}_{2} + \overline{\rho}_{1}}{2}$ is the density of the fluid mixture, $\der \bm{v} = \frac{1}{2} (\nabla \bm{v} + (\nabla \bm{v})^{\top})$ is the symmetric gradient, $p$ denotes the pressure, $\eta(\varphi) = \frac{\eta_{2} - \eta_{1}}{2} \varphi + \frac{\eta_{2} + \eta_{1}}{2}$ is the viscosity of the mixture, $\sigma$ is a constant related to the surface energy density, $\eps > 0$ is a (small) parameter related to the thickness of the interfacial regions, $\Psi'$ is the derivative of a potential $\Psi$ which has equal minima at $\pm 1$, $\mu$ is the chemical potential, $m(\varphi)$ is a non-negative mobility, $\md \varphi = \pd_{t}\varphi + \nabla \varphi \cdot \bm{v}$ is the material derivative of $\varphi$, and $\bm{G} = \bm{G}(\rho(\varphi))$ denotes a body force which may depend on the density.  The example we have in mind refers to the gravitational force and reads:
\begin{align}\label{BodyForceChoice}
\bm{G}(\varphi) = \rho(\varphi) g \hat{\bm{g}},
\end{align}
where the unit vector $\hat{\bm{g}}$ indicates the direction of gravity and $g$ is the modulus.

The model \eqref{GravNS} consists of the Navier--Stokes equations coupled with a Cahn--Hilliard system.  The capillary forces due to surface tension are modeled by the term $\sigma \eps \div (\nabla \varphi \otimes \nabla \varphi)$, and the term $\frac{\overline{\rho}_{2} - \overline{\rho}_{1}}{2} \div (m(\varphi) \bm{v} \otimes \nabla \mu)$ accounts for the effects of non-matched fluid densities.  We point out that the simple form for continuity equation \eqref{GravNS:div} is due to the choice of $\bm{v}$ as the volume-averaged velocity, when compared for instance to the approach of Antanovskii \cite{Antanovskii} and Lowengrub and Truskinovsky \cite{Lowengrub}, where a mass-averaged velocity is used and leads to a more complex expression for the continuity equation.

Furthermore, \eqref{GravNS} satisfies the energy equality
\begin{align*}
\frac{\dd}{\dt} \int_{\Omega} \left ( \frac{\rho}{2} \abs{\bm{v}}^{2} + \frac{\sigma}{\eps}\Psi(\varphi) + \frac{\sigma \eps}{2} \abs{\nabla \varphi}^{2} \right ) \dx + \int_{\Omega} \left ( 2 \eta \abs{\der \bm{v}}^{2} + m \abs{\nabla \mu}^{2} \right ) \dx = \int_{\Omega} \bm{G} \cdot \bm{v} \dx,
\end{align*}
when we complement \eqref{GravNS} with the boundary conditions
\begin{align*}
\pdnu \varphi = \nabla \varphi \cdot \bm{\nu} = 0, \quad \pdnu \mu = 0, \quad \bm{v} = \bm{0} 
\end{align*}
on the boundary $\Gamma$ of the bounded domain $\Omega \subset \R^{d}$, $d = 1,2,3$, under consideration.  Here $\frac{\rho}{2} \abs{\bm{v}}^{2}$ denotes the kinetic energy of the fluid mixture, $\frac{1}{\eps} \Psi(\varphi) + \frac{\eps}{2} \abs{\nabla \varphi}^{2}$ is the Ginzburg--Landau energy density, and its product with $\sigma$ approximates the surface energy density in the limit $\eps \to 0$, cf. \cite{Modica87}.  The total energy (consisting of the kinetic energy and the surface energy) is dissipated by viscous stress and diffusion, given by the second integral on the left-hand side.  We also obtain energy contributions via the body force $\bm{G}$ in the form of the right-hand side.  For the existence of weak solutions to \eqref{GravNS} we refer to the work of Abels, Depner and Garcke \cite{ADG, ADG:deg}.

\subsection{Nondimensionalization and the Hele--Shaw approximation}
We now follow the procedure outlined in \cite[Chapter 4]{Ockendon}, and consider the Navier--Stokes--Cahn--Hilliard equations \eqref{GravNS} with the body force $\bm{G}$ given as in \eqref{BodyForceChoice} in a domain $\Omega \subset \R^{3}$ which occupies a region in between two rigid walls, one at $\{x_{3} = 0\}$ and one at $\{x_{3} = H\}$, for some $H > 0$.  To be precise, we assume that $\Omega = \Omega' \times (0,H)$ with a domain $\Omega' \subset \R^{2}$.

We consider a characteristic length $L$ and a characteristic velocity $V$.  We denote by $\delta = \frac{H}{L} \ll 1$ the ratio between the height $H$ and the characteristic length $L$ in the $(x_{1},x_{2})$-directions.  We set $T = \frac{V}{L}$ as the characteristic time scale and, due to the geometry of the domain under consideration, we rescale the third component of the spatial variable and the third component of the velocity by $\delta$.  That is,
\begin{equation*}
\begin{alignedat}{3}
x_{i} &= L x_{i,*}, && \quad v_{i} && = V v_{i,*}, \text{ for } i = 1,2, \\
x_{3} & = \delta L x_{3,*}, && \quad v_{3} && = \delta V v_{3,*},
\end{alignedat}
\end{equation*}
where the variables with $*$-subscript denote nondimensionalized variables.  In the following, we use the notation $\pd_{i} := \pd_{x_{i,*}}$ for $i = 1,2,3$, and $\nabla_{*} = (\pd_{1}, \, \pd_{2}, \, \pd_{3} \,)^{\top}$.  Let us consider a constant mobility $m(\varphi) = m$ and define
\begin{align*}
\eps = \eps_{*} L, \quad \mu = \frac{\sigma}{L} \mu_{*}, \quad \mathrm{Pe} = \frac{V L^{2}}{\sigma m},
\end{align*}
where $\mathrm{Pe}$ is the P\'{e}lect number.  Then, the Cahn--Hilliard part \eqref{GravNS:varphi}-\eqref{GravNS:mu} and the Neumann boundary conditions become
\begin{subequations}\label{Nondim:CH}
\begin{alignat}{3}
\pd_{t_{*}} \varphi + \nabla_{*} \varphi \cdot \bm{v}_{*} & = \frac{1}{\mathrm{Pe}} \left ( \pd_{1}^{2} \mu_{*} + \pd_{2}^{2} \mu_{*} + \frac{1}{\delta^{2}} \pd_{3}^{2} \mu_{*} \right )  && \text{ in } \Omega, \label{Nondim:CH:varphi} \\
\mu_{*} & = \frac{1}{\eps_{*}} \Psi'(\varphi) - \eps_{*} \left ( \pd_{1}^{2} \varphi + \pd_{2}^{2} \varphi + \frac{1}{\delta^{2}} \pd_{3}^{2} \varphi \right ) && \text{ in } \Omega, \label{Nondim:CH:mu} \\
0 & = \pd_{1} \varphi \nu_{1} + \pd_{2} \varphi \nu_{2} + \frac{1}{\delta} \pd_{3} \varphi \nu_{3} && \text{ on } \Gamma, \label{Nondim:CH:varphi:bdy}  \\
0 & = \pd_{1} \mu_{*} \nu_{1} + \pd_{2} \mu_{*} \nu_{2} + \frac{1}{\delta} \pd_{3} \mu_{*} \nu_{3} && \text{ on } \Gamma. \label{Nondim:CH:mu:bdy}
\end{alignat}
\end{subequations}
Since $\bm{v}_{*}$, $\varphi$, and $\mu_{*}$ depend on $\delta$ via the third spatial component, we assume that there exists an asymptotic expansion in $\delta$, i.e.,
\begin{align*}
v_{j,*} & = v_{j,0} + \delta v_{j,1} + \delta^{2} v_{j,2} + \text{ h.o.t.}, \quad \text{ for } j = 1,2,3, \\
\varphi & = \varphi_{0} + \delta \varphi_{1} + \delta^{2} \varphi_{2} + \text{ h.o.t.}, \\
\mu_{*} & = \mu_{0} + \delta \mu_{1} + \delta^{2} \mu_{2} + \text{ h.o.t.}.
\end{align*}
We will substitute these expansions into \eqref{Nondim:CH} and solve them order by order.  On the surfaces $\{x_{3,*} = 0\}$ and $\{x_{3,*} = 1 \}$, as $\nu_{1} = \nu_{2} = 0$ we obtain from \eqref{Nondim:CH:varphi:bdy}-\eqref{Nondim:CH:mu:bdy} for all orders $j = 0, 1, 2 \dots$,
\begin{equation}\label{Expansion:CH:bdy}
\begin{aligned}
\pd_{3}\varphi_{j}(x_{1,*},x_{2,*},0) & = \pd_{3}\mu_{j}(x_{1,*},x_{2,*},0) = 0, \\
\pd_{3}\varphi_{j}(x_{1,*},x_{2,*},1) & = \pd_{3} \mu_{j}(x_{1,*},x_{2,*},1) = 0.
\end{aligned}
\end{equation}
Meanwhile, to orders $\OO(\frac{1}{\delta^{2}})$ and  $\OO(\frac{1}{\delta})$, we obtain from \eqref{Nondim:CH:varphi}-\eqref{Nondim:CH:mu},
\begin{align*}
\pd_{3}^{2} \mu_{0} = \pd_{3}^{2} \varphi_{0} = 0, \quad \pd_{3}^{2} \mu_{1} = \pd_{3}^{2} \varphi_{1} = 0.
\end{align*}
Upon integrating with respect to $x_{3,*}$ and using the conditions \eqref{Expansion:CH:bdy} we have that $\mu_{0}$, $\mu_{1}$, $\varphi_{0}$, and $\varphi_{1}$ are independent of $x_{3,*}$.  Then, to order $\OO(1)$ we obtain from \eqref{Nondim:CH:varphi}-\eqref{Nondim:CH:mu},
\begin{align*}
\pd_{t_{*}} \varphi_{0} + \pd_{1} \varphi_{0} v_{1,0} + \pd_{2} \varphi_{0} v_{2,0} & = \frac{1}{\mathrm{Pe}} \left ( \pd_{1}^{2} \mu_{0} + \pd_{2}^{2} \mu_{0} + \pd_{3}^{2} \mu_{2} \right ), \\
\mu_{0} & = \frac{1}{\eps_{*}} \Psi'(\varphi_{0}) - \eps_{*} \left ( \pd_{1}^{2}\varphi_{0} + \pd_{2}^{2} \varphi_{0} + \pd_{3}^{2}\varphi_{2} \right ).
\end{align*}
Integrating the above equations with respect to $x_{3,*}$ from $0$ to $1$, and using the condition \eqref{Expansion:CH:bdy} leads to 
\begin{subequations}\label{Limit:deltato0:CHpart}
\begin{align}
\pd_{t_{*}} \varphi_{0} + \pd_{1} \varphi_{0} \overline{v_{1,0}} + \pd_{2} \varphi_{0} \overline{v_{2,0}} & = \frac{1}{\mathrm{Pe}} \left ( \pd_{1}^{2} \mu_{0} + \pd_{2}^{2} \mu_{0} \right ), \\
\mu_{0} & = \frac{1}{\eps_{*}} \Psi'(\varphi_{0}) - \eps_{*} \left ( \pd_{1}^{2}\varphi_{0} + \pd_{2}^{2} \varphi_{0} \right ),
\end{align}
\end{subequations}
where
\begin{align}\label{defn:meanvelo}
\overline{v_{j,0}}(x_{1,*},x_{2,*}) := \int_{0}^{1} v_{j,0}(x_{1,*},x_{2,*},s) \ds , \quad j = 1,2,
\end{align}
denotes the components of the mean velocity $\overline{\bm{v}} = (\overline{v_{1,0}}, \overline{v_{2,0}})^{\top}$.  In particular, in the limit $\delta \to 0$, we obtain a two-dimensional Cahn--Hilliard system convected by the mean velocity $\overline{\bm{v}}$ and complemented with Neumann boundary conditions on $\pd \Omega'$ from \eqref{Nondim:CH:varphi:bdy}-\eqref{Nondim:CH:mu:bdy}.

For the Navier--Stokes part \eqref{GravNS:div}-\eqref{GravNS:velo}, the continuity equation after the transformation becomes
\begin{align}\label{Nondim:Continuity}
\pd_{1} v_{1,*} + \pd_{2} v_{2,*} + \pd_{3} v_{3,*} = 0.
\end{align}
From the above computation with the Cahn--Hilliard part, we expect that a scale factor of $\frac{1}{\delta^{2}}$ will appear from the term $\div (2 \eta(\varphi) \der \bm{v})$ in \eqref{GravNS:velo}.  Thus, in order to retain the pressure, the body force and the capillary term in the limit $\delta \to 0$, we set
\begin{align*}
\rho(\varphi)  = \overline{\rho}_{2} \rho_{*}(\varphi), \quad \eta(\varphi) = \eta_{2} \eta_{*}(\varphi), \quad p = \frac{\eta_{2} V}{L \delta^{2}} p_{*}, \quad  \bm{G} = \overline{\rho}_{2} g \rho_{*}(\varphi) \hat{\bm{g}},
\end{align*}
and define
\begin{align*}
\mathrm{Ca} & = \frac{V \eta_{2}}{\delta^{2} \sigma}, \quad \mathrm{Re} = \frac{\overline{\rho}_{2} V L}{\eta_{2}}, \quad \mathrm{Bo} = \frac{\delta^{2} \overline{\rho}_{2} g L^{2}}{\eta_{2} V}
\end{align*}
where the capillary number $\mathrm{Ca}$ is the ratio between viscous forces and surface tension,  the Reynolds number $\mathrm{Re}$ is the ratio between inertial forces and viscous forces, and the Bond number $\mathrm{Bo}$ is the ratio between gravitational forces and viscous forces.  We now nondimensionalize the first component of the momentum equation \eqref{GravNS:velo}:
\begin{equation}\label{Nondim:NS:comp:1}
\begin{aligned}
0 & = \mathrm{Re} \left ( \pd_{t_{*}} (\rho_{*} v_{1,*}) + (\bm{v}_{*} \cdot \nabla_{*} )(\rho_{*} v_{1,*}) \right ) + \frac{1}{\delta^{2}} \pd_{1} p_{*} - \frac{\mathrm{Bo}}{\delta^{2}} \rho_{*} \hat{g}_{1} \\
& + \frac{\eps_{*}}{\delta^{2} \mathrm{Ca}} \left (  \pd_{1} (\pd_{1}\varphi \pd_{1}\varphi) + \pd_{2}(\pd_{1}\varphi \pd_{2}\varphi) + \frac{1}{\delta^{2}} \pd_{3}(\pd_{1} \varphi \pd_{3}\varphi) \right ) \\
& - \left ( \pd_{1} ( 2\eta_{*} \pd_{1} v_{1,*}) + \pd_{2}(\eta_{*}(\pd_{1}v_{2,*} + \pd_{2} v_{1,*} ) + \pd_{3} \left ( \eta_{*} \left ( \pd_{1} v_{3,*} + \frac{1}{\delta^{2}} \pd_{3} v_{1,*} \right ) \right ) \right ) \\
& - \frac{\mathrm{Re}}{\mathrm{Pe}} \frac{1-\lambda_{\rho}}{2} \left ( \pd_{1}(v_{1,*} \pd_{1} \mu_{*}) + \pd_{2}(v_{1,*} \pd_{2} \mu_{*}) + \frac{1}{\delta^{2}} \pd_{3}(v_{1,*} \pd_{3}\mu_{*}) \right ),
\end{aligned}
\end{equation}
where $\lambda_{\rho} = \frac{\overline{\rho}_{1}}{\overline{\rho}_{2}}$ denotes the density ratio.  We point out that, in the case $\overline{\rho}_{2} \geq \overline{\rho}_{1}$, i.e., fluid 2 is the heavier fluid, then the Atwood number $\mathrm{A} := \frac{\overline{\rho}_{2} - \overline{\rho}_{1}}{\overline{\rho}_{2} + \overline{\rho}_{1}}$ can be expressed as $\mathrm{A} = \frac{1-\lambda_{\rho}}{1+\lambda_{\rho}}$.  Similarly, for the second component of the momentum equation \eqref{GravNS:velo} we obtain
\begin{equation}\label{Nondim:NS:comp:2}
\begin{aligned}
0 & = \mathrm{Re} \left ( \pd_{t_{*}} (\rho_{*} v_{2,*}) + (\bm{v}_{*} \cdot \nabla_{*} )(\rho_{*} v_{2,*}) \right ) + \frac{1}{\delta^{2}} \pd_{2} p_{*} -  \frac{\mathrm{Bo}}{\delta^{2}} \rho_{*} \hat{g}_{2} \\
& + \frac{\eps_{*}}{\delta^{2} \mathrm{Ca}} \left (  \pd_{1} (\pd_{2}\varphi \pd_{1}\varphi) + \pd_{2}(\pd_{2}\varphi \pd_{2}\varphi) + \frac{1}{\delta^{2}} \pd_{3}(\pd_{2} \varphi \pd_{3}\varphi) \right ) \\
& - \left ( \pd_{1} ( \eta_{*} (\pd_{2} v_{1,*} + \pd_{1} v_{2,*})) + \pd_{2}(2 \eta_{*} \pd_{2}v_{2,*} ) + \pd_{3} \left ( \eta_{*} \left ( \pd_{2} v_{3,*} + \frac{1}{\delta^{2}} \pd_{3} v_{2,*} \right ) \right ) \right ) \\
& - \frac{\mathrm{Re}}{\mathrm{Pe}} \frac{1-\lambda_{\rho}}{2} \left ( \pd_{1}(v_{2,*} \pd_{1} \mu_{*}) + \pd_{2}(v_{2,*} \pd_{2} \mu_{*}) + \frac{1}{\delta^{2}} \pd_{3}(v_{2,*} \pd_{3}\mu_{*}) \right ).
\end{aligned}
\end{equation}
Meanwhile, for the third component of the momentum equation \eqref{GravNS:velo} we have
\begin{equation}\label{Nondim:NS:comp:3}
\begin{aligned}
0 & = \delta \mathrm{Re} \left ( \pd_{t_{*}} (\rho_{*} v_{3,*}) + (\bm{v}_{*} \cdot \nabla_{*} )(\rho_{*} v_{3,*}) \right ) + \frac{1}{\delta^{3}} \pd_{3} p_{*} -  \frac{\mathrm{Bo} }{\delta^{2}} \rho_{*} \hat{g}_{3} \\
& + \frac{\eps_{*}}{\delta^{3} \mathrm{Ca}} \left (  \pd_{1} (\pd_{3}\varphi \pd_{1}\varphi) + \pd_{2}(\pd_{3}\varphi \pd_{2}\varphi) + \frac{1}{\delta^{2}} \pd_{3}(\pd_{3} \varphi \pd_{3}\varphi) \right ) \\
& - \left ( \pd_{1} \left ( \eta_{*} \left  ( \frac{1}{\delta} \pd_{3} v_{1,*} +  \delta \pd_{1} v_{3,*} \right ) \right ) + \pd_{2} \left ( \eta_{*} \left ( \frac{1}{\delta} \pd_{3}v_{2,*} + \delta \pd_{2} v_{3,*} \right ) \right )  +  \frac{1}{\delta} \pd_{3} \left ( 2  \eta_{*} \pd_{3} v_{3,*} \right ) \right ) \\
& - \frac{\mathrm{Re}}{\mathrm{Pe}} \frac{1-\lambda_{\rho}}{2} \left ( \delta \pd_{1}(v_{3,*} \pd_{1} \mu_{*}) + \delta \pd_{2}(v_{3,*} \pd_{2} \mu_{*}) + \frac{1}{\delta} \pd_{3}(v_{3,*} \pd_{3}\mu_{*}) \right ).
\end{aligned}
\end{equation}
The no-slip boundary condition becomes
\begin{align*}
v_{1,*} = 0, \quad v_{2,*} = 0, \quad v_{3,*} = 0 \text{ on } \Gamma,
\end{align*}
and thus on the surfaces $\{x_{3,*} = 0 \}$ and $\{x_{3,*} = 1\}$ we have
\begin{align}\label{Expansion:velo:bdy}
v_{3,0}(x_{1,*},x_{2,*},0) = v_{3,0}(x_{1,*},x_{2,*},1) = 0.
\end{align}
The procedure to obtain a set of equations from the Navier--Stokes part in the limit $\delta \to 0$ is similar to what we have performed for the Cahn--Hilliard part.  In the following, we will only sketch the details.  Let $p_{*} = p_{0} + \delta p_{1} + \delta^{2} p_{2} + \text{ h.o.t.}$ denote an asymptotic expansion of the pressure.  Due to the fact that $\varphi_{0}$ and $\varphi_{1}$ are independent of $x_{3,*}$, to order $\OO(\frac{1}{\delta^{3}})$ we find that \eqref{Nondim:NS:comp:3} yields
\begin{align*}
\pd_{3}p_{0} = 0,
\end{align*}
and thus $p_{0}$ is independent of $x_{3,*}$.  Similarly, thanks to the fact that $\pd_{3}\varphi_{0} = \pd_{3}\varphi_{1} = \pd_{3}\mu_{0} = 0$, to order $\OO(\frac{1}{\delta^{2}})$ we obtain from \eqref{Nondim:NS:comp:1} and \eqref{Nondim:NS:comp:2},
\begin{align*}
0 & = \pd_{i} p_{0} - \mathrm{Bo} \, \rho_{*} \hat{g}_{i} + \frac{\eps_{*}}{\mathrm{Ca}} \left ( \pd_{1}(\pd_{1}\varphi_{0} \pd_{i}\varphi_{0}) + \pd_{2}(\pd_{2}\varphi_{0} \pd_{i}\varphi_{0}) + \pd_{3}(\pd_{3}\varphi_{2} \pd_{i}\varphi_{0}) \right ) + \eta_{*} \pd_{3}^{2} v_{i,0}
\end{align*}
for $i = 1,2$.  Integrating the above equation with respect to $x_{3,*}$ from $0$ to $1$, and using the conditions \eqref{Expansion:CH:bdy} and \eqref{Expansion:velo:bdy} leads to 
\begin{align*}
\eta_{*} v_{i,0}(x_{1,*},x_{2,*},s) = \frac{1}{2} s(s-1) \left ( \pd_{i} p_{0} - \mathrm{Bo} \, \rho_{*} \hat{g}_{i} + \frac{\eps_{*}}{\mathrm{Ca}} \left ( \pd_{1}(\pd_{1}\varphi_{0} \pd_{i}\varphi_{0}) + \pd_{2} (\pd_{2}\varphi_{0} \pd_{i} \varphi_{0}) \right ) \right ),
\end{align*}
for $i = 1,2$.  Dividing by $\eta_{*}$ and integrating over $s$ from $0$ to $1$ leads to the equation for the mean velocity $\overline{\bm{v}} = (\overline{v_{1,0}}, \overline{v_{2,0}})$, (recall \eqref{defn:meanvelo}):
\begin{align*}
\overline{v_{i,0}} = -\frac{1}{12 \eta_{*}}\left ( \pd_{i} p_{0} - \mathrm{Bo} \, \rho_{*} \hat{g}_{i} + \frac{\eps_{*}}{\mathrm{Ca}} \left ( \pd_{1}(\pd_{1}\varphi_{0} \pd_{i}\varphi_{0}) + \pd_{2} (\pd_{2}\varphi_{0} \pd_{i} \varphi_{0}) \right ) \right )
\end{align*}
for $i = 1,2$.  Furthermore, thanks to \eqref{Nondim:Continuity} and the condition \eqref{Expansion:velo:bdy}, we obtain
\begin{align*}
0 = \int_{0}^{1} \pd_{3} v_{3,0}(x_{1,*},x_{2,*},s) \ds = -\int_{0}^{1} \sum_{i=1,2} \pd_{i} v_{i,0}(x_{1,*},x_{2,*},s) \ds = - \pd_{1} \overline{v_{1,0}} - \pd_{2} \overline{v_{2,0}}.
\end{align*}
Thus, from the Navier--Stokes part \eqref{Nondim:Continuity}-\eqref{Nondim:NS:comp:3} we obtain 
\begin{subequations}\label{Limit:deltato0:NSpart}
\begin{align}
\overline{\div} \overline{\bm{v}} & = 0, \\
\overline{\bm{v}} & = - \frac{1}{12 \eta_{*}(\varphi_{0})} \left ( \overline{\nabla} p_{0} - \mathrm{Bo} \, \rho_{*}(\varphi_{0}) \hat{\bm{g}} + \frac{\eps_{*}}{\mathrm{Ca}} \overline{\div} \left ( \overline{\nabla} \varphi_{0} \otimes \overline{\nabla} \varphi_{0} \right ) \right ),
\end{align}
\end{subequations}
where $\overline{\nabla} f = (\pd_{1}f, \pd_{2}f)^{\top}$ denotes the two-dimensional gradient of a scalar function $f$, and $\overline{\div} \bm{f} = \pd_{1}f_{1} + \pd_{2} f_{2}$ denotes the two-dimensional divergence of a vector function $\bm{f}$.  Here, we reuse the notation $\hat{\bm{g}} = (\hat{g}_{1}, \hat{g}_{2})^{\top}$.

Dropping the subscripts and combining \eqref{Limit:deltato0:CHpart} and \eqref{Limit:deltato0:NSpart} leads to the following nondimensionalized Hele--Shaw--Cahn--Hilliard model:
\begin{subequations}\label{HSCHnew}
\begin{align}
\div \overline{\bm{v}} & = 0, \label{Nondim:HSCH:div} \\
12 \eta(\varphi) \overline{\bm{v}} & = - \nabla p + \mathrm{Bo} \, \rho(\varphi) \hat{\bm{g}} - \frac{\eps}{\mathrm{Ca}} \div (\nabla \varphi \otimes \nabla \varphi), \label{Nondim:HSCH:velo} \\
\pd_{t} \varphi + \nabla \varphi \cdot \overline{\bm{v}}  & =  \frac{1}{\mathrm{Pe}} \Laplace \mu, \label{Nondim:NSCH:varphi} \\
\mu & = \frac{1}{\eps} \Psi'(\varphi) - \eps_{*} \Laplace \varphi, \label{Nondim:HSCH:mu}
\end{align}
\end{subequations}
where $\div \cdot$, $\nabla \cdot$, and $\Laplace \cdot$ are to be interpreted as the two-dimensional divergence, gradient and Laplace operators, respectively.  Using the identity
\begin{align*}
\nabla \left ( \frac{1}{\eps} \Psi(\varphi) + \frac{\eps}{2} \abs{\nabla \varphi}^{2} \right ) = \left ( \frac{1}{\eps} \Psi'(\varphi) - \eps \Laplace \varphi \right ) \nabla \varphi + \eps \div \left ( \nabla \varphi \otimes \nabla \varphi \right )
\end{align*}
and defining the modified pressures
\begin{align*}
q = p + \frac{1}{\eps} \Psi(\varphi) + \frac{\eps}{2} \abs{\nabla \varphi}^{2}, \quad r = p +  \frac{1}{\eps} \Psi(\varphi) + \frac{\eps}{2} \abs{\nabla \varphi}^{2} + \mu \varphi,
\end{align*}
we obtain two variants of \eqref{Nondim:HSCH:velo}:
\begin{subequations}
\begin{align}
12 \eta(\varphi) \overline{\bm{v}} & = - \nabla q + \mathrm{Bo} \, \rho(\varphi) \hat{\bm{g}} + \frac{1}{\mathrm{Ca}} \mu \nabla \varphi, \label{Nondim:HSCH:velo:alternate:q}\\
12 \eta(\varphi) \overline{\bm{v}} & = - \nabla r + \mathrm{Bo} \, \rho(\varphi) \hat{\bm{g}} - \frac{1}{\mathrm{Ca}} \varphi \nabla \mu. \label{Nondim:HSCH:velo:alternate:r}
\end{align}
\end{subequations}
In the case where there is no density contrast, i.e., $\overline{\rho}_{1} = \overline{\rho}_{2}$, and the gravitational forces are neglected, the model \eqref{HSCHnew} with \eqref{Nondim:HSCH:velo:alternate:r} has been studied by Wang and Zhang in \cite{WangZhang} concerning strong well-posedness globally in time for two dimensions and locally in time for three dimensions, and by Wang and Wu in \cite{WangWu} concerning long-time behavior and well-posedness in three dimensions with well-prepared data.  

If, in addition, there is no viscosity contrast, i.e., $\eta_{1} = \eta_{2}$, then Feng and Wise established the global existence of weak solutions in two and three dimensions via the convergence of a fully discrete and energy stable  implicit finite element scheme in \cite{FengWise12}.  Uniqueness of weak solutions can be shown if additional regularity assumptions on the solutions are imposed, see \cite[Thm. 2.4]{FengWise12}, and the error analysis of the numerical scheme is performed in \cite{LiuCHHS}.  For the convergence analysis of finite difference schemes, we refer the reader to \cite{ChenCHHS,ChenLiuWangWise,Wise}. 

Meanwhile, Bosia, Conti and Grasselli proved that weak solutions to the Cahn--Hilliard--Brinkman model converge to a weak solution of the Hele--Shaw--Cahn--Hilliard  model in \cite{BosiaContiGrasselli}.  The Cahn--Hilliard--Brinkman model is a related system where an addition term of the form $-\div( \nu \der \overline{\bm{v}})$ is added to the left-hand side of \eqref{Nondim:HSCH:velo:alternate:r}.  Here, $\der \overline{\bm{v}} := \frac{1}{2} (\nabla \overline{\bm{v}} + (\nabla \overline{\bm{v}})^{\top})$ is the rate of deformation tensor and $\nu  > 0$ is the approximation parameter.  Error estimates in terms of $\nu$ between the Cahn--Hilliard--Brinkman model and the Hele--Shaw--Cahn--Hilliard have also been derived in two dimensions.  A nonlocal version of the results of \cite{BosiaContiGrasselli} has been recently established in \cite{DellaPortaGrasselli}.

Recently, the asymptotic behavior $\eps \to 0$ of global weak solutions \footnote{We point out that the $L^{2}$ temporal regularity for the time derivative $\pd_{t}\varphi$ (written as $\pd_{t}c^{\eps}$) in \cite{Fei} may be a typo, cf. Theorem \ref{thm:Existence} below.} to the Hele--Shaw--Cahn--Hilliard model \eqref{HSCHnew} with \eqref{Nondim:HSCH:velo:alternate:q}, and the particular scaling $\frac{1}{\mathrm{Pe}} = \eps^{\alpha}$ for $0 \leq \alpha < 1$ and $\mathrm{Bo} = 0$ has been studied by Fei in \cite{Fei}, which employs the varifold approach of Chen \cite{Chen}; see also \cite{Abels, GKwak, MelchionnaRocca} and \cite[Appendix A]{AbelsRoger}.  In Section \ref{sec:Existence} below we will establish the global in time existence of weak solutions to \eqref{HSCHnew} (with the variant \eqref{Nondim:HSCH:velo:alternate:q} and a general body force $\bm{G}(\varphi)$ replacing $\mathrm{Bo} \, \rho(\varphi) \hat{\bm{g}}$) for two and three dimensions.

\subsection{Comparison with the Lee--Lowengrub--Goodman model}\label{sec:LLGModel}
In this section, we compare the model \eqref{HSCHnew} with the model of Lee, Lowengrub and Goodman \cite{LLG01}.  In the sequel, we will denote the mass-averaged velocity by $\bm{w}$.  Let $c$ denote an order parameter distinguishing the two fluid phases, with $\Omega_{1} := \{ c = 1\}$ and $\Omega_{2} = \{ c = 0 \}$.  Recalling $\rho_{i}$ and $\bm{v}_{i}$ as the actual mass density and individual velocity of fluid $i$, $i = 1,2$, the total density $\rho(c)$ and mass-averaged velocity $\bm{w}$ are defined as
\begin{align*}
\rho(c) = \frac{1}{\rho_{1}^{-1} c + \rho_{2}^{-1} (1-c)}, \quad \rho(c) \bm{w} = \rho_{1} \bm{v}_{1} + \rho_{2} \bm{v}_{2}.
\end{align*}
Let $\eta(c) = \eta_{1} c + \eta_{2}(1-c)$ denote the interpolation of the two viscosities.  We introduce the coefficient
\begin{align}\label{defn:alpha}
\alpha := \frac{1}{\rho_{1}} - \frac{1}{\rho_{2}} = -\frac{\rho'(c)}{(\rho(c))^{2}},
\end{align}
and let $g$ denote the modulus of the gravity vector $\bm{g} = g \hat{\bm{g}}$ with unit vector $\hat{\bm{g}}$.  Then, the nondimensionalized Hele--Shaw--Cahn--Hilliard equations of \cite[Equ. (2.18)-(2.21)]{LLG01} are 
\begin{subequations}\label{LLG}
\begin{alignat}{3}
\div \bm{w} - \frac{\alpha}{\mathrm{Pe}} \Laplace \mu & = 0, \label{LLG:div} \\
\rho(c) (\pd_{t}c + \nabla c \cdot \bm{w}) - \frac{1}{\mathrm{Pe}} \Laplace \mu & = 0, \label{LLG:phase} \\
\bm{w} + \frac{1}{12 \eta(c)} \left ( \nabla p + \frac{\mathrm{Ch}}{\mathrm{Ma}} \div(\rho(c) \nabla c \otimes \nabla c)  - \rho(c) \hat{\bm{g}} \right ) & = \bm{0}, \label{LLG:velo} \\
\mu - f'_{0}(c) +  \frac{\mathrm{Ch}}{\rho(c)} \div (\rho(c) \nabla c) -  \mathrm{Ma} \,  \alpha p& = 0, \label{LLG:chem}
\end{alignat}
\end{subequations}
where $f_{0} = c^{2}(1-c^{2})$ has two minima at $c = 0$ and $c = 1$, and the dimensionless constants $\mathrm{Pe}$, $\mathrm{Ch}$ and $\mathrm{Ma}$ are the P\'{e}lect number, the Cahn number and the Mach number, respectively.  

Here we point out that the continuity equation \eqref{Nondim:HSCH:div} and the equation for the chemical potential \eqref{Nondim:HSCH:mu} in the volume-averaged model \eqref{HSCHnew} are considerably simpler than their counterparts \eqref{LLG:div} and \eqref{LLG:chem} in the mass-averaged model \eqref{LLG}.  In particular, the pressure appears explicitly in \eqref{LLG:chem} and compressibility effects may be introduced as the mass-averaged velocity $\bm{w}$ need not be solenoidal.  In contrast, these features are not present in \eqref{HSCHnew}.

\section{Analysis of the volume-averaged model}\label{sec:analysis}
\subsection{Sharp interface asymptotics}\label{sec:asymp}
We now consider the sharp interface asymptotics of the nondimensional model \eqref{HSCHnew} (using $\bm{v}$ to denote the averaged velocity $\overline{\bm{v}}$ and $\sigma$ to denote the reciprocal of the capillary number $\mathrm{Ca}$) in the following setting:
\begin{assump}\label{assump:SharpInterfaceAsym}
\
\begin{itemize}
\item We set $\mathrm{Pe} = \frac{1}{\eps}$ and consider a more general function $\bm{G}$ replacing the term $\mathrm{Bo} \, \rho(\varphi) \hat{\bm{g}}$, where $\bm{G}$ depends only on $\varphi$ but not high order derivatives.
\item We assume that there is a family $(\varphi_{\eps}, \bm{v}_{\eps}, p_{\eps}, \mu_{\eps})_{\eps > 0}$ of solutions to \eqref{HSCHnew}, which are sufficiently smooth.  For small $\eps$, the domain $\Omega$ can be divided into two open subdomains $\Omega^{\pm}(\eps)$, separated by an interface $\Sigma(\eps)$, given as the zero-level set of $\varphi_{\eps}$, that does not intersect with $\pd \Omega = \Gamma$.  
\item We assume that $(\varphi_{\eps}, \bm{v}_{\eps}, p_{\eps}, \mu_{\eps})_{\eps > 0}$ have an asymptotic expansion in $\eps$ in the bulk regions away from $\Sigma(\eps)$ (the outer expansion), and another expansion in the interfacial region close to $\Sigma(\eps)$ (the inner expansion). 
\item We assume that the zero level sets of $\varphi_{\eps}$ converge to a limiting hypersurface $\Sigma$ moving with normal velocity $\mathcal{V}$ as $\eps \to 0$.
\item We rescale the potential $\Psi$ such that
\begin{align}\label{rescaled:potential}
\int_{-1}^{1}  \sqrt{2 \Psi(s)} \ds = 1.
\end{align}
For example, the classical quartic double-well potential $\Psi(s) = \frac{1}{4}(1-s^{2})^{2}$ is rescaled to $\Psi(s) = \frac{3}{2 \sqrt{2}} \frac{1}{4} (1-s^{2})^{2}$.
\end{itemize}
\end{assump}
The equations we study are
\begin{subequations}\label{Asymp:HSCH}
\begin{align}
\div \bm{v} & = 0 \label{HSCH:div}, \\
12 \eta(\varphi) \bm{v} & = - \nabla p + \bm{G}(\varphi) - \sigma \eps \div (\nabla \varphi \otimes \nabla \varphi ) \label{HSCH:pressure}, \\
\pd_{t}\varphi + \nabla \varphi \cdot \bm{v} & = \eps \Laplace \mu \label{HSCH:varphi}, \\
\mu & = \frac{1}{\eps} \Psi'(\varphi) - \eps \Laplace \varphi \label{HSCH:mu}.
\end{align}
\end{subequations}
The idea of the method is to plug the outer and inner expansions in the model equations and solve them order by order, and in addition we have to define a suitable region where these expansions should match up.  For $\alpha = -2, -1, 0, 1, \dots$, we will use the notation $\eqref{HSCH:div}_{O}^{\alpha}$ and $\eqref{HSCH:div}_{I}^{\alpha}$ to denote the terms resulting from the order $\alpha$ outer and inner expansions of \eqref{HSCH:div}, respectively.

\subsubsection{Outer expansion}\label{sec:OuterExp}
We assume that $(\bm{v}_{\eps}, p_{\eps}, \varphi_{\eps}, \mu_{\eps})$ have the following outer expansions
\begin{equation*}
\begin{alignedat}{4}
\bm{v}_{\eps} & = \bm{v}_{0} + \eps \bm{v}_{1} + \text{ h.o.t.}, && \quad p_{\eps} && = \frac{1}{\eps} p_{-1} + p_{0} + \text{ h.o.t.}, \\
\varphi_{\eps} & = \varphi_{0} + \eps \varphi_{1} + \text{ h.o.t.}, && \quad \mu_{\eps} && = \mu_{0} + \eps \mu_{1} + \text{ h.o.t.}.
\end{alignedat}
\end{equation*}
To leading order $\eqref{HSCH:pressure}_{O}^{-1}$ we obtain 
\begin{align}\label{outer:-1:p-1}
\bm{0} = \nabla p_{-1},
\end{align}
and so $p_{-1}$ is constant in the bulk regions.  Meanwhile $\eqref{HSCH:mu}_{O}^{-1}$ gives
\begin{align*}
\Psi'(\varphi_{0}) = 0.
\end{align*}
The stable solutions to the above equation are the minima of $\Psi$, which yields that $\varphi_{0} = \pm 1$.  This allows us to define the bulk fluid domains $\Omega_{1} := \{ \varphi(x) = -1\}$ and $\Omega_{2} := \{ \varphi(x) = 1\}$.  To leading order we obtain from $\eqref{HSCH:div}_{O}^{0}$
\begin{align*}
\div \bm{v}_{0} = 0,
\end{align*} 
and to first order we obtain from $\eqref{HSCH:pressure}_{O}^{0}$
\begin{align*}
12 \eta(\varphi_{0}) \bm{v}_{0} = - \nabla p_{0} + \bm{G}(\varphi_{0}).
\end{align*}

\subsubsection{Inner expansions}
By assumption, $\Sigma$ is the limiting hypersurface of the zero level sets of $\varphi_{\eps}$.  In order to study the limiting behavior close to $\Sigma$ we introduce a new coordinate system, which involves the signed distance function $d(x)$ to $\Sigma$.  Setting $z = \frac{d}{\eps}$ as the rescaled distance variable to $\Sigma$, and using the convention that $d(x) < 0$ in $\Omega_{1}$, and $d(x) > 0$ in $\Omega_{2}$, we see that the gradient $\nabla d$ points from $\Omega_{1}$ to $\Omega_{2}$, and we may use $\nabla d$ on $\Sigma$ to denote the unit normal of $\Sigma$, pointing from $\Omega_{1}$ to $\Omega_{2}$.

Let $\para(t,s)$ denote a parametrization of $\Sigma$ with tangential coordinates $s$, and let $\bm{\nu}$ denote the unit normal of $\Sigma$, pointing into $\Omega_{2}$.  Then, in a tubular neighborhood of $\Sigma$, for a sufficiently smooth function $f(x)$, we have
\begin{align*}
f(x) = f(\para(t,s) + \eps z \bm{\nu}(\para(t,s))) =: F(t,s,z).
\end{align*}
In this new $(t,s,z)$-coordinate system, the following change of variables apply, see \cite{GarckeStinner06},
\begin{align*}
\pd_{t} f & = -\frac{1}{\eps} \mathcal{V} \pd_{z} F + \text{ h.o.t.}, \\
\nabla_{x} f & = \frac{1}{\eps}\pd_{z} F \bm{\nu} + \surf F + \text{ h.o.t.},
\end{align*}
where $\mathcal{V}$ is the normal velocity of $\Sigma$, $\surf g$ denotes the surface gradient of $g$ on $\Sigma$ and h.o.t. denotes higher order terms with respect to $\eps$.  In particular, we have
\begin{align*}
\Laplace f  = \div_{x} (\nabla_{x} f) & = \frac{1}{\eps^{2}} \pd_{zz}F + \frac{1}{\eps}\underbrace{\div_{\Sigma} (\pd_{z}F \bm{\nu})}_{= - \kappa \pd_{z}F} + \text{ h.o.t.},
\end{align*}
where $\kappa = - \div_{\Sigma} \bm{\nu}$ is the mean curvature of $\Sigma$.  If $\bm{v}$ is a vector-valued function with $\bm{V}(t,s,z) = \bm{v}(x)$ for $x$ in a tubular neighborhood of $\Sigma$, then we obtain
\begin{align*}
\div_{x} \bm{v} = \frac{1}{\eps}\pd_{z} \bm{V} \cdot \bm{\nu} + \div_{\Sigma} \bm{V} + \text{ h.o.t.}.
\end{align*}
The inner variables of $(\bm{v}_{\eps}, p_{\eps}, \varphi_{\eps}, \mu_{\eps})$ are denoted as $(\bm{V}_{\eps}, P_{\eps}, \Phi_{\eps}, \Xi_{\eps})$ with the inner expansion
\begin{equation}\label{InnerExpansion}
\begin{aligned}
F_{\eps}(t,s,z) & = F_{0}(t,s,z) + \eps F_{1}(t,s,z) + \text{ h.o.t.}, \quad \text{ for } F_{\eps} \in \{ \bm{V}_{\eps}, \Phi_{\eps}, \Xi_{\eps} \}, \\
P_{\eps}(t,s,z) &= \frac{1}{\eps} P_{-1}(t,s,z) + P_{0}(t,s,z) + \text{ h.o.t.}.
\end{aligned}
\end{equation} 
Since the zero level sets of $\varphi_{\eps}$ converge to $\Sigma$, we additionally impose that
\begin{align}\label{Phi0:z=0}
\Phi_{0}(t,s, z=0) = 0.
\end{align}
In order to match the inner expansions valid in the interfacial region to the outer expansions of Section \ref{sec:OuterExp} we employ the matching conditions, see \cite{GarckeStinner06},
\begin{align}
\label{MatchingCond1}
\lim_{z \to \pm \infty} F_{0}(t,s,z) &= f_{0}^{\pm}(t,x), \\
\label{MatchingCond2}
\lim_{z \to \pm \infty} \pd_{z}F_{0}(t,s,z) &= 0 ,\\
\label{MatchingCond3}
\lim_{z \to \pm \infty} \pd_{z} F_{1}(t,s,z) &= \pd_{\bm{\nu}} f_{0}^{\pm}(t,x) ,
\end{align}
where $f_{0}^{\pm}(t,x):= \lim_{\delta \searrow 0} f_{0}(t,x \pm \delta \bm{\nu})$ for $x \in \Sigma$.  For the pressure, we have
\begin{align}
\lim_{z \to \pm \infty} P_{-1}(t,s,z) &= p_{-1}^{\pm}(t,x), \label{Matching1:Pressure} \\
\lim_{z \to \pm \infty} \left ( P_{0}(t,s,z) - z \pdnu p_{-1}^{\pm}(t,x) \right ) & = p_{0}^{\pm}(t,x). \label{Matching2:Pressure}
\end{align}
We will employ the following notation:  Let $\delta > 0$ and for $x \in \Sigma$ with $x - \delta \bm{\nu} \in \Omega_{1}$ and $x + \delta \bm{\nu} \in \Omega_{2}$, we denote the jump of a quantity $f$ across the interface by
\begin{align}\label{defn:jump}
\jump{f} := \lim_{\delta \searrow 0} f(t,x + \delta \bm{\nu}) - \lim_{\delta \searrow 0} f(t,x - \delta \bm{\nu}).
\end{align}
Then, the expansions of \eqref{HSCH:div}, \eqref{HSCH:varphi} and \eqref{HSCH:mu} in terms of the inner variables are
\begin{subequations}
\begin{align}
\frac{1}{\eps} \pd_{z} \bm{V} \cdot \bm{\nu} + \div_{\Sigma} \bm{V} + \text{ h.o.t.} & = 0, \label{inner:div} \\
 \frac{1}{\eps} \left ( -\mathcal{V} + \bm{V} \cdot \bm{\nu} \right ) \pd_{z}\Phi  - \pd_{zz} \Xi  + \text{ h.o.t.}  & = 0, \label{inner:conv} \\
 \Xi - \frac{1}{\eps} \Psi'(\Phi) + \frac{1}{\eps} \pd_{zz}\Phi - \kappa \pd_{z}\Phi + \text{ h.o.t.} & = 0. \label{inner:mu}
\end{align}
\end{subequations}
For the tensor product $\eps \div (\nabla \varphi \otimes \nabla \varphi)$ we obtain the formula
\begin{align*}
\div ( \eps \nabla \varphi \otimes \nabla \varphi) & = \frac{1}{\eps^{2}} \pd_{z}( (\pd_{z}\Phi)^{2} \bm{\nu}) + \frac{1}{\eps}  \pd_{z}(\pd_{z}\Phi \surf \Phi) + \frac{1}{\eps} \div_{\Sigma} ( (\pd_{z}\Phi)^{2} \bm{\nu} \otimes \bm{\nu})\\
& + \div_{\Sigma} ( \pd_{z}\Phi (\bm{\nu} \otimes \surf \Phi + \surf \Phi \otimes \bm{\nu} )) + \text{ h.o.t.},
\end{align*}
so that the expansion of \eqref{HSCH:pressure} becomes
\begin{equation}\label{inner:pressure}
\begin{aligned}
& 12 \eta(\Phi) \bm{V} + \left ( \frac{1}{\eps} \pd_{z}P \nu + \surf P \right )  - \bm{G}(\Phi) \\
& + \frac{1}{\eps^{2}} \pd_{z}( \sigma (\pd_{z}\Phi)^{2} \bm{\nu}) + \frac{1}{\eps} \sigma \pd_{z}(\pd_{z}\Phi \surf \Phi) + \frac{1}{\eps} \div_{\Sigma}( \sigma (\pd_{z}\Phi)^{2} \bm{\nu} \otimes \bm{\nu}) \\
& + \div_{\Sigma} ( \sigma \pd_{z}\Phi (\bm{\nu} \otimes \surf \Phi + \sigma \surf \Phi \otimes \bm{\nu} )) + \text{ h.o.t.} = \bm{0}.
\end{aligned}
\end{equation}
\subsubsection{Expansions to leading order}
To leading order we obtain from $\eqref{HSCH:mu}_{I}^{-1}$
\begin{align}\label{ODE}
\Psi'(\Phi_{0}) - \pd_{zz} \Phi_{0} = 0.
\end{align}
This is a second order equation in $z$ and together with the conditions $\lim_{z \to \pm \infty} \Phi_{0}(t,s,z) = \pm 1$, and $\Phi_{0}(t,s,0) = 0$ we obtain a unique solution $\Phi_{0}(z)$ to \eqref{ODE} that is independent of $s$ and $t$, i.e., \eqref{ODE} can be viewed as an ordinary differential equation in $z$.  For the double-well potential $\Psi(s) = \frac{1}{4}(1-s^{2})^{2}$, the unique solution is given by $\Phi_{0}(z) = \tanh \left ( \frac{z}{\sqrt{2}} \right )$.  Furthermore, multiplying \eqref{ODE} by $\Phi_{0}'$, integrating and applying matching conditions \eqref{MatchingCond1} and \eqref{MatchingCond2} to $\Phi_{0}$ leads to the so-called equipartition of energy
\begin{align*}
\frac{1}{2} \abs{\Phi_{0}'(z)}^{2} = \Psi(\Phi_{0}(z)) \quad \forall z \in \R.
\end{align*}
By \eqref{rescaled:potential}, we see that
\begin{align}\label{equipartition:constant}
\int_{\R} \abs{\Phi_{0}'(z)}^{2} \dz = \int_{\R} 2 \Psi(\Phi_{0}(z)) \dz = \int_{-1}^{1}  \sqrt{2 \Psi(s)} \ds =1.
\end{align}
Then, to leading order $\eqref{HSCH:div}_{I}^{-1}$, we obtain
\begin{align}\label{pdzV0}
\pd_{z} \bm{V}_{0} \cdot \bm{\nu} = 0,
\end{align}
which implies that $\bm{V}_{0} \cdot \bm{\nu}$ is independent of $z$.  Integrating and applying the matching condition \eqref{MatchingCond1} to $\bm{V}_{0}$ yields
\begin{align*}
\jump{\bm{v}_{0}} \cdot \bm{\nu} = 0.
\end{align*}
Meanwhile, from $\eqref{HSCH:pressure}_{I}^{-2}$ we have
\begin{align*}
\pd_{z} P_{-1} \bm{\nu} +  \sigma \pd_{z}(\Phi_{0}')^{2} \bm{\nu} = \bm{0}.
\end{align*}
Taking the scalar product with $\bm{\nu}$ and upon integrating with respect to $z$ leads to
\begin{align*}
P_{-1}(t,s,z) = \hat{P}(t,s) - \sigma  (\Phi_{0}'(z))^{2},
\end{align*}
for some function $\hat{P}$ independent of $z$.  Sending $z \to \pm \infty$ and applying the matching condition \eqref{Matching1:Pressure} to $P_{-1}$ and \eqref{MatchingCond2} to $\Phi_{0}$, we see that 
\begin{align*}
p_{-1}^{-} = \hat{P}(t,s) = p_{-1}^{+}.
\end{align*}
In particular, the constant values of $p_{-1}$ in the bulk phase (see \eqref{outer:-1:p-1}) should match.  We take $p_{-1}^{\pm} = 0$ so that $P_{-1}$ is a function only in $z$ and 
\begin{align}\label{Inner:P-1}
P_{-1}(z) = - \sigma (\Phi_{0}'(z))^{2}.
\end{align}
To leading order $\eqref{HSCH:mu}_{I}^{-1}$ gives
\begin{align*}
\left ( -\mathcal{V} + \bm{V}_{0} \cdot \bm{\nu} \right ) \Phi_{0}' = \pd_{zz} \Xi_{0}.
\end{align*}
By \eqref{pdzV0}, $\bm{V}_{0} \cdot \bm{\nu}$ is independent of $z$, and so upon integrating and apply matching conditions \eqref{MatchingCond1} to $\Phi_{0}$ and \eqref{MatchingCond2} to $\Xi_{0}$, we obtain
\begin{align*}
2(-\mathcal{V} + \bm{v}_{0} \cdot \bm{\nu} ) = (-\mathcal{V} + \bm{v}_{0} \cdot \bm{\nu} ) \int_{\R} \Phi_{0}' \dz = \int_{\R} \pd_{zz} \Xi_{0} \dz = 0.
\end{align*}
This implies that
\begin{align}\label{pdzXi0}
\mathcal{V} = \bm{v}_{0} \cdot \bm{\nu}, \quad \pd_{z} \Xi_{0} = 0.
\end{align}

\subsubsection{Expansions to first order}
To first order, we obtain from $\eqref{HSCH:mu}_{I}^{0}$
\begin{align*}
\Xi_{0} = \Psi''(\Phi_{0}) \Phi_{1} - \sigma \pd_{zz} \Phi_{1} + \kappa \Phi_{0}'.
\end{align*}
Multiplying by $\Phi_{0}'$, integrating over $\R$ with respect to $z$ leads to
\begin{align}\label{Inner:mu:0:int}
\int_{-\infty}^{\infty} \Xi_{0}(t,s) \Phi_{0}'(z) \dz = \int_{-\infty}^{\infty} (\Psi'(\Phi_{0}))' \Phi_{1} - \pd_{zz} \Phi_{1} \Phi_{0}' + \kappa\abs{\Phi_{0}'}^{2} \dz.
\end{align}
Integration by parts, applying the matching conditions \eqref{MatchingCond1} and \eqref{MatchingCond2} applied to $\Phi_{0}$, and using that $\Psi'(\pm 1) = 0$, we see that
\begin{align*}
\int_{-\infty}^{\infty} (\Psi'(\Phi_{0}))' \Phi_{1} - \pd_{zz} \Phi_{1} \Phi_{0}' \dz & = \underbrace{[\Psi'(\Phi_{0}) \Phi_{1} - \pd_{z} \Phi_{1} \Phi_{0}']_{-\infty}^{\infty}}_{=0 \text{ by } (\ref{MatchingCond1}), (\ref{MatchingCond2})} - \int_{-\infty}^{\infty} \pd_{z} \Phi_{1} \underbrace{(\Psi'(\Phi_{0}) - \Phi_{0}'')}_{=0 \text{ by } (\ref{ODE})} \dz,
\end{align*}
and so the first two terms on the right-hand side of \eqref{Inner:mu:0:int} are zero.  Then, using \eqref{equipartition:constant} and \eqref{pdzXi0}, we obtain from \eqref{Inner:mu:0:int},
\begin{align}
\label{Inner:mu:0:int:interfacelaw}
2 \mu_{0} =  \kappa.
\end{align}
Next, using that $P_{-1}$ and $\Phi_{0}$ depend only on $z$, to first order we obtain from $\eqref{HSCH:pressure}_{I}^{-1}$
\begin{align*}
\bm{0} = \pd_{z}P_{0} \bm{\nu} + \sigma \pd_{z}(2 \Phi_{0}' \pd_{z}\Phi_{1}) \bm{\nu} + \sigma \div_{\Sigma} ((\Phi_{0}')^{2} \bm{\nu} \otimes \bm{\nu}).
\end{align*}
Taking the scalar product with $\bm{\nu}$, integrating and applying the matching condition \eqref{Matching2:Pressure} and using \eqref{equipartition:constant} leads to
\begin{align*}
0 = \jump{p_{0}}+ \sigma \left [ 2 \Phi_{0}' \pd_{z}\Phi_{1} \right ]_{-\infty}^{\infty} + \sigma \div_{\Sigma} \left ( \bm{\nu} \otimes \bm{\nu} \right ) \bm{\nu} = \jump{p_{0}}  - \sigma \kappa,
\end{align*}
where we used that $\div_{\Sigma} (\bm{\nu} \otimes \bm{\nu}) = - \kappa \bm{\nu}$.  Hence, the sharp interface limit of \eqref{Asymp:HSCH} is
\begin{subequations}\label{SIM:HSCHnew}
\begin{alignat}{3}
\div \bm{v}_{0} &= 0 && \text{ in } \left ( \Omega_{1} \cup \Omega_{2} \right ) \setminus \Sigma, \\
12 \eta(\varphi_{0}) \bm{v}_{0}  &= - \nabla p_{0} + \bm{G}(\varphi_{0}) && \text{ in }  \left ( \Omega_{1} \cup \Omega_{2} \right ) \setminus \Sigma, \\
\jump{\bm{v}_{0}} \cdot \bm{\nu} &= 0 && \text{ on } \Sigma, \\
\jump{p_{0}} &= \sigma \kappa  && \text{ on } \Sigma, \label{YoungLaplace} \\
\mathcal{V}&= \bm{v}_{0} \cdot \bm{\nu} && \text{ on } \Sigma.
\end{alignat}
\end{subequations}

\begin{remark}
We point out that the formal asymptotic analysis performed with the degenerate mobility 
\begin{align*}
m(\varphi) = (1-\varphi^{2})_{+}
\end{align*}
will yield the same sharp interface limit \eqref{SIM:HSCHnew}.  For more details, we refer to \cite{AGG,GLS}.
\end{remark}

\begin{remark}
If we use the variant \eqref{Nondim:HSCH:velo:alternate:q} of the velocity equation instead of \eqref{HSCH:pressure}, i.e.,
\begin{align}\label{HSCH:pressure:alt}
12 \eta(\varphi) \bm{v} = - \nabla q + \bm{G}(\varphi) + \sigma \mu \nabla \varphi,
\end{align}
then the outer and inner expansions of the pressure $q$ do not require a term scaling with $\frac{1}{\eps}$.  That is, we can consider
\begin{align*}
q_{\eps} = q_{0} + \eps q_{1} + \text{ h.o.t.}, \quad q_{\eps} = Q_{0} + \eps Q_{1} + \text{ h.o.t.}
\end{align*}
as the corresponding outer and inner expansions, respectively.  While the analysis for the outer expansions remains unchanged, from the leading order inner expansion $\eqref{HSCH:pressure:alt}_{I}^{-1}$ we obtain after taking the scalar product with $\bm{\nu}$ and integrating, and using \eqref{Inner:mu:0:int} and \eqref{Inner:mu:0:int:interfacelaw},
\begin{align*}
0 = \int_{-\infty}^{\infty} \pd_{z} Q_{0} - \sigma \Xi_{0} \Phi_{0}' \dz = \jump{q_{0}} - 2 \mu_{0} = \jump{q_{0}} - \sigma \kappa,
\end{align*}
which is the nondimensionalized Young--Laplace law \eqref{YoungLaplace} for the modified pressure $q$.
\end{remark}

\subsection{Sharp interface limit for the mass-averaged model}
It turns out that in choosing 
\begin{align*}
\mathrm{Pe} = \frac{1}{\eps}, \quad \mathrm{Ch} = \eps^{2}, \quad \mathrm{Ma} = \frac{\Sigma_{r}}{\sigma} \eps,\quad \Sigma_{r} := \int_{0}^{1} \sqrt{2} \rho(s) \sqrt{f_{0}(s)} \ds,
\end{align*}
and the rescaling $\mu \mapsto \frac{1}{\eps} \mu$ in the mass-averaged model \eqref{LLG}, that is, 
\begin{subequations}\label{LLG:rescaled}
\begin{alignat}{3}
\div \bm{w} - \alpha \eps^{2} \Laplace \mu & = 0, \label{LLG:rescaled:div} \\
\rho(c) (\pd_{t}c + \nabla c \cdot \bm{w}) - \eps^{2} \Laplace \mu & = 0, \label{LLG:rescaled:phase} \\
\bm{w} + \frac{1}{12 \eta(c)} \left ( \nabla p + \frac{\sigma}{\Sigma_{r}} \eps \div(\rho(c) \nabla c \otimes \nabla c) - \rho(c) \hat{\bm{g}} \right ) & = \bm{0}, \label{LLG:rescaled:velo} \\
\mu - \frac{1}{\eps} f'_{0}(c) +  \frac{\eps}{\rho(c)} \div (\rho(c) \nabla c) - \alpha \frac{\Sigma_{r}}{\sigma} p & = 0, \label{LLG:rescaled:chem}
\end{alignat}
\end{subequations}
will result in a sharp interface limit that coincides with \eqref{SIM:HSCHnew} when we consider $\bm{G}(\varphi) = \rho(\varphi) \hat{\bm{g}}$. We will briefly sketch the details below.
\begin{itemize}
\item We consider an outer expansion for the pressure $p = p_{0} + \eps p_{1} + \dots$, that is, $p_{-1} = 0$.  Then, one obtains to leading order $\eqref{LLG:rescaled:chem} _{O}^{-1}$ that $f'_{0}(c_{0}) = 0$, which yields the solutions $c_{0} = 0$ or $1$, and the bulk domains can be defined as $\Omega_{1} = \{ c_{0} = 1\}$ and $\Omega_{2} = \{c_{0} = 0\}$.  Then, from $\eqref{LLG:rescaled:div}_{O}^{0}$ and $\eqref{LLG:rescaled:velo}_{O}^{0}$ we obtain
\begin{align*}
\div \bm{w}_{0} = 0, \quad \bm{w}_{0} = -\frac{1}{12 \eta(c_{0})} \left ( \nabla p_{0} - \rho(c_{0})\hat{\bm{g}} \right ) \text{ in } \left ( \Omega_{1} \cup \Omega_{2} \right ) \setminus \Sigma.
\end{align*}
\item For the inner expansions, we denote the inner variable of $c$ and $\bm{w}$ by $C$ and $\bm{W}$, respectively, and assume that the $\frac{1}{2}$-level sets of $c_{\eps}$ converges to $\Sigma$, which implies that
\begin{align*}
C_{0}(t,s,z=0) = \frac{1}{2}.
\end{align*}
Furthermore, we assume that the inner expansion for the pressure $P_{\eps}$ is given as in \eqref{InnerExpansion}, and we alter the matching conditions \eqref{Matching1:Pressure}, \eqref{Matching2:Pressure} to
\begin{align*}
\lim_{z \to \pm \infty} P_{-1}(t,s,z) = 0, \quad \lim_{z \to \pm \infty} P_{0}(t,s,z) = p_{0}^{\pm}(t,x).
\end{align*}
\item To leading order $\eqref{LLG:rescaled:div}_{I}^{-1}$ we obtain $\jump{\bm{w}_{0}} \cdot \bm{\nu} = 0$, and to leading order $\eqref{LLG:rescaled:phase}_{I}^{-1}$ we obtain $\mathcal{V} = \bm{w}_{0} \cdot \bm{\nu}$ whenever $\rho > 0$ and $\pd_{z}C_{0} \neq 0$.
\item To leading order $\eqref{LLG:rescaled:velo}_{I}^{-2}$ we obtain
\begin{align*}
\pd_{z}P_{-1} + \frac{\sigma}{\Sigma_{r}} \pd_{z}(\rho(C_{0}) (\pd_{z}C_{0})^{2}) = 0.
\end{align*}
Integrating and applying the matching conditions for $P_{-1}$ and $\pd_{z}C_{0}$ yields that $P_{-1}(t,s,z) = -\frac{\sigma}{\Sigma_{r}} (\rho(C_{0}) (\pd_{z}C_{0})^{2})(t,s,z)$.  Then, substituting this into $\eqref{LLG:rescaled:chem}_{I}^{-1}$ gives
\begin{align}\label{LLG:ODE}
0 = f_{0}'(C_{0}) - \frac{1}{\rho(C_{0})} \pd_{z}( \rho(C_{0}) \pd_{z}C_{0}) - \alpha \rho(C_{0})(\pd_{z}C_{0})^{2}.
\end{align}
Together with the conditions $\lim_{z \to \infty} C_{0}(t,s,z) = 0$, $\lim_{z \to -\infty} C_{0}(t,s,z) = 1$, and $C_{0}(t,s,0) = \frac{1}{2}$ this yields a second order ODE in $z$, which implies that we can choose $C_{0}$ to be a function depending only on $z$, and thus $P_{-1}$ only depends on $z$.  Multiplying \eqref{LLG:ODE} by $C_{0}'$, applying the product rule to the second term and using the definition of $\alpha$ leads to $0 = (f_{0}(C_{0}) - \frac{1}{2} \abs{C_{0}'}^{2} )'$, and upon integrating yields the equipartition of energy
\begin{align}\label{LLG:Equipartition}
\frac{1}{2}\abs{C_{0}'(z)}^{2} = f_{0}(C_{0}(z)) \quad \forall z \in \R.
\end{align}
\item Lastly, using the fact that $C_{0}$, $P_{-1}$ are independent of $s$ and $t$, we obtain from $\eqref{LLG:rescaled:velo}_{I}^{-1}$
\begin{align*}
\pd_{z}P_{0} \bm{\nu} + \frac{\sigma}{\Sigma_{r}} \pd_{z} \left ( 2 \rho(C_{0}) C_{0}' \pd_{z}C_{1} + \rho'(C_{0}) C_{1} (C_{0}')^{2} \right ) - \frac{\sigma}{\Sigma_{r}} \kappa (\rho(C_{0}) (C_{0}')^{2}) \bm{\nu} = \bm{0}.
\end{align*}
Taking the scalar product with $\bm{\nu}$, integrating with respect to $z$ and applying the matching conditions for $C_{0}'$, we obtain with the help of the equiparition of energy \eqref{LLG:Equipartition} and a change of variables $s = C_{0}'(z)$,
\begin{align*}
\jump{p_{0}} = \frac{\sigma \kappa}{\Sigma_{r}} \int_{\R} \rho(C_{0}(z)) \abs{C_{0}'(z)}^{2} \dz = \frac{\sigma \kappa}{\Sigma_{r}} \int_{0}^{1} \sqrt{2} \rho(s) \sqrt{f_{0}(s)} \ds = \sigma \kappa.
\end{align*}
\end{itemize}

\subsection{Global existence of weak solutions}\label{sec:Existence}
In this section, we investigate the existence of weak solutions to the Hele--Shaw--Cahn--Hilliard model \eqref{Asymp:HSCH} with the parameters $\eps = \sigma = 1$, and rescaling the viscosity by a factor of $\frac{1}{12}$.  For a bounded domain $\Omega \subset \R^{d}$, $d = 2,3$, with boundary $\Gamma$ and an arbitrary but fixed terminal time $T > 0$, we consider
\begin{subequations}\label{HSCHnew:Existence}
\begin{alignat}{3}
\div \bm{v} & = 0 && \text{ in } \Omega \times (0,T) =: Q, \label{HSCHnew:exist:div} \\
\eta(\varphi) \bm{v} & = - \nabla q + \bm{G}(\varphi) + \mu \nabla \varphi  && \text{ in } Q , \label{HSCHnew:exist:pressure} \\
\pd_{t}\varphi + \div (\varphi \bm{v}) & = \Laplace \mu && \text{ in } Q, \label{HSCHnew:exist:varphi} \\
\mu & = \Psi'(\varphi) - \Laplace \varphi && \text{ in } Q, \label{HSCHnew:exist:mu} \\
0 & = \pdnu \varphi = \pdnu \mu && \text{ on } \Gamma \times (0,T), \\
0 & = \bm{v} \cdot \bm{\nu} + b(h - a q) && \text{ on } \Gamma \times (0,T), \label{HSCHnew:exist:velo:bdy} \\
\varphi(0) & = \varphi_{0} && \text{ in } \Omega.
\end{alignat}
\end{subequations}
Here $a > 0 ,b \geq 0$ are constants, $h$ is a prescribed boundary function.  Although \eqref{HSCHnew} is derived as a model in two dimensions, we include in our analysis the existence theory for three dimensions, which is applicable to the situation of fluid flow in a porous medium.  We also point out that, in the case $b = 0$, the pressure $q$ is determined up to a constant, and therefore we prescribe in addition that $\int_{\Omega} q \dx = 0$ for the case $b = 0$.  Before presenting the existence result we introduce the notation and useful preliminaries for this section.

\paragraph{Notation.}  We set $H := L^{2}(\Omega)$, $V := H^{1}(\Omega)$, $H_{\Gamma} := L^{2}(\Gamma)$.  For a (real) Banach space $X$ its dual is denoted as $X'$ and $\inner{\cdot}{\cdot}_{X}$ denotes the duality pairing between $X$ and $X'$.  The $L^{2}$-inner product on $\Omega$ and on $\Gamma$ will be denoted by $(\cdot,\cdot)$ and $(\cdot, \cdot)_{\Gamma}$, respectively.  For convenience, we use the notation $L^{p} := L^{p}(\Omega)$ and $W^{k,p} := W^{k,p}(\Omega)$ for any $p \in [1,\infty]$, $k > 0$ to denote the standard Lebesgue spaces and Sobolev spaces equipped with the norms $\norm{\cdot}_{L^{p}}$ and $\norm{\cdot}_{W^{k,p}}$.  In the case $p = 2$ we use notation $\norm{\cdot}_{H} := \norm{\cdot}_{L^{2}}$, $\norm{\cdot}_{H_{\Gamma}} := \norm{\cdot}_{L^{2}(\Gamma)}$, and $\norm{\cdot}_{V} := \norm{\cdot}_{H^{1}}$.  We denote $\R^{d}$-valued functions and spaces consisting of $\R^{d}$-valued functions in boldface, that is $\bm{H} := (L^{2}(\Omega))^{d}$ and $\bm{V} := (H^{1}(\Omega))^{d}$.  The mean of an integrable function $f: \Omega \to \R$ is defined as $\mean{f} := \frac{1}{\abs{\Omega}} \int_{\Omega} f \dx$, and we denote 
\begin{align*}
L^{2}_{0} := \{ f \in H : \mean{f} = 0 \}, \quad V'_{0} := \{ f \in V' : \inner{f}{1}_{V} = 0 \}, \quad H^{2}_{N} := \{ f \in H^{2} : \pdnu f = 0 \text{ on } \Gamma \}.
\end{align*}
For the velocity, we introduce the space
\begin{align*}
\bm{H}_{\div} := \overline{\{ \bm{f} \in (C^{\infty}_{0}(\Omega))^{d} : \div \bm{f} = 0 \text{ in } \Omega \}}^{\norm{\cdot}_{\bm{L}^{2}}},
\end{align*}
i.e., $\bm{H}_{\div}$ is the closure of the space of all divergence free vector fields in $(C^{\infty}_{0}(\Omega))^{d}$ in the $\bm{L}^{2}$-norm.  Integration with respect to the Hausdorff measure on $\Gamma$ will be denoted by $\dHaus$.

\paragraph{Useful preliminaries.} 
We have the Sobolev embedding $V \subset L^{r}$ for any $r \in [1,\infty)$ in two dimensions and $r \in [1,6]$ in three dimensions, and the following compact embeddings in dimension $d$ (see \cite[Thm. 6.3]{AdamsFournier} and \cite[Thm. 11.2, p. 31]{Friedman})
\begin{align*}
H^{j+1} := W^{j+1,2} \subset \subset W^{j,q}\quad \forall j \geq 0, j \in \Z, 
\end{align*} 
for any $q \in [1, \infty)$ in two dimensions and $q \in [1,6)$ in three dimensions.  We state the Gagliardo--Nirenberg interpolation inequality in dimension $d$ (see \cite[Thm. 10.1, p. 27]{Friedman}, \cite[Thm. 2.1]{DiBenedetto} and \cite[Thm. 5.8]{AdamsFournier}):  Let $\Omega$ be a bounded domain with Lipschitz boundary, and $f \in W^{m,r}(\Omega) \cap L^{q}(\Omega)$, $1 \leq q,r \leq \infty$.  For any integer $j$, $0 \leq j < m$, suppose there is $\alpha \in \R$ such that
\begin{align*}
\frac{1}{p} = \frac{j}{d} + \left ( \frac{1}{r} - \frac{m}{d} \right ) \alpha + \frac{1-\alpha}{q}, \quad \frac{j}{m} \leq \alpha \leq 1.
\end{align*}
If $r \in (1,\infty)$ and $m-j - \frac{d}{r}$ is a nonnegative integer, we in addition assume $\alpha \neq 1$.  Under these assumptions, there exists a positive constant $C$ depending only on $\Omega$, $m$, $j$, $q$, $r$, and $\alpha$ such that
\begin{align}
\label{GagNirenIneq}
\norm{D^{j} f}_{L^{p}} \leq C \norm{f}_{W^{m,r}}^{\alpha} \norm{f}_{L^{q}}^{1-\alpha}.
\end{align}
We recall the Poincar\'{e} inequalities (see for instance \cite[Equ. (1.35), (1.37a) and (1.37c)]{Temam}): There exist positive constants $C_{p}$ depending only on $\Omega$ such that, for all $f \in V$,
\begin{align}
\bignorm{f - \overline{f}}_{H} & \leq C_{p} \norm{\nabla f}_{\bm{H}}, \label{Poincare}\\
\norm{f}_{H} & \leq C_{p} \left (\norm{\nabla f}_{\bm{H}} + \norm{f}_{H_{\Gamma}} \right ). \label{boundary:Poincare}
\end{align}
For fixed $b > 0$ and a given function $\varphi$, we introduce the operators $\mathcal{N}_{b,\varphi}: V \to V'$ and $\mathcal{N}_{0,\varphi} : V \cap L^{2}_{0} \to V'_{0}$ by
\begin{equation}\label{Operators:Robin+Neumann}
\begin{aligned}
\inner{\mathcal{N}_{b,\varphi}(f)}{\zeta}_{V} & := \int_{\Omega} \tfrac{1}{\eta(\varphi)} \nabla f \cdot \nabla \zeta \dx + \int_{\Gamma} b a f \zeta \dHaus, \\
\inner{\mathcal{N}_{0,\varphi}(f)}{\zeta}_{V} & := \int_{\Omega} \tfrac{1}{\eta(\varphi)} \nabla f \cdot \nabla \zeta \dx.
\end{aligned}
\end{equation}
Under a boundedness assumption on $\eta$ (see \ref{assump:eta:G} below), the Lax--Milgram theorem and the Poincar\'{e} inequality \eqref{boundary:Poincare} yield that the inverse operator $\mathcal{N}_{b,\varphi}^{-1}$ is well-defined and stable under perturbations.  I.e., for any $g \in V'$, there exists a unique $u \in V$ such that
\begin{align*}
u = \mathcal{N}_{b,\varphi}^{-1}(g) \text{ with } \norm{u}_{V} \leq C \norm{g}_{V'},
\end{align*} 
for some positive constant $C$ not depending on $g$ and $u$.  Furthermore, given $g_{1}, g_{2} \in V'$ and the corresponding unique solution $u_{1}, u_{2} \in V$ it holds that
\begin{align*}
\norm{u_{1} - u_{2}}_{V} \leq C \norm{g_{1} - g_{2}}_{V'}.
\end{align*}
Similarly, using the Poincar\'{e} inequality \eqref{Poincare} with zero mean, the inverse operator $\mathcal{N}_{0,\varphi}^{-1} : V'_{0} \to V \cap L^{2}_{0}$ is also well-defined and stable under perturbations.  

\begin{assump}\label{assump:HSCH}
\
\begin{enumerate}[label=$(\mathrm{A \arabic*})$, ref = $(\mathrm{A \arabic*})$]
\item We assume that $\Omega \subset \R^{d}$, $d = 2,3$, is a bounded domain with $C^{3}$-boundary $\Gamma$.
\item \label{assump:eta:G} We assume that $\eta \in C^{0}(\R)$, $\bm{G} \in C^{0}(\R; \R^{d})$ and
\begin{align*}
\eta_{0} \leq \eta(s) \leq \eta_{1}, \quad \abs{\bm{G}(s)} \leq G_{0} \abs{s} + G_{1} \quad \forall s \in \R,
\end{align*}
for some positive constants $\eta_{0}$, $\eta_{1}$, $G_{0}$ and $G_{1}$.
\item \label{assump:initialbdy} We assume that $h \in L^{2}(0,T;H_{\Gamma})$ and $\varphi_{0} \in V$.
\item \label{assump:Psi} The potential $\Psi \in C^{2}(\R)$ is nonnegative and satisfies
\begin{align*}
\Psi(s) \geq c_{0} \abs{s}^{2} - c_{1}, \quad \abs{\Psi'(s)} \leq c_{2} \Psi(s) + c_{3}, \quad \abs{\Psi''(s)} \leq c_{4} \abs{s}^{r-2} + c_{5}
\end{align*}
for positive constants $c_{0}$, $c_{1}$, $c_{2}$, $c_{3}$, $c_{4}$, $c_{5}$, and exponent $r \geq 2$ for two dimensions and $r \in [2,6)$ for three dimensions.
\end{enumerate}
\end{assump} 

\begin{thm}[Existence of weak solutions]\label{thm:Existence}
Under Assumption \ref{assump:HSCH}, for
\begin{align*}
\frac{4}{3} \leq p < 2 \text{ in two dimensions}, \quad p = \frac{8}{5} \text{ in three dimensions},
\end{align*}
there exists a quadruple of functions $(\varphi, \mu, q, \bm{v})$ with
 \begin{equation*}
\begin{alignedat}{4}
\varphi & \in L^{2}(0,T;H^{3}) \cap L^{\infty}(0,T;V) \cap W^{1,p}(0,T;V'),  \\
\mu & \in L^{2}(0,T;V), \quad \bm{v} \in L^{2}(0,T;\bm{H}_{\div}), \\
q & \in L^{p}(0,T;V) \text{ with a trace in } L^{2}(0,T;H_{\Gamma}) \text{ for } b > 0, \\
q & \in L^{p}(0,T;V \cap L^{2}_{0}) \text{ for } b = 0
\end{alignedat}
\end{equation*}
such that $\varphi(0) = \varphi_{0}$ and
\begin{subequations}\label{weaksoln:HSCH}
\begin{align}
(\eta(\varphi) \bm{v} + \nabla q - \bm{G}(\varphi) - \mu \nabla \varphi, \bm{\zeta}) & = 0, \label{Weak:velo} \\
( \eta(\varphi)^{-1} (\nabla q - \bm{G}(\varphi) - \mu \nabla \varphi) , \nabla \phi ) + b( aq - h, \phi)_{\Gamma} & = 0 , \label{Weak:Darcy} \\
\inner{\pd_{t}\varphi}{\phi}_{V} + (\nabla \mu,\nabla \phi) + ( \bm{v} \cdot \nabla \varphi, \phi) & = 0, \\
(\mu, \phi) - (\Psi'(\varphi), \phi) - (\nabla \varphi, \nabla \phi) & = 0
\end{align}
\end{subequations}
for a.e. $t \in (0,T)$, and for all $\phi \in V$ and $\bm{\zeta} \in \bm{H}$.
\end{thm}
Note that by the compact embedding
\begin{align*}
L^{\infty}(0,T;V) \cap W^{1,1}(0,T;V') \subset \subset C^{0}([0,T];H),
\end{align*}
the initial value $\varphi(0)$ makes sense as a function in $H$ and thus the initial condition $\varphi_{0}$ is attained.  Furthermore, the boundary condition \eqref{HSCHnew:exist:velo:bdy} can be attained by choosing $\bm{\zeta} = \eta(\varphi)^{-1} \nabla \phi$ in \eqref{Weak:velo}, leading to
\begin{align*}
(\bm{v}, \nabla \phi) = (-\nabla q + \bm{G}(\varphi) + \mu \nabla \varphi, \eta(\varphi)^{-1} \nabla \phi) = b(aq - h, \phi)_{\Gamma}.
\end{align*} 
We further point out that the temporal regularity for $\pd_{t}\varphi$ and the pressure $q$ have been similarly observed in the work of \cite{BosiaContiGrasselli, GarckeLamDarcy, JiangWuZheng}.

\begin{proof}
The proof is based on a Galerkin approximation.  We consider the set of eigenfunctions of the Neumann-Laplacian $\{ w_{i} \}_{i \in \N}$ which forms an orthonormal basis of $H$.  In \cite[\S 3]{GarckeLamDarcy} it has been shown that $\{w_{i}\}_{i \in \N}$ is also a basis of $H^{2}_{N}$.  Let $W_{k} := \mathrm{span} \{ w_{1}, \dots, w_{k} \}$ denote the finite dimensional subspace spanned by the first $k$ eigenfunctions, and let $\Pi_{k}$ denote the orthogonal projection into $W_{k}$.  We consider a Galerkin ansatz  $(\varphi_{k}, \mu_{k}, q_{k}, \bm{v}_{k})_{k \in \N}$ which satisfy $\varphi_{k} = \sum_{i=1}^{k} \alpha_{ik}(t) w_{i} \in W_{k}$, 
\begin{align}
\pd_{t}\varphi_{k} & = \Laplace \mu_{k} - \Pi_{k}\left ( \bm{v}_{k} \cdot \nabla \varphi_{k} \right ), \quad \varphi_{k}(0) = \Pi_{k}(\varphi_{0}), \label{Galerkin:varphi} \\
\mu_{k} & = - \Laplace \varphi_{k} + \Pi_{k} \left ( \Psi'(\varphi_{k}) \right ), \label{Galerkin:mu} \\
\bm{v}_{k} & = - \eta_{k}^{-1} \left ( \nabla q_{k} - \bm{G}(\varphi_{k}) - \mu_{k} \nabla \varphi_{k} \right ), \label{Galerkin:velo}
\end{align}
where $\eta_{k} := \eta(\varphi_{k})$, and $q_{k}$ satisfies an elliptic problem whose weak formulation reads as
\begin{align}\label{Galerkin:pressure}
( \eta_{k}^{-1} \nabla q_{k} , \nabla \zeta ) + ba (q_{k}, \zeta)_{\Gamma} = ( \eta_{k}^{-1} \left ( \bm{G}(\varphi_{k}) + \mu_{k} \nabla \varphi_{k} \right ) , \nabla \zeta ) +  b (h ,\zeta )_{\Gamma} \quad \forall \zeta \in V.
\end{align}
Let us define the linear functionals $F_{b,\varphi_{k}}, F_{0,\varphi_{k}} \in V'$ by
\begin{align*}
\inner{F_{b,\varphi_{k}}}{\zeta}_{V} & :=  ( \eta_{k}^{-1} \left ( \bm{G}(\varphi_{k}) + \mu_{k} \nabla \varphi_{k} \right ) , \nabla \zeta ) +  b (h ,\zeta )_{\Gamma}, \\
\inner{F_{0,\varphi_{k}}}{\zeta}_{V}  & := ( \eta_{k}^{-1} \left ( \bm{G}(\varphi_{k}) + \mu_{k} \nabla \varphi_{k} \right ), \nabla \zeta)
\end{align*}
for all $\zeta \in V$, where $\mu_{k}$ is as defined in \eqref{Galerkin:mu}.  Then, we may express $q_{k}$ as
\begin{align}\label{Nonlocalform:qk}
q_{k} = \mathcal{N}_{b,\varphi_{k}}^{-1} \left ( F_{b,\varphi_{k}} \right ) \text{ if } b > 0, \text{ or } q_{k} = \mathcal{N}_{0,\varphi_{k}}^{-1} \left ( F_{0,\varphi_{k}} \right ) \text{ if } b = 0,
\end{align}
where the operators $\mathcal{N}_{b,\varphi_{k}}$ and $\mathcal{N}_{0,\varphi_{k}}$ are defined in \eqref{Operators:Robin+Neumann}.
Taking the inner product of \eqref{Galerkin:varphi} with $w_{j}$, $j = 1, \dots, k$, and substituting \eqref{Galerkin:mu}, \eqref{Galerkin:velo} and \eqref{Nonlocalform:qk} leads to a system of nonlinear ODEs for the coefficients $\{\alpha_{ik}(t)\}_{1 \leq i \leq k}$.  The right-hand side depends continuously on the coefficients $\{\alpha_{ik}(t)\}_{1 \leq i \leq k}$.  Applying the theory of ordinary differential equations yields the existence of $t_{k} \in (0,T]$ such that the resulting ODE system has a solution $\bm{\alpha}_{k} = (\alpha_{ik})_{1 \leq i \leq k} \in C^{0}([0,t_{k}); \R^{k})$ that is absolutely continuous.  We may define $\mu_{k}$ by the equation \eqref{Galerkin:mu}, then $q_{k}$ is defined by \eqref{Nonlocalform:qk} and $\bm{v}_{k}$ is defined by \eqref{Galerkin:velo}.  

We now derive a priori estimates for the Galerkin ansatz $(\varphi_{k}, \mu_{k}, q_{k}, \bm{v}_{k})$.  In the following, the constant $C > 0$ may vary line to line, but it is independent of $k$.  For convenience, we denote 
\begin{align*}
\EE_{k}(t) := \int_{\Omega} \Psi(\varphi_{k}(t)) + \frac{1}{2} \abs{\nabla \varphi_{k}(t)}^{2} \dx.
\end{align*}
Note that by the assumption $\varphi_{0} \in V$, the growth assumptions \ref{assump:Psi} on $\Psi$ and the Sobolev embedding $V \subset L^{p}$ for $p \geq 2$ in two dimensions and $p \in [1,6]$ for three dimensions, there exists a constant $C$ such that
\begin{align*}
\EE_{k}(0) \leq C \left ( \norm{\varphi_{0}}_{V}^{\max(r,2)} + 1 \right ).
\end{align*}

\paragraph{First estimate.} Substituting $\zeta = q_{k}$ in \eqref{Galerkin:pressure}, and taking the inner product of \eqref{Galerkin:varphi} with $\mu_{k}$, the inner product of \eqref{Galerkin:mu} with $\pd_{t}\varphi_{k}$, and the inner product of \eqref{Galerkin:velo} with $\bm{v}_{k}$, summing and integrating from $0$ to $s \in (0,T]$ leads to
\begin{equation}\label{FirstAprioriEst:1}
\begin{aligned}
&  \EE_{k}(s) + \int_{0}^{s} \norm{\nabla \mu_{k}}_{\bm{H}}^{2} + \norm{\sqrt{\eta_{k}} \bm{v}_{k}}_{\bm{H}}^{2} + ba \norm{q_{k}}_{H_{\Gamma}}^{2}\dt \\
& \quad  = \int_{0}^{s} (\bm{G}(\varphi_{k}), \bm{v}_{k}) + b(h, q_{k})_{\Gamma} \dt + \EE_{k}(0).
\end{aligned}
\end{equation}
By the growth conditions for $\bm{G}$ in \ref{assump:eta:G} we see that
\begin{align*}
\abs{\int_{0}^{s} (\bm{G}(\varphi_{k}), \bm{v}_{k}) \dt} & \leq \int_{0}^{s} G_{1} \norm{\varphi_{k}}_{H} \norm{\bm{v}_{k}}_{\bm{H}} + G_{2} \abs{\Omega}^{\frac{1}{2}} \norm{\bm{v}_{k}}_{\bm{H}} \dt \\
& \leq \int_{0}^{s} \frac{1}{2} \eta_{0} \norm{\bm{v}_{k}}_{\bm{H}}^{2} + C(\eta_{0}, G_{1}, G_{2}, \abs{\Omega}) \left ( \norm{\varphi_{k}}_{H}^{2} + 1 \right ) \dt.
\end{align*}
Using the lower bound for $\Psi$ in \ref{assump:Psi}, we have
\begin{align*}
\int_{\Omega} \Psi(\varphi_{k}(s)) \dx \geq c_{1} \norm{\varphi_{k}(s)}_{H}^{2} - c_{2} \abs{\Omega}^{\frac{1}{2}},
\end{align*}
and thus, by the lower bound on the viscosity $\eta$, we obtain from \eqref{FirstAprioriEst:1}
\begin{equation}\label{FirstAprioriEst:2}
\begin{aligned}
& \frac{1}{2}\norm{\Psi(\varphi_{k}(s))}_{L^{1}} + \frac{c_{1}}{2} \norm{\varphi_{k}(s)}_{H}^{2} + \frac{1}{2} \norm{\nabla \varphi_{k}(s)}_{\bm{H}}^{2} \\
& \quad + \norm{\nabla \mu_{k}}_{L^{2}(0,s;\bm{H})}^{2} + \frac{\eta_{0}}{2} \norm{\bm{v_{k}}}_{L^{2}(0,s;\bm{H})}^{2} + ba \norm{q_{k}}_{L^{2}(0,s;H_{\Gamma})}^{2} \\
& \quad \leq C \norm{\varphi_{k}}_{L^{2}(0,s;H)}^{2} + C + \int_{0}^{s} b \norm{h}_{H_{\Gamma}} \norm{q_{k}}_{H_{\Gamma}} \ds.
\end{aligned}
\end{equation}
Applying Young's inequality to the last term on the right-hand side of \eqref{FirstAprioriEst:2}, and then applying Gronwall's inequality (see \cite[Lem. 3.1]{GarckeLamNeumann}) leads to
\begin{equation}\label{FirstAprioriEst:3}
\begin{aligned}
& \norm{\Psi(\varphi_{k}(s))}_{L^{1}} + \norm{\varphi_{k}(s)}_{H}^{2} +  \norm{\nabla \varphi_{k}(s)}_{\bm{H}}^{2} \\
& \quad + C \left ( \norm{ \nabla \mu_{k}}_{L^{2}(0,s;\bm{H})}^{2} +  \norm{\bm{v_{k}}}_{L^{2}(0,s;\bm{H})}^{2} + ba \norm{q_{k}}_{L^{2}(0,s;H_{\Gamma})}^{2} \right ) \\
& \quad \leq C \left (1 + \frac{b}{a} \norm{h}_{L^{2}(0,T;H_{\Gamma})}^{2} \right ) \quad \forall s \in (0,T].
\end{aligned}
\end{equation}
For the case $b = 0$, we obtain \eqref{FirstAprioriEst:2} without the terms in the $H_{\Gamma}$-norm.  Furthermore, the a priori estimate \eqref{FirstAprioriEst:3} guarantees that we can extend the Galerkin ansatz to the whole of $[0,T]$, and thus $t_{k} = T$ for all $k \in \N$.

\paragraph{Second estimate.}
Integrating \eqref{Galerkin:mu} and using  \ref{assump:Psi} leads to
\begin{align*}
\abs{\int_{\Omega} \mu_{k} \dx} \leq \int_{\Omega} \abs{\Psi'(\varphi_{k})} \dx \leq c_{2} \norm{\Psi(\varphi_{k})}_{L^{1}} + c_{3} \abs{\Omega}.
\end{align*}
By \eqref{FirstAprioriEst:3} we have that the mean $\mean{\mu_{k}}$ is bounded uniformly in $L^{\infty}(0,T)$, and thus by the Poincar\'{e} inequality \eqref{Poincare} and the boundedness of $\norm{\nabla \mu_{k}}_{L^{2}(0,T;\bm{H})}$, we have
\begin{align*}
\norm{\mu_{k}}_{L^{2}(0,T;H)} \leq C.
\end{align*}

\paragraph{Third estimate.}
We may view \eqref{Galerkin:mu} as an elliptic equation for $\varphi_{k}$:
\begin{subequations}
\begin{alignat}{3}
-\Laplace \varphi_{k} + \varphi_{k} & = \mu_{k} - \Pi_{k}(\Psi'(\varphi_{k})) + \varphi_{k} && \text{ in } \Omega, \label{Elliptic:varphi} \\
\pdnu \varphi_{k} & = 0 && \text{ on } \Gamma.
\end{alignat}
\end{subequations}
Then, the argument in \cite[\S 4.2]{GarckeLamDarcy} yields that $\{\varphi_{k}\}_{k \in \N}$ is bounded uniformly in $L^{2}(0,T;H^{3})$.  We will omit the details and refer the reader to  \cite{GarckeLamDarcy}.

\paragraph{Fourth estimate.}
Substituting $\zeta = q_{k}$ in \eqref{Galerkin:pressure} leads to
\begin{align*}
& \int_{\Omega} \frac{1}{\eta_{k}} \abs{\nabla q_{k}}^{2} \dx + ba \norm{q_{k}}_{H_{\Gamma}}^{2} = \int_{\Omega} \frac{1}{\eta_{k}} \left ( \bm{G}(\varphi_{k}) + \mu_{k} \nabla \varphi_{k} \right ) \cdot \nabla q_{k} \dx + \int_{\Gamma} b h q_{k} \dHaus \\
& \quad \leq \int_{\Omega} \frac{1}{2 \eta_{k}} \abs{\nabla q_{k}}^{2} + \frac{1}{2 \eta_{0}} \left ( \abs{\bm{G}(\varphi_{k})}^{2} + \abs{\mu_{k} \nabla \varphi_{k}}^{2} \right ) \dx + b \norm{h}_{H_{\Gamma}} \norm{q_{k}}_{H_{\Gamma}}.
\end{align*}
\paragraph{Case (i) $b > 0$.}
For this case, Young's inequality gives $b \norm{h}_{H_{\Gamma}} \norm{q_{k}}_{H_{\Gamma}} \leq \frac{ba}{2} \norm{q_{k}}_{H_{\Gamma}}^{2} + \frac{b}{2a} \norm{h}_{H_{\Gamma}}^{2}$, which leads to
\begin{equation}\label{Pressure:Est:ba>0}
\begin{aligned}
\frac{1}{\eta_{1}} \norm{\nabla q_{k}}_{\bm{H}}^{2} + ba \norm{q_{k}}_{H_{\Gamma}}^{2} \leq \frac{1}{\eta_{0}} \left ( \norm{\bm{G}(\varphi_{k})}_{\bm{H}}^{2} + \norm{\mu_{k} \nabla \varphi_{k}}_{\bm{H}}^{2} \right ) + \frac{b}{a} \norm{h}_{H_{\Gamma}}^{2}.
\end{aligned}
\end{equation}
\paragraph{Case (ii) $b = 0$.} For this case, we obtain
\begin{equation}\label{Pressure:Est:b=0}
\begin{aligned}
\frac{\eta_{0}}{\eta_{1}} \norm{\nabla q_{k}}_{\bm{H}}^{2} \leq  \norm{\bm{G}(\varphi_{k})}_{\bm{H}}^{2} + \norm{\mu_{k} \nabla \varphi_{k}}_{\bm{H}}^{2}.
\end{aligned}
\end{equation}
For both cases, we obtain an a priori estimate of the form
\begin{align}\label{Pressure:Est:PreIntegration}
\norm{\nabla q_{k}}_{\bm{H}} \leq C \left ( \norm{\bm{G}(\varphi_{k})}_{\bm{H}} + \norm{\mu_{k} \nabla \varphi_{k}}_{\bm{H}} + \sqrt{b} \norm{h}_{H_{\Gamma}} \right ).
\end{align}
By \ref{assump:eta:G} and \ref{assump:initialbdy}, we have that $\bm{G}(\varphi_{k}) \in L^{\infty}(0,T;\bm{H})$ and $h \in L^{2}(0,T;H_{\Gamma})$.  Thus, we expect that the temporal regularity of $\nabla q_{k}$ will be no greater than the temporal regularity of the product $\mu_{k} \nabla \varphi_{k}$.  Let $s \in [1,\infty)$ for two dimensions, then by the Gagliardo--Nirenberg inequality \eqref{GagNirenIneq}, we see that
\begin{equation}\label{GNineq:nablavarphi:L3}
\begin{alignedat}{3}
\norm{\nabla \varphi_{k}}_{\bm{L}^{\frac{2s}{s-1}}} & \leq C \norm{\nabla \varphi_{k}}_{\bm{H}^{2}}^{\frac{1}{2s}} \norm{\nabla \varphi_{k}}_{\bm{H}}^{1 - \frac{1}{2s}} && \text{ for two dimensions}, \\
\norm{\nabla \varphi_{k}}_{\bm{L}^{3}} & \leq C \norm{\nabla \varphi_{k}}_{\bm{H}^{2}}^{\frac{1}{4}} \norm{\nabla \varphi_{k}}_{\bm{H}}^{\frac{3}{4}} && \text{ for three dimensions}.
\end{alignedat}
\end{equation}
By H\"{o}lder's inequality and Sobolev embedding, we obtain for two dimensions,
\begin{align*}
\int_{0}^{T} \norm{\mu_{k} \nabla \varphi_{k}}_{\bm{H}}^{\frac{4s}{2s+1}} \dt & \leq \int_{0}^{T} \norm{\mu_{k}}_{L^{2s}}^{\frac{4s}{2s+1}} \norm{\nabla \varphi_{k}}_{\bm{L}^{\frac{2s}{s-1}}}^{\frac{4s}{2s+1}} \dt \\
&  \leq C \norm{\varphi_{k}}_{L^{\infty}(0,T;V)}^{2 \frac{2s-1}{2s+1}} \int_{0}^{T} \norm{\mu_{k}}_{V}^{\frac{4s}{2s+1}} \norm{\varphi_{k}}_{H^{3}}^{\frac{2}{2s+1}} \dt \\
&  \leq C \norm{\varphi_{k}}_{L^{\infty}(0,T;V)}^{2\frac{2s-1}{2s+1}} \norm{\mu_{k}}_{L^{2}(0,T;V)}^{\frac{4s}{2s+1}} \norm{\varphi_{k}}_{L^{2}(0,T;H^{3})}^{\frac{2s}{2s+1}},
\end{align*}
and so $\mu_{k} \nabla \varphi_{k} \in L^{r}(0,T;\bm{H})$ for $\frac{4}{3} \leq r < 2$ in two dimensions.  For three dimensions, we obtain analogously
\begin{align*}
\int_{0}^{T} \norm{\mu_{k} \nabla \varphi_{k}}_{\bm{H}}^{\frac{8}{5}} \dt \leq \int_{0}^{T} \norm{\mu_{k}}_{L^{6}}^{\frac{8}{5}} \norm{\nabla \varphi_{k}}_{\bm{L}^{3}}^{\frac{8}{5}} \dt \leq C \norm{\varphi_{k}}_{L^{\infty}(0,T;V)}^{\frac{6}{5}} \norm{\mu_{k}}_{L^{2}(0,T;V)}^{\frac{8}{5}} \norm{\varphi_{k}}_{L^{2}(0,T;H^{3})}^{\frac{2}{5}},
\end{align*}
and so $\mu_{k} \nabla \varphi_{k} \in L^{\frac{8}{5}}(0,T;\bm{H})$.  Thus, from \eqref{Pressure:Est:PreIntegration}, and using the Poincar\'{e} inequality \eqref{boundary:Poincare} for the case $b > 0$ or the condition $\mean{q_{k}} = 0$ and the Poincar\'{e} inequality \eqref{Poincare} for the case $b = 0$, we obtain that
\begin{align*}
\{q_{k}\}_{k \in \N} \text{ is bounded in } \begin{cases}
L^{p}(0,T;V), \; \frac{4}{3} \leq p < 2 & \text{ in two dimensions}, \\
L^{\frac{8}{5}}(0,T;V) & \text{ in three dimensions}.
\end{cases}
\end{align*}

\paragraph{Fifth estimate.}
Using \eqref{GNineq:nablavarphi:L3}, in three dimensions, for an arbitrary test function $\zeta \in L^{\frac{8}{3}}(0,T;V)$ we have
\begin{align*}
& \abs{\int_{Q} \Pi_{k}(\bm{v}_{k} \cdot \nabla \varphi_{k}) \zeta \dx \dt} = \abs{\int_{Q} \bm{v}_{k} \cdot \nabla \varphi_{k} \Pi_{k}(\zeta) \dx \dt} \leq \int_{0}^{T} \norm{\bm{v}_{k}}_{\bm{L}^{2}} \norm{\nabla \varphi_{k}}_{\bm{L}^{3}} \norm{\Pi_{k}(\zeta)}_{L^{6}} \dt \\
& \quad \leq C \norm{\bm{v}_{k}}_{L^{2}(0,T;\bm{H})} \norm{\varphi_{k}}_{L^{\infty}(0,T;V)}^{\frac{3}{4}} \norm{\varphi_{k}}_{L^{2}(0,T;H^{3})}^{\frac{1}{4}} \norm{\zeta}_{L^{\frac{8}{3}}(0,T;V)}.
\end{align*}
This implies that $\{\Pi_{k}(\bm{v}_{k} \cdot \nabla \varphi_{k})\}_{k \in \N}$ is bounded in $L^{\frac{8}{5}}(0,T;V')$.  Then, from \eqref{Galerkin:varphi} we have that $\{\pd_{t}\varphi_{k}\}_{k \in \N}$ is bounded in $L^{\frac{8}{5}}(0,T;V')$.  For two dimensions, we have for any $s \in [1,\infty)$,
\begin{align*}
& \abs{\int_{Q} \bm{v}_{k} \cdot \nabla \varphi_{k}  \Pi_{k} (\zeta) \dx \dt} \leq \int_{0}^{T} \norm{\bm{v}_{k}}_{\bm{L}^{2}} \norm{\nabla \varphi_{k}}_{\bm{L}^{\frac{2s}{s-1}}} \norm{\Pi_{k} (\zeta)}_{L^{2s}} \dt \\
& \quad \leq C \norm{\varphi_{k}}_{L^{\infty}(0,T;V)}^{1 - \frac{1}{2s}} \norm{\bm{v}_{k}}_{L^{2}(0,T;\bm{H})} \norm{\varphi_{k}}_{L^{2}(0,T;H^{3})}^{\frac{1}{2s}} \norm{\zeta}_{L^{\frac{4s}{2s-1}}(0,T;V)}
\end{align*}
and so $\{\Pi_{k}(\bm{v}_{k} \cdot \nabla \varphi_{k})\}_{k \in \N}$ and $\{\pd_{t}\varphi_{k}\}_{k \in \N}$ are bounded in $L^{p}(0,T;V')$ for $\frac{4}{3} \leq p < 2$.

\paragraph{Compactness.} The above a priori estimates and the application of \cite[\S 8, Corollary 4]{Simon86} yield the existence of a relabelled subsequence $(\bm{v}_{k}, q_{k}, \varphi_{k}, \mu_{k})_{k \in \N}$ such that
\begin{alignat*}{3}
\varphi_{k} & \rightarrow \varphi && \quad \text{ weakly-}* && \quad \text{ in } L^{\infty}(0,T;V) \cap L^{2}(0,T;H^{3}) \cap W^{1,p}(0,T;V'), \\
\varphi_{k} & \rightarrow \varphi && \quad \text{ strongly } && \quad \text{ in } C^{0}([0,T];L^{s}) \cap L^{2}(0,T;W^{2,s}) \text{ and a.e. in } Q, \\
\mu_{k} & \rightarrow \mu && \quad \text{ weakly } && \quad \text{ in } L^{2}(0,T;V), \\
q_{k} & \rightarrow q && \quad \text{ weakly } && \quad \text{ in } L^{p}(0,T;V) \text{ and also in } L^{2}(0,T;H_{\Gamma}) \text{ if } b > 0, \\
\bm{v}_{k} & \rightarrow \bm{v} && \quad \text{ weakly } && \quad \text{ in } L^{2}(0,T;\bm{H}_{\div}), \\
\Pi_{k}(\bm{v}_{k} \cdot \nabla \varphi_{k} ) & \rightarrow \xi && \quad \text{ weakly } && \quad \text{ in } L^{p}(0,T;V'),
\end{alignat*}
for some function $\xi \in L^{p}(0,T;V')$ and
\begin{align*}
\frac{4}{3} \leq p < 2, \quad 1 \leq s < \infty \text{ in two dimensions}, \quad p = \frac{8}{5}, \quad 1 \leq s < 6 \text{ in three dimensions}.
\end{align*}

To deduce that $(\varphi, \mu, q, \bm{v})$ is a weak solution of \eqref{HSCHnew} that satisfies \eqref{weaksoln:HSCH}, we argue as follows: Fix $j \in \N$ and $\delta \in C^{\infty}_{c}(0,T)$, multiplying \eqref{Galerkin:varphi}, \eqref{Galerkin:mu} with $\delta w_{j}$, and integrating in time leads to
\begin{equation}\label{timeintegrated:weakform:varphi:mu}
\begin{aligned}
0 & = \int_{0}^{T} \delta(t) \left [ (\pd_{t}\varphi_{k},w_{j}) + (\nabla  \mu_{k}, \nabla w_{j}) + (\Pi_{k}(\bm{v}_{k} \cdot \nabla \varphi_{k}),w_{j}) \right ] \dt, \\
0 & = \int_{0}^{T} \delta(t) \left [ (\mu_{k},w_{j}) - (\nabla  \varphi_{k}, \nabla w_{j}) - (\Psi'(\varphi_{k}),w_{j}) \right ] \dt,
\end{aligned} 
\end{equation}
where we used $(\Pi_{k}(\Psi'(\varphi_{k})),w_{j}) = (\Psi'(\varphi_{k}), w_{j})$.  On one hand we see that
\begin{align}\label{LimitId:Xi:Part1}
\int_{0}^{T} \delta(t) (\bm{v}_{k} \cdot \nabla \varphi_{k}, w_{j}) \dt  = \int_{0}^{T} \delta(t) (\Pi_{k}(\bm{v}_{k} \cdot \nabla \varphi_{k}), w_{j}) \dt \to \int_{0}^{T} \delta (t) \inner{\xi}{w_{j}}_{V} \dt.
\end{align}
On the other hand, the strong convergence of $\nabla \varphi_{k}$ to $\nabla \varphi$ in $L^{2}(0,T;W^{1,s})$ and the fact that $w_{j} \in H^{2}$ shows that
\begin{align*}
\int_{Q} \abs{\delta(\nabla \varphi_{k} - \nabla \varphi) w_{j}}^{2} \dx \dt \leq \norm{\delta}_{L^{\infty}(0,T)}^{2} \norm{\nabla \varphi_{k} - \nabla \varphi}_{L^{2}(0,T;\bm{H})}^{2} \norm{w_{j}}_{L^{\infty}}^{2} \to 0,
\end{align*}
and so $\delta w_{j} \nabla \varphi_{k}  \to \delta w_{j} \nabla \varphi$ strongly in $L^{2}(0,T;\bm{H})$.  Together with the weak convergence of $\bm{v}_{k}$ in $L^{2}(0,T;\bm{H})$, we obtain
\begin{align}\label{LimitID:Xi:Part2}
\int_{0}^{T} \delta(t) (\bm{v}_{k} \cdot \nabla \varphi_{k}, w_{j}) \dt \to \int_{0}^{T} \delta (t) (\bm{v} \cdot \nabla \varphi, w_{j})  \dt.
\end{align}
Equating \eqref{LimitId:Xi:Part1} and \eqref{LimitID:Xi:Part2} leads to
\begin{align*}
\int_{0}^{T} \delta(t) \inner{\xi}{w_{j}}_{V} \dt = \int_{0}^{T} \delta(t) ( \bm{v} \cdot \nabla \varphi, w_{j}) \dt.
\end{align*}
Passing to the limit $k \to \infty$ in \eqref{timeintegrated:weakform:varphi:mu}, using the above weak/weak* convergences yields
\begin{align}
0 & = \int_{0}^{T} \delta(t) \left [ \inner{\pd_{t}\varphi}{w_{j}}_{V} + (\nabla  \mu, \nabla w_{j}) + ( \bm{v} \cdot \nabla \varphi, w_{j}) \right ] \dt, \label{Passtolimit:varphi} \\
0 & = \int_{0}^{T} \delta(t) \left [ (\mu,w_{j}) - (\nabla  \varphi, \nabla w_{j}) - (\Psi'(\varphi),w_{j}) \right ] \dt. \label{Passtolimit:mu}
\end{align}
We refer to \cite[\S 3.1.2]{GarckeLamDirichlet} for the details on how to pass to the limit in the term with $\Psi'$.  Meanwhile, substituting $\zeta = w_{j}$ in \eqref{Galerkin:pressure}, then multiplying with $\delta$ and integrating over time, we obtain
\begin{align}\label{timeintegrated:weakform:pressure}
0 & = \int_{0}^{T} \delta(t) \left [ (\nabla q_{k} - \bm{G}(\varphi_{k}) - \mu_{k} \nabla \varphi_{k}, \eta_{k}^{-1} \nabla w_{j}) + b (a q_{k} - h, w_{j})_{\Gamma} \right ] \dt.
\end{align}
Due to the a.e. convergence of $\varphi_{k}$ to $\varphi$ in $Q$, and the continuity of $\eta$ and $\bm{G}$, we have that $\eta(\varphi_{k})^{-1} \to \eta(\varphi)^{-1}$ and $\bm{G}(\varphi_{k}) \to \bm{G}(\varphi)$ a.e. in $Q$.  Furthermore, by the boundedness of $\eta$,  applying Lebesgue's dominated convergence theorem yields
\begin{align}\label{passtolimit:eta-1:strong}
\eta(\varphi_{k})^{-1} \delta \nabla w_{j} \to \eta(\varphi)^{-1} \delta \nabla w_{j} \text{ strongly in } L^{m}(0,T;\bm{L}^{m}) \text{ for } m \in [1,6].
\end{align}
Meanwhile, from the strong convergence $\varphi_{k} \to \varphi$ in $L^{2}(0,T;H)$ we find that
\begin{align*}
G_{0} \abs{\varphi_{k}}^{2} + G_{1} \to G_{0} \abs{\varphi}^{2} + G_{1} \text{ strongly in } L^{1}(Q).
\end{align*}
Then, using the growth assumption \ref{assump:eta:G} for $\bm{G}$ and the generalized Lebesgue dominated convergence theorem (\cite[Thm. 1.9, p. 89]{Royden}, \cite[Thm. 3.25, p. 60]{Alt}), it holds that
\begin{align}\label{passtolimit:G:L2H:strong}
\bm{G}(\varphi_{k}) \to \bm{G}(\varphi) \text{ strongly in } L^{2}(0,T;\bm{H}).
\end{align}
By the Gagliardo--Nirenberg inequality \eqref{GagNirenIneq} we find that
\begin{align*}
L^{\infty}(0,T;H) \cap L^{2}(0,T;H^{2}) \subset \begin{cases}
L^{12}(0,T;L^{3}) & \text{ in two dimensions}, \\
L^{8}(0,T;L^{3}) & \text{ in three dimensions}.
\end{cases}
\end{align*}
Thus, from the boundedness of $\{\varphi_{k}\}_{k \in \N}$ in $L^{\infty}(0,T;V) \cap L^{2}(0,T;H^{3})$, we see that $\{\nabla \varphi_{k}\}_{k \in \N}$ is bounded in $L^{3}(0,T;\bm{L}^{3})$.  Furthermore, using the strong convergence of $\varphi_{k}$ to $\varphi$ in $L^{2}(0,T;W^{2,s})$ for $s \in [1,6)$, and \eqref{passtolimit:eta-1:strong} for $m = 6$, we obtain
\begin{align*}
& \int_{Q} \abs{\delta \left [ (\eta(\varphi_{k})^{-1} - \eta(\varphi)^{-1}) \nabla w_{j} \cdot \nabla \varphi_{k} + \eta(\varphi)^{-1} \nabla w_{j} \cdot \nabla (\varphi_{k} - \varphi) \right ]}^{2} \dx \dt \\
& \quad \leq \norm{\delta (\eta(\varphi_{k})^{-1} - \eta(\varphi)^{-1}) \nabla w_{j}}_{L^{6}(0,T;\bm{L}^{6})}^{2} \norm{\nabla \varphi_{k}}_{L^{3}(0,T;\bm{L}^{3})}^{2} \\
& \quad +  \frac{1}{\eta_{0}^{2}} \norm{\delta}_{L^{\infty}(0,T)}^{2} \norm{\nabla w_{j}}_{\bm{L}^{6}}^{2} \norm{\nabla (\varphi_{k} - \varphi)}_{L^{2}(0,T;\bm{L}^{3})}^{2} \to 0.
\end{align*}
This implies that
\begin{align}\label{strong:eta-1nablaproduct:L2H}
\delta \eta(\varphi_{k})^{-1} \nabla w_{j} \cdot \nabla \varphi_{k} \to \delta \eta(\varphi)^{-1} \nabla w_{j} \cdot \nabla \varphi \text{ strongly in } L^{2}(0,T;H).
\end{align}
Then, combining \eqref{passtolimit:eta-1:strong}, \eqref{passtolimit:G:L2H:strong}, \eqref{strong:eta-1nablaproduct:L2H} and the weak convergences for $q_{k}$ and $\mu_{k}$, after passing to the limit $k \to \infty$ in \eqref{timeintegrated:weakform:pressure} we obtain
\begin{align}\label{Passtolimit:pressure}
0 & = \int_{0}^{T} \delta(t) \left [ \eta(\varphi)^{-1} (\nabla q - \bm{G}(\varphi) - \mu \nabla \varphi) ,\nabla w_{j}) + b (a q - h, w_{j})_{\Gamma} \right ] \dt.
\end{align}
Next, multiplying \eqref{Galerkin:velo} by $\delta \eta(\varphi_{k}) (w_{j_{1}}, \dots, w_{j_{d}})^{\top} =: \delta \eta(\varphi_{k}) \bm{\zeta}_{j}$ for $1 \leq j_{1}, \dots, j_{d} \leq k$, passing to the limit $k \to \infty$ yields
\begin{align}\label{Passtolimit:div}
\int_{0}^{T} \delta(t) (\eta(\varphi) \bm{v}, \bm{\zeta}_{j}) \dt  = \int_{0}^{T} \delta(t) (-\nabla q + \bm{G}(\varphi) + \mu \nabla \varphi, \bm{\zeta}_{j}) \dt.
\end{align}
Since \eqref{Passtolimit:varphi}, \eqref{Passtolimit:mu}, \eqref{Passtolimit:pressure} and \eqref{Passtolimit:div} hold for arbitrary $\delta \in C^{\infty}_{c}(0,T)$, we infer that $(\varphi, \mu, q, \bm{v})$ satisfies \eqref{weaksoln:HSCH} with $\phi = w_{j}$ and $\bm{\zeta} = (w_{j_{1}}, \dots, w_{j_{d}})^{\top}$.  Using that $\{w_{j}\}_{j \in \N}$ is basis of $H^{2}_{N}$ and $H^{2}_{N}$ is dense in $V$, we deduce that \eqref{weaksoln:HSCH} holds for arbitrary $\phi \in V$ and $\bm{\zeta} \in \bm{H}$.  This concludes the proof.
\end{proof}

\section{Numerical approximation}\label{sec:approx}
We briefly describe the numerical approximation of the Hele--Shaw--Cahn--Hilliard problem~\eqref{HSCHnew} with the variant \eqref{Nondim:HSCH:velo:alternate:q}.  In particular, by recalling that $\mu = \mu(\varphi)= \frac{1}{\eps} \Psi'(\varphi) - \eps \Laplace \varphi$ from \eqref{Nondim:HSCH:mu} and $\bm{v}(q,\varphi) = - \frac{1}{12 \eta(\varphi)} \left( \nabla q - \mathrm{Bo} \, \rho(\varphi) \hat{\bm{g}} - \frac{1}{\mathrm{Ca}} \mu(\varphi) \nabla \varphi\right)$ from \eqref{Nondim:HSCH:velo:alternate:q}, we reformulate the dimensionless problem~\eqref{HSCHnew} in terms of the (modified) pressure $q$ and order parameter $\varphi$ and endow it with suitable initial and boundary conditions as:
\begin{subequations}\label{eq:HSCHstrong4th}
\begin{alignat}{3}
- \div \left( \frac{1}{12 \eta(\varphi)} \left( \nabla q - \mathrm{Bo} \, \rho(\varphi) \hat{\bm{g}} - \frac{1}{\mathrm{Ca}} \mu(\varphi) \nabla \varphi \right) \right) & = 0 \quad && \text{ in } \Omega \times (0,T), \label{eq:HSCHstrong4th.q} \\
\pd_{t} \varphi  + \bm{v}(q,\varphi) \cdot \nabla \varphi - \eps \Laplace \mu(\varphi) & = 0 \quad && \text{ in } \Omega \times (0,T), \label{eq:HSCHstrong4th.varphi}\\
\frac{1}{12\eta(\varphi)}\left(\nabla q \cdot \bm{\nu} - \mathrm{Bo} \, \rho(\varphi) \hat{\bm{g}} \cdot \bm{\nu}\right) & = f_{N} \quad && \text{ on } \Gamma_{N} \times (0,T), \\
q &= 0 \quad && \text{ on } \Gamma_{D} \times (0,T), \\\nabla \mu(\varphi) \cdot \bm{\nu}  & =  0 \quad && \text{ on } \Gamma \times (0,T), \\
\nabla \varphi \cdot \bm{\nu}  &= 0 \quad && \text{ on } \Gamma \times (0,T), \\
\varphi(t=0) & = \varphi_{0} \quad && \text{ in } \Omega,
\end{alignat}
\end{subequations}
where $\overline{\Gamma}_{N} \cup \overline{\Gamma}_{D} = \Gamma \equiv \pd \Omega$, $\mathring{\Gamma}_{N} \cap \mathring{\Gamma}_{D} = \emptyset$, and $f_{N}$ is a suitable function.  We remark that Problem \eqref{eq:HSCHstrong4th} is time-dependent, nonlinear, and it involves a fourth order differential operator in  \eqref{eq:HSCHstrong4th.varphi}.  Then, we rewrite for convenience the dimensionless density and viscosity as $\rho(\varphi) = \Theta_{1} \,\varphi + \Theta_{2}$ and $\eta(\varphi) = \Lambda_{1} \,\varphi + \Lambda_{2}$, respectively, where $\Theta_{1} := \frac{1}{2} \left( 1 - \frac{\overline{\rho}_{1}}{\overline{\rho}_{2}} \right )$, $\Theta_{2} := \frac{1}{2} \left( 1 + \frac{\overline{\rho}_{1}}{\overline{\rho}_{2}} \right )$, $\Lambda_{1} := \frac{1}{2} \left( 1 - \frac{\eta_{1}}{\eta_{2}} \right)$, and $\Lambda_{2} := \frac{1}{2} \left( 1 + \frac{\eta_{1}}{\eta_{2}} \right)$. We recall that for $\varphi = -1$, we obtain the pure phase labeled ``$1$'', while $\varphi = 1$ refers instead to the pure phase ``$2$''.

Let us now introduce the function spaces $\mathcal{V} = \left\{ w \in H^{1}(\Omega) \ : \ w = 0 \ \text{ on } \Gamma_{D} \right \}$ and $\mathcal{H} := H^{2}_{N} = \left\{ w \in H^{2}(\Omega) \ : \pdnu w = 0 \ \text{ on } \Gamma \right\} $, then, by suitably using integration by parts, the weak formulation of \eqref{eq:HSCHstrong4th} reads as follows:  Find, for all $t \in (0,T)$, $q \in \mathcal{V}$ and $\varphi \in \mathcal{H}$, such that
\begin{subequations}\label{eq:HSCHweak}
\begin{align}
\int_{\Omega} \nabla \psi \cdot \left(\frac{1}{12 \eta(\varphi)} \nabla q \right) \dx & - \mathrm{Bo} \int_{\Omega} \nabla \psi \cdot \hat{\bm{g}} \frac{\rho(\varphi)}{12 \eta(\varphi)} \dx \notag \\
& - \frac{1}{\mathrm{Ca}} \int_{\Omega} \nabla \psi \cdot \left( \frac{\mu(\varphi)}{12 \eta(\varphi)} \nabla \varphi \right) \dx = \int_{\Gamma_N} \psi \, f_N \dHaus,\\
\int_{\Omega} \vartheta \, \pd_{t} \varphi \dx &+ \int_{\Omega} \vartheta \, \bm{v}(q, \varphi) \cdot \nabla \varphi \dx + \int_{\Omega} \nabla \vartheta \cdot \left( \Psi''(\varphi) \nabla \varphi \right) \dx \notag \\ 
&+ \eps^{2} \int_{\Omega} \Laplace \vartheta \,  \Laplace \varphi \dx = 0, \label{eq:HSCHweak.varphi} 
\end{align}
\end{subequations}
hold for all $\psi \in \mathcal{V}$ and $\vartheta \in \mathcal{H}$ with $\varphi(t=0) = \varphi_{0}$ in $\Omega$.

\subsection{Spatial approximation}\label{sec:approx.space}
For the spatial approximation of \eqref{eq:HSCHweak} we use NURBS-based Isogeometric Analysis (IGA) \cite{Cottrell09, Hughes05}.  Indeed, in \eqref{eq:HSCHweak} we look for a solution $\varphi \in \mathcal{H} \subset H^{2}(\Omega)$ for all $t \in (0,T)$, i.e., we need $H^{2}(\Omega)$-conformal finite dimensional test and trial function spaces, say $\mathcal{H}_{h}$, which are comprised of globally $C^{1}$-continuous basis functions.  This requirement can be straightforwardly fulfilled by using B-splines (or NURBS) basis functions \cite{Piegl97} of degree $p \geq 2$; we refer the interested reader to \cite{Bartezzaghi15, DedeBordenHughes, Gomez08, LiuDede, Tagliabue} for an overview of high order PDEs -- including phase field problems -- solved by means of NURBS-based IGA.

We introduce the bivariate B-splines basis $\left\{ N_{A}(\mathrm{\mathbf{x}}) \right\}_{A=1}^{n_{bf}}$ and we write the approximate pressure and order parameter as 
\begin{align*}
q_{h}(\mathrm{\mathbf{x}},t) = \sum_{A=1}^{n_{bf}} N_{A}(\mathrm{\mathbf{x}}) \, q_{A}(t) \text{ and }\varphi_{h}(\mathrm{\mathbf{x}},t) = \sum_{A=1}^{n_{bf}} N_{A}(\mathrm{\mathbf{x}})\, \varphi_{A}(t),
\end{align*}
respectively, with the control variables $\left\{ q_{A}(t) \right\}_{A=1}^{n_{bf}}$ and $\left\{ \varphi_{A}(t) \right\}_{A=1}^{n_{bf}}$ being time-dependent.  By introducing the B-splines space $\mathcal{N}_{h} = \text{span}\left\{N_{A}, \ \ A = 1, \ldots, n_{bf} \right\}$, we define the finite dimensional spaces $\mathcal{V}_{h} := \mathcal{V} \bigcap \mathcal{N}_{h}$ and $\mathcal{H}_{h} := \mathcal{H} \bigcap \mathcal{N}_{h}$.  Then, the semi-discrete formulation of \eqref{eq:HSCHweak} reads as follows:  Find, for all $t \in (0,T)$, $q_{h} \in \mathcal{V}_{h}$, $\varphi_{h} \in \mathcal{H}_{h}$, such that
\begin{subequations}\label{eq:HSCHweakIGA}
\begin{alignat}{3}
\int_{\Omega} \nabla \psi_{h} \cdot \left( \frac{1}{12 \eta(\varphi_{h})} \nabla q_{h} \right) \dx & - \mathrm{Bo} \, \int_{\Omega} \nabla \psi_{h} \cdot \hat{\bm{g}} \frac{\rho(\varphi_{h})}{12 \eta(\varphi_{h})} \dx  \notag \\
& - \frac 1 {\mathrm{Ca}} \int_{\Omega} \nabla \psi_{h} \cdot \left( \frac{\mu(\varphi_{h})}{12 \eta(\varphi_{h})} \nabla \varphi_{h} \right) \dx = \int_{\Gamma_{N}} \psi_{h} \, f_{N} \dHaus,\\
\int_{\Omega} \vartheta_{h} \, \pd_{t} \varphi_{h} \dx &+ \int_{\Omega} \vartheta_{h} \, \bm{v}(q_{h},\varphi_{h}) \cdot \nabla \varphi_{h} \dx + \int_{\Omega} \nabla \vartheta_{h} \cdot \left( \Psi''(\varphi_{h}) \nabla \varphi_{h} \right) \dx \notag \\ 
& + \eps^{2} \int_{\Omega} \Laplace \vartheta_{h} \, \Laplace \varphi_{h} \dx = 0, \label{eq:HSCHweakIGA:phase}
\end{alignat}
\end{subequations}
hold for all $\psi_{h} \in \mathcal{V}_{h}$ and $\vartheta_{h} \in \mathcal{H}_{h}$ with $\varphi_{h}(t=0) = \varphi_{0,h}$ in $\Omega$, where $\varphi_{0,h}$ is the $L^{2}(\Omega)$ projection of the initial condition $\varphi_{0}$ onto the space $\mathcal{N}_{h}$.

\subsection{Time discretization}\label{sec:approx.time}
The time discretization of \eqref{eq:HSCHweakIGA} is based on Backward Differentiation Formulas (BDF)  \cite{GSV06, QSS07} with equal order temporal extrapolations based on Newton--Gregory backward polynomials \cite{CK06, Rao09}.  Using this semi-implicit formulation yields a fully discrete problem which can be solved in a computationally efficient and accurate manner; see for example \cite{FortiDede} and \cite{Bartezzaghi16} for the use of the BDF scheme together with NURBS-based IGA spatial approximations of the PDEs.

We partition the time interval $[0,T]$ into $N_{t}$ subintervals of equal size $\Delta t = \frac{T}{N_{t}}$ yielding the discrete time instances $t_{n} = n\,\Delta t$ for $n=0, \ldots, N_{t}$.  Then, we denote with $q_{h}^{n}$ and $\varphi_{h}^{n}$ the approximations of the pressure $q_{h}$ and order parameter $\varphi_{h}$ at the time $t_{n}$.  The approximation of $\pd_{t} \varphi_{h}$ in \eqref{eq:HSCHweakIGA} by a $\sigma$-order BDF scheme is 
\begin{align*}
\pd_{t} \varphi_{h} \approx \frac{\alpha_{\sigma} \varphi_{h}^{n+1} - \varphi_{h}^{n,\text{BDF}\sigma}}{\Delta t}.
\end{align*}
For example for $\sigma = 1$, we have $\alpha_{\sigma} = 1$ and $\varphi_{h}^{n,\text{BDF}\sigma} = \varphi_{h}^{n}$ for $n \geq 0$; instead, for $\sigma = 2$, $\alpha_{\sigma} = \frac{3}{2}$ and $\varphi_{h}^{n,\text{BDF} \sigma}= 2 \varphi_{h}^{n} - \frac{1}{2}\varphi_{h}^{n-1}$ for $n \geq 1$.  Then, replacing the derivative $\pd_{t} \varphi_{h}$ in \eqref{eq:HSCHweakIGA:phase} with the $\sigma$-order BDF approximation, while the other time dependent terms are evaluated at the time instance $t_{n+1}$ (i.e., terms involving $q_{h}^{n+1}$ and $\varphi_{h}^{n+1}$), yields a nonlinear fully discrete problem at each time instance (for example, for $\sigma = 1$ we have the backward Euler scheme).

In order to obtain a semi-implicit fully discrete problem, the nonlinear terms depending on $q_{h}^{n+1}$ and $\varphi_{h}^{n+1}$ are replaced by extrapolations of order $\sigma$ by means of the Newton--Gregory backward polynomials, say $q_{h}^{n+1,\sigma}$ and $\varphi_{h}^{n+1,\sigma}$, respectively.  For example, for $\sigma = 1$, these are $q_{h}^{n+1,\sigma} = q_{h}^{n}$ and $\varphi_{h}^{n+1,\sigma} = \varphi_{h}^{n}$ for $n\geq 0$; while, for $\sigma = 2$, we have $q_{h}^{n+1,\sigma} = 2\, q_{h}^{n} - q_{h}^{n-1}$ and $\varphi_{h}^{n+1,\sigma}= 2\, \varphi_{h}^{n} -\varphi_{h}^{n-1}$ for $n \geq 1$.  For a BDF scheme of order $\sigma$, the semi-implicit formulation of the fully discrete problem reads as follows:  Find, for all $n \geq \sigma - 1$, $q_{h}^{n+1} \in \mathcal{V}_{h}$, $\varphi_{h}^{n+1} \in \mathcal{H}_{h}$, such that
\begin{subequations}\label{eq:HSCHweakBDF}
\begin{alignat}{3}
\int_{\Omega} \nabla \psi_{h} \cdot \left( \frac{1}{12 \eta \left (\varphi_{h}^{n+1,\sigma} \right )} \nabla q_{h} \right) \dx & - \mathrm{Bo} \, \Theta_{1} \int_{\Omega} \nabla \psi_{h} \cdot \hat{\bm{g}} \frac{1}{12 \eta \left (\varphi_{h}^{n+1,\sigma} \right )} \varphi_{h}^{n+1} \dx  \notag \\
& - \frac 1 {\mathrm{Ca}} \int_{\Omega} \nabla \psi_{h} \cdot \left( \frac{\mu \left (\varphi_{h}^{n+1,\sigma} \right )}{12 \eta \left (\varphi_{h}^{n+1,\sigma} \right )} \nabla \varphi_{h} \right) \dx \notag \\
& =  \mathrm{Bo} \, \Theta_{2} \int_{\Omega} \nabla \psi_{h} \cdot \hat{\bm{g}} \frac{1}{12 \eta \left (\varphi_{h}^{n+1,\sigma} \right )} \dx + \int_{\Gamma_{N}} \psi_{h} \, f_N \dHaus,\\
\frac{\alpha_{\sigma}}{\Delta t} \int_{\Omega} \vartheta_{h} \, \varphi_{h}^{n+1} \dx & - \int_{\Omega} \vartheta_{h} \left( \frac{1}{12 \eta \left (\varphi_{h}^{n+1,\sigma} \right )} \nabla \varphi_{h}^{n+1,\sigma} \right) \cdot \nabla q_{h}^{n+1} \dx \notag \\ 
&+\mathrm{Bo}\, \Theta_{1} \int_{\Omega} \vartheta_{h} \left( \frac{1}{12 \eta \left (\varphi_{h}^{n+1,\sigma} \right )} \nabla \varphi_{h}^{n+1,\sigma} \cdot \hat{\bm{g}} \right ) \varphi_{h}^{n+1} \dx \notag \\ 
&+\frac{1}{\mathrm{Ca}} \int_{\Omega} \vartheta_{h} \, \frac{\mu \left (\varphi_{h}^{n+1,\sigma} \right )}{12 \eta \left (\varphi_{h}^{n+1,\sigma} \right )}  \nabla \varphi_{h}^{n+1,\sigma} \cdot \nabla \varphi_{h}^{n+1} \dx \notag \\ 
& + \int_{\Omega} \nabla \vartheta_{h} \cdot \left( \Psi'' \left (\varphi_{h}^{n+1,\sigma} \right ) \nabla \varphi_{h}^{n+1} \right) \dx + \eps^2 \int_{\Omega} \Delta \vartheta_{h} \,  \Delta \varphi_{h}^{n+1} \dx \notag \\ 
& = \frac{1}{\Delta t} \int_{\Omega} \vartheta_{h} \, \varphi_{h}^{n, \text{BDF} \sigma} \dx \notag\\
&- \mathrm{Bo} \, \Theta_{2} \int_{\Omega} \vartheta_{h} \, \nabla \varphi_{h}^{n+1,\sigma} \cdot \hat{\bm{g}} \frac{1}{12 \eta \left (\varphi_{h}^{n+1,\sigma} \right )} \dx,
\end{alignat}
\end{subequations}
hold for all $\psi_{h} \in \mathcal{V}_{h}$ and $\vartheta_{h} \in \mathcal{H}_{h}$ with $\varphi_{h}^{0} = \varphi_{0,h}$ in $\Omega$.

\section{Numerical results}\label{sec:reso}
We present some numerical results for the Hele--Shaw--Cahn--Hilliard model for incompressible flows.  Specifically, we solve two benchmark problems: the \emph{rising bubble} test, for which a less dense fluid rises into a more dense one in presence of a gravitational field as e.g. in \cite{ Hysing, LLG01}, and the \emph{viscous fingering} test, for which a less viscous fluid is injected into a more viscous one \cite{Saffman}.

For both the tests, we use the (dimensionless) computational domain $\Omega = (0,0.5) \times (0,1)$.  For the spatial approximation, we consider NURBS-based IGA with globally $C^{1}$-continuous B-splines basis functions of degree $p=2$ -- as described in Section \ref{sec:approx.space} -- with $32,768$ equally-sized mesh elements, yielding the dimensionless mesh size $h = \frac{1}{256}$ and a total of $n_{bf}=33,540$ B-splines basis functions.
For the time discretization, we use the semi-implicit formulation \eqref{eq:HSCHweakBDF} with the BDF scheme of order $\sigma = 2$; the time step size $\Delta t$ and $T$ are specified later for the two tests.

\begin{figure}
\begin{center}
\begin{tabular}{ccc}
\scalebox{0.24}{\includegraphics[angle=0, trim=10 10 10 10]{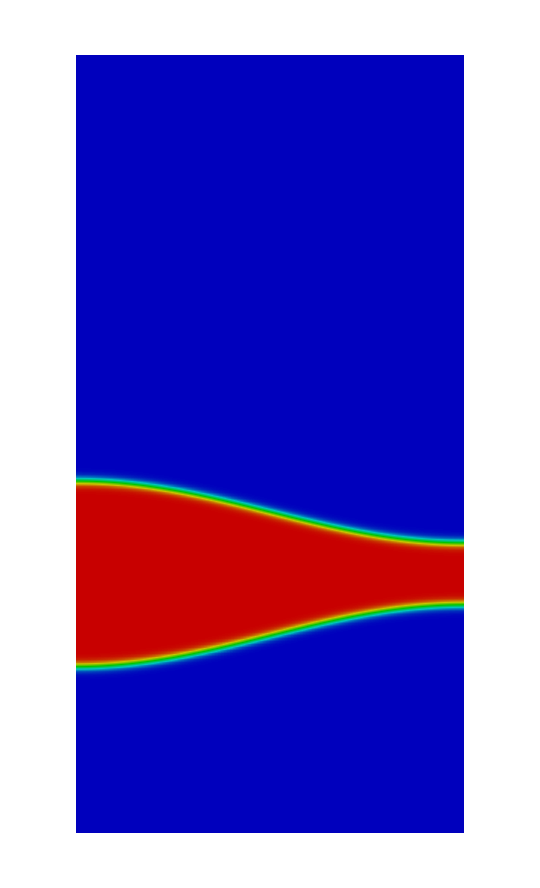}}  &
\scalebox{0.24}{\includegraphics[angle=0, trim=10 10 10 10]{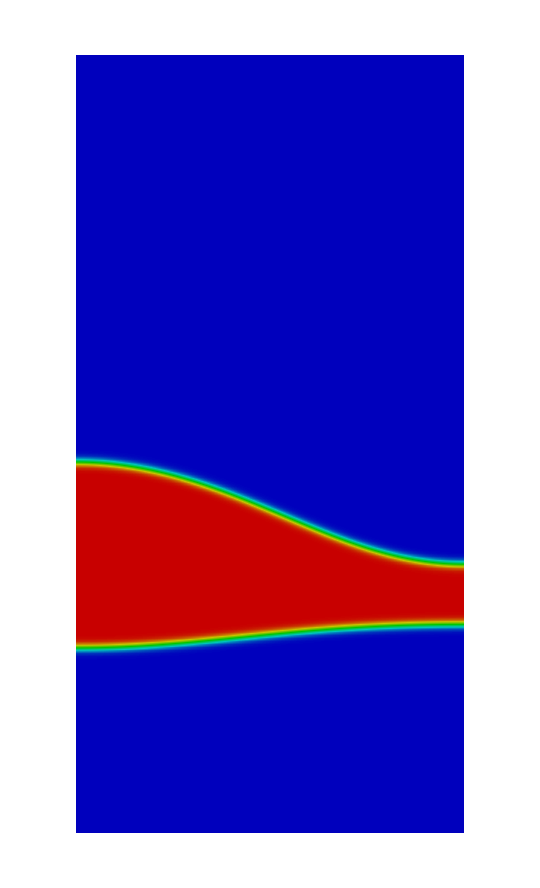}} &
\scalebox{0.24}{\includegraphics[angle=0, trim=10 10 10 10]{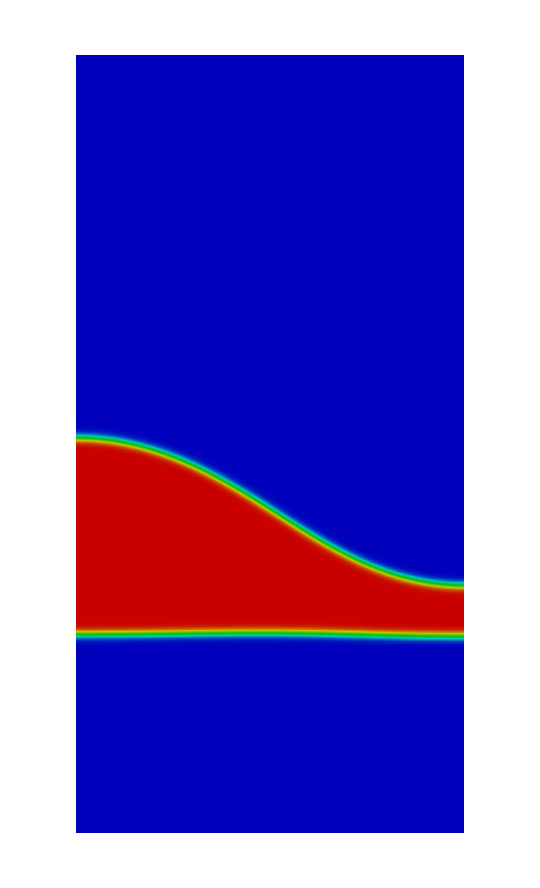}} \\[-0.3cm]
$t = 0.0$ & $t = 5.000 \cdot 10^{-2}$ & $t = 1.000 \cdot 10^{-1}$ \\
\scalebox{0.24}{\includegraphics[angle=0, trim=10 10 10 10]{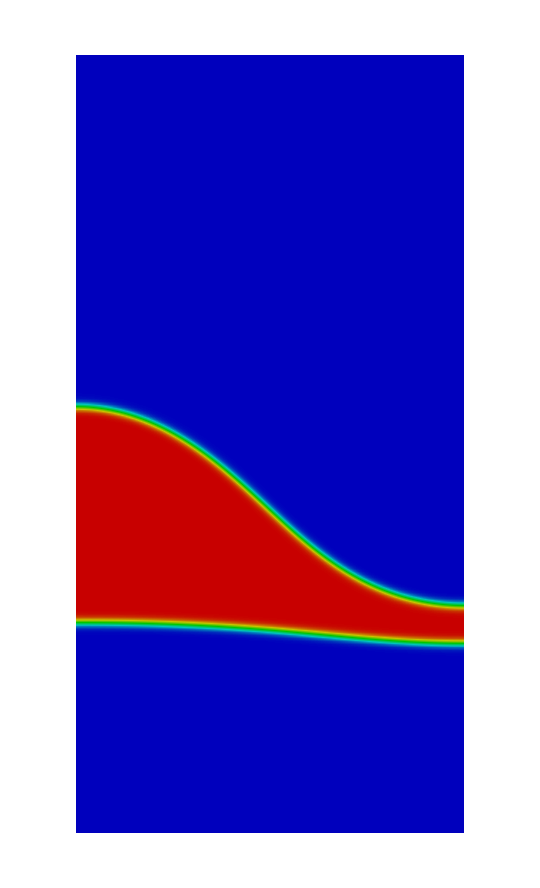}} &
\scalebox{0.24}{\includegraphics[angle=0, trim=10 10 10 10]{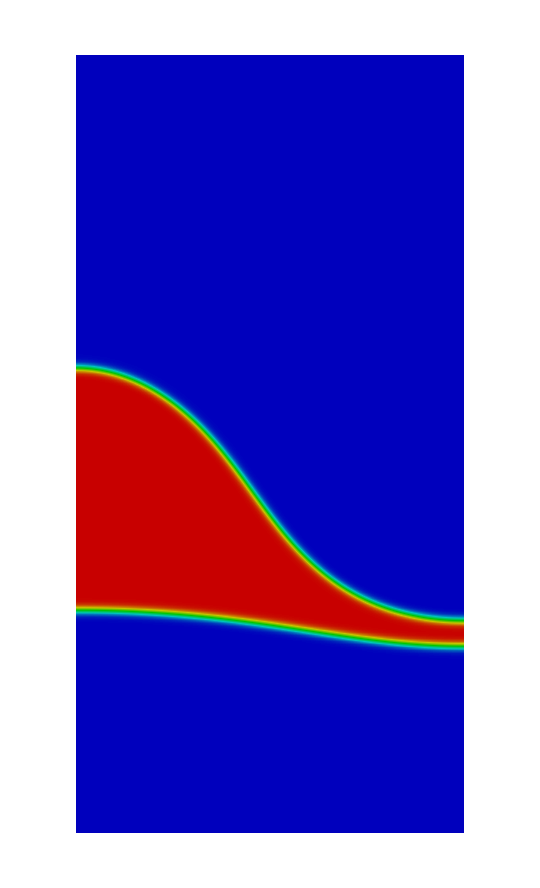}} &
\scalebox{0.24}{\includegraphics[angle=0, trim=10 10 10 10]{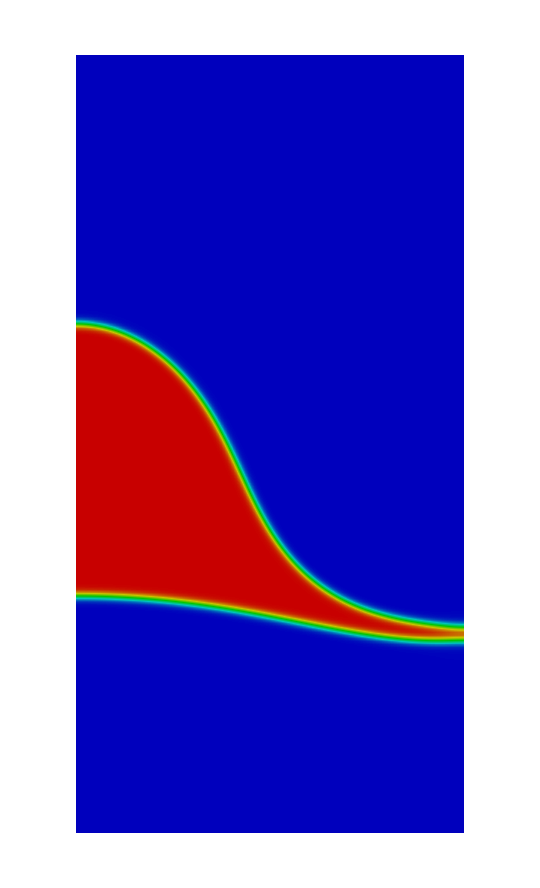}}  \\ [-0.3cm]
$t = 1.500 \cdot 10^{-1}$ & $t = 2.000 \cdot 10^{-1}$ & $t = 2.490 \cdot 10^{-1}$ \\
\scalebox{0.24}{\includegraphics[angle=0, trim=10 10 10 10]{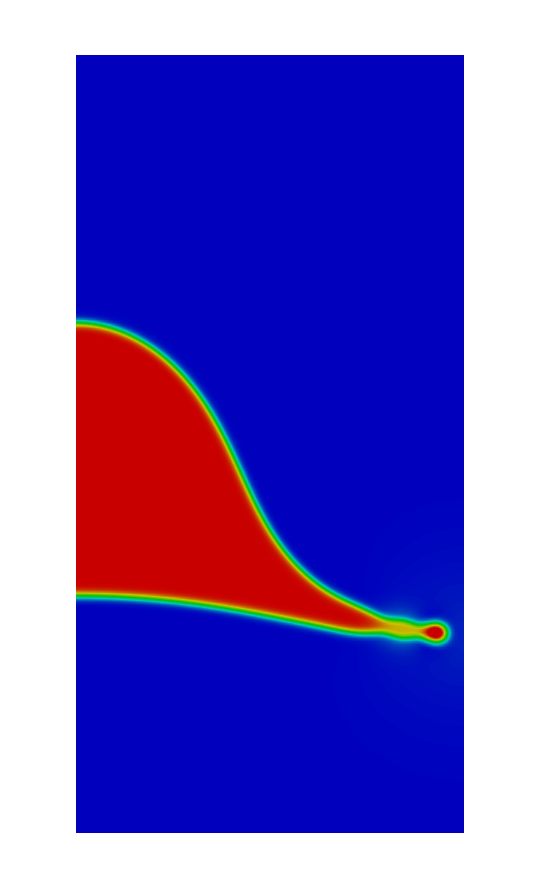}} &
\scalebox{0.24}{\includegraphics[angle=0, trim=10 10 10 10]{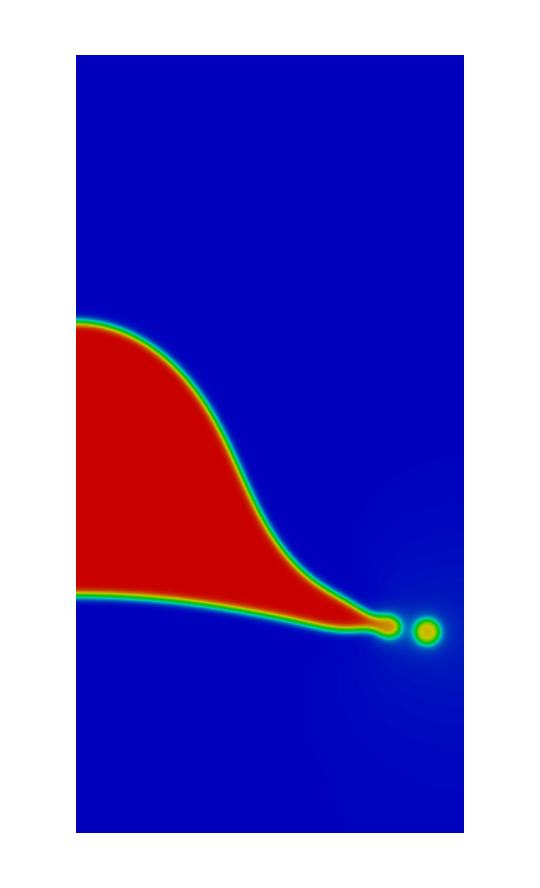}}  &
\scalebox{0.24}{\includegraphics[angle=0, trim=10 10 10 10]{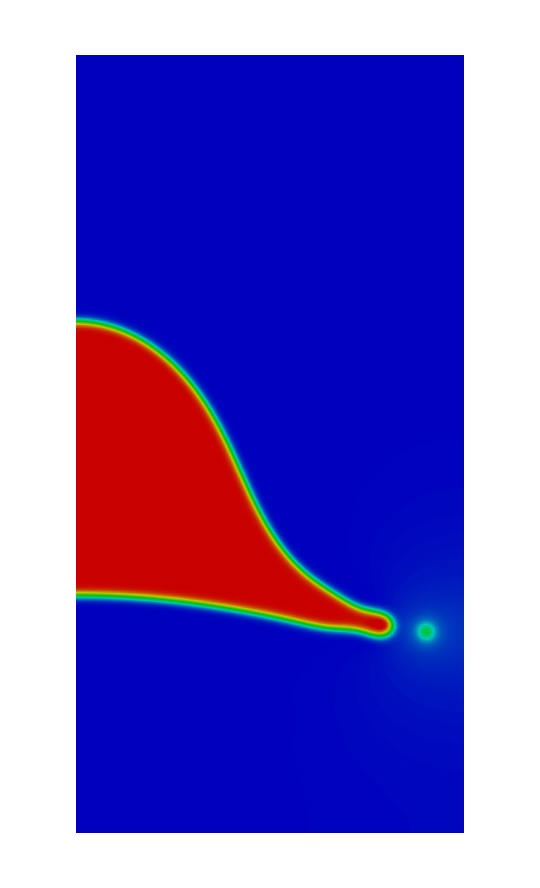}}  \\ [-0.3cm]
$t = 2.500 \cdot 10^{-1}$ & $t = 2.510 \cdot 10^{-1}$ & $t = 2.515 \cdot 10^{-1}$ \\
\end{tabular}
\caption{Test~$1$. Rising bubble for $\rho_{1}=5$ and $\rho_{2}=1$. Phases evolution at different time instances.}
\label{fig:G1_phase_1}
\end{center}
\end{figure}

\begin{figure}
\begin{center}
\begin{tabular}{ccc}
\scalebox{0.24}{\includegraphics[angle=0, trim=10 10 10 10]{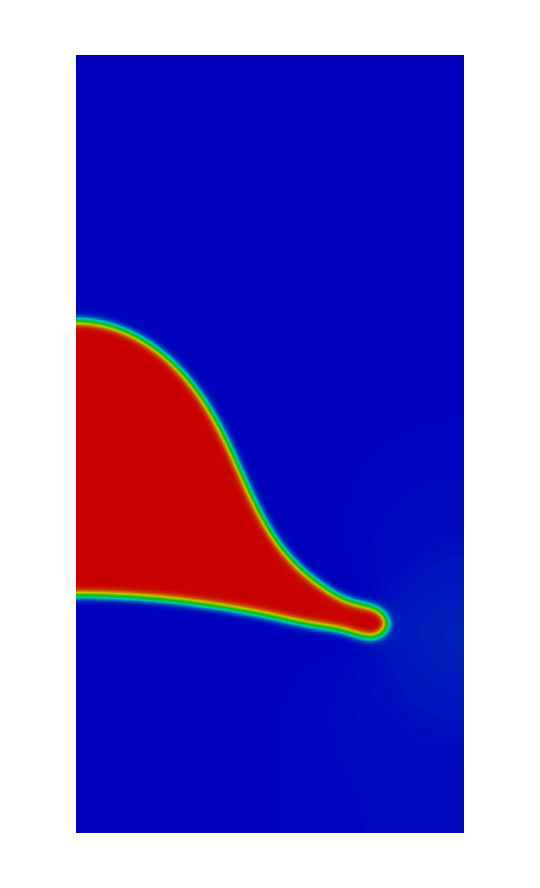}} & 
\scalebox{0.24}{\includegraphics[angle=0, trim=10 10 10 10]{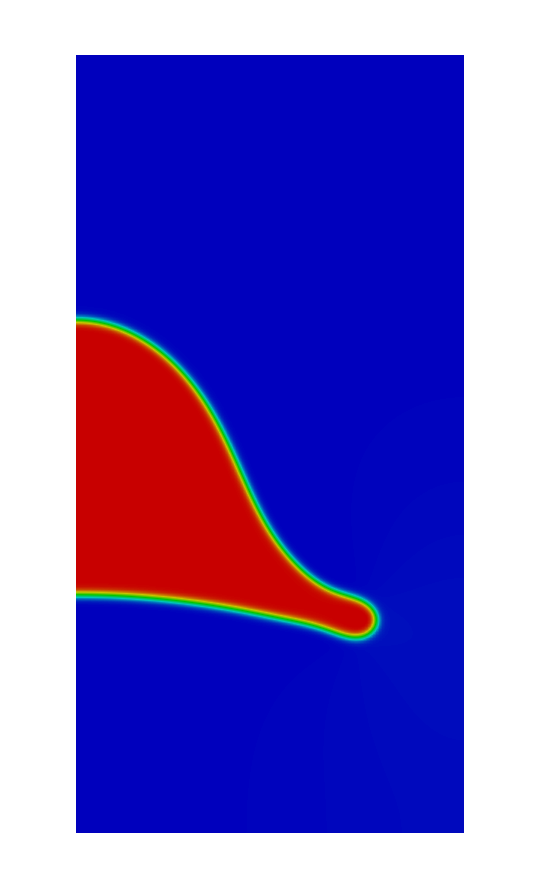}}  &
\scalebox{0.24}{\includegraphics[angle=0, trim=10 10 10 10]{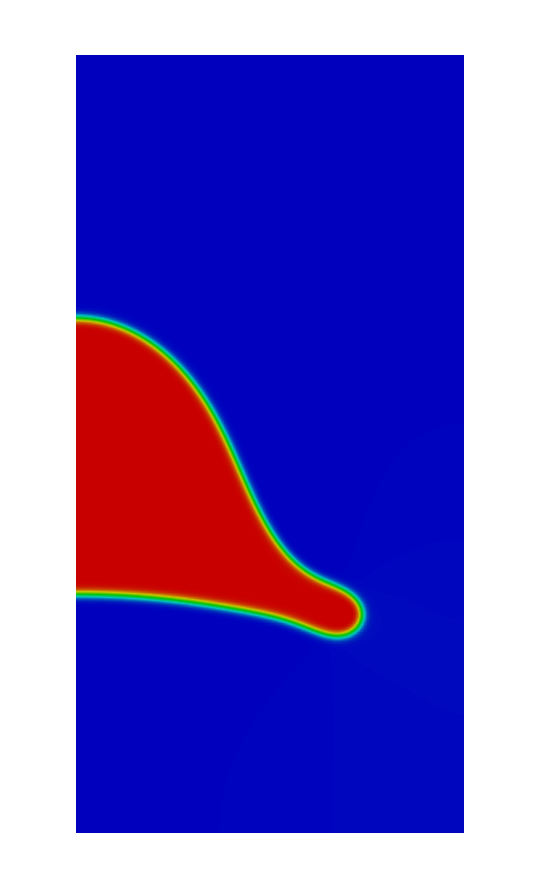}} \\ [-0.3cm]
$t = 2.520 \cdot 10^{-1}$ & $t = 2.530 \cdot 10^{-1}$ & $t = 2.550 \cdot 10^{-1}$ \\
\scalebox{0.24}{\includegraphics[angle=0, trim=10 10 10 10]{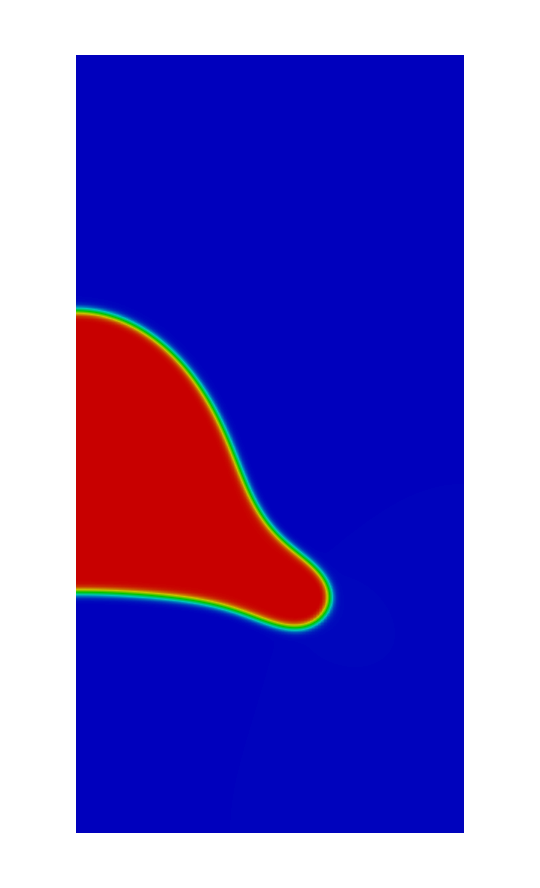}}  &
\scalebox{0.24}{\includegraphics[angle=0, trim=10 10 10 10]{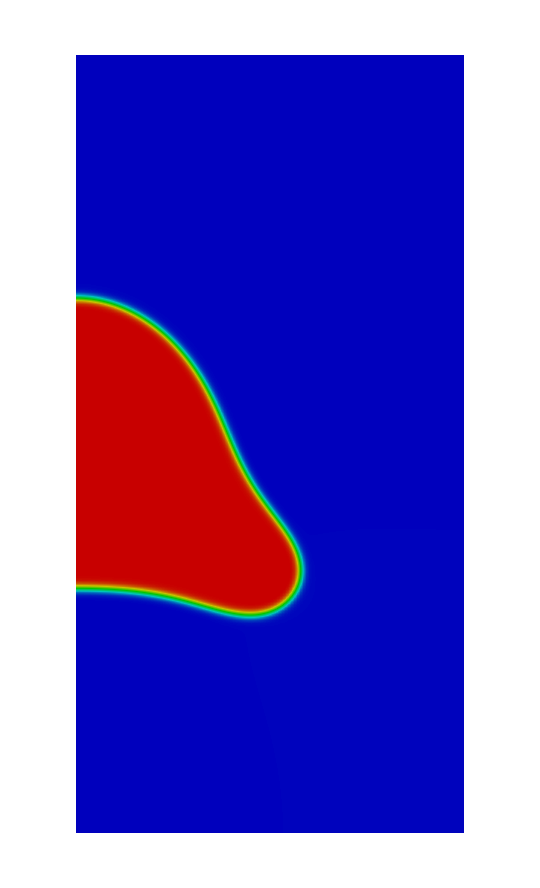}} &
\scalebox{0.24}{\includegraphics[angle=0, trim=10 10 10 10]{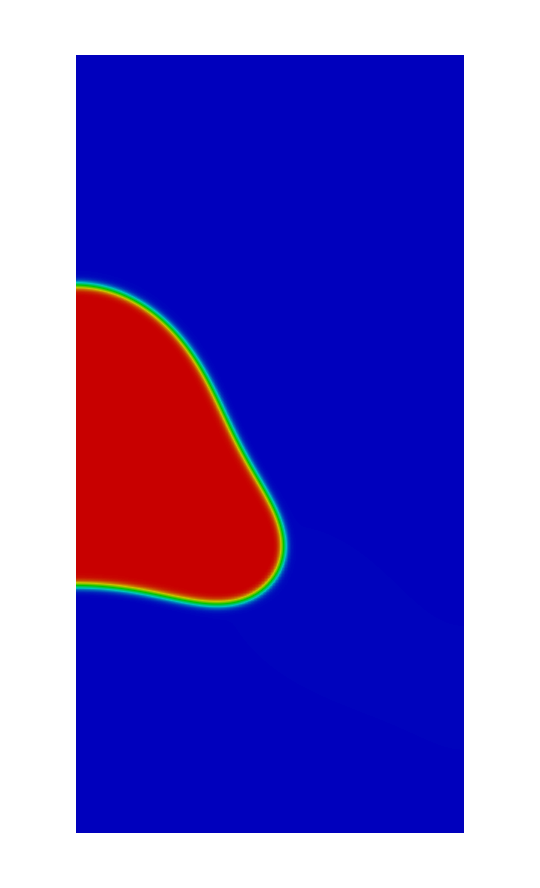}} \\[-0.3cm]
$t=2.625 \cdot 10^{-1}$ & $t=2.750 \cdot 10^{-1}$ & $t=2.875 \cdot 10^{-1}$ \\ 
\scalebox{0.24}{\includegraphics[angle=0, trim=10 10 10 10]{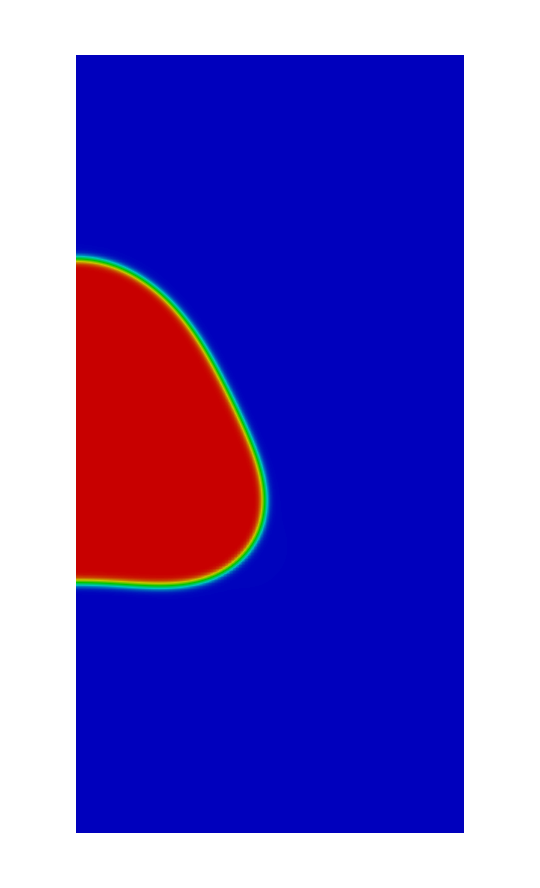}} &
\scalebox{0.24}{\includegraphics[angle=0, trim=10 10 10 10]{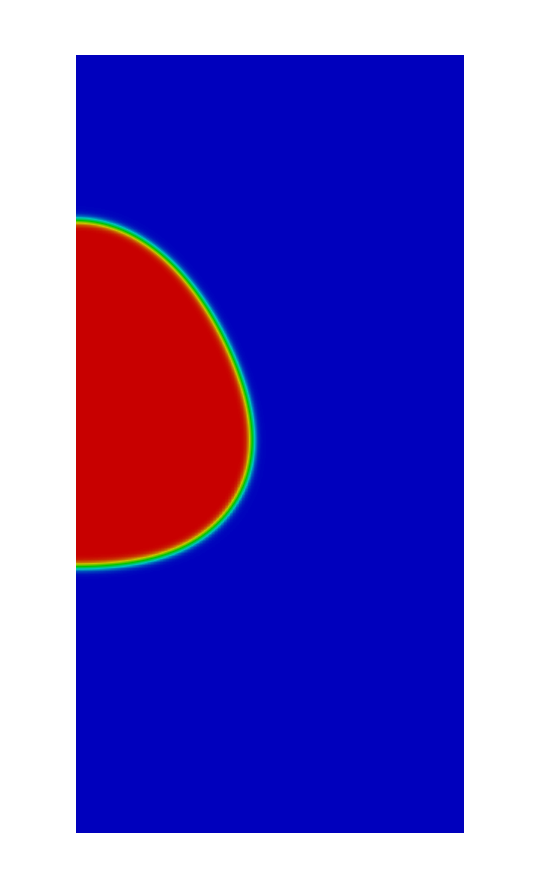}}  &
\scalebox{0.24}{\includegraphics[angle=0, trim=10 10 10 10]{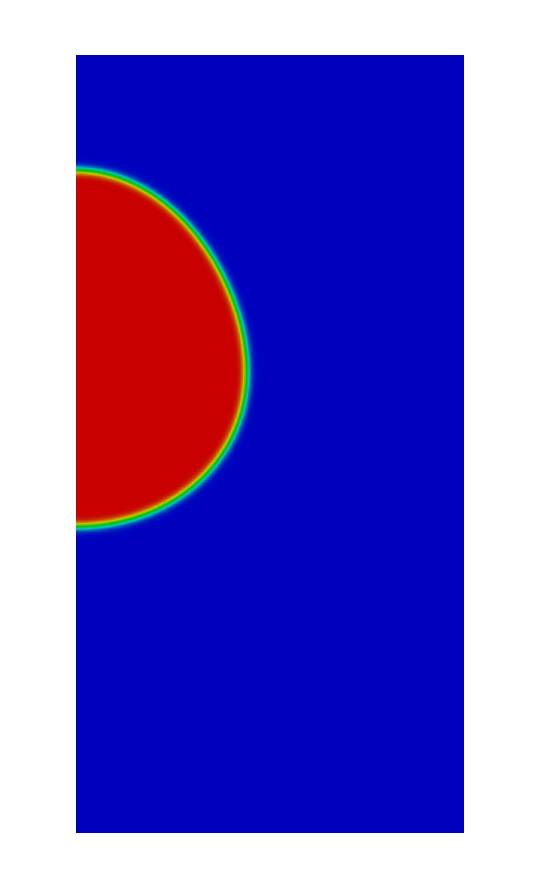}} \\[-0.3cm]
$t = 3.125 \cdot 10^{-1}$ & $t = 3.500 \cdot 10^{-1}$ & $t = 4.000 \cdot 10^{-1}$ \\
\end{tabular}
\caption{Test~$1$. Rising bubble for $\rho_{1}=5$ and $\rho_{2}=1$. Phases evolution at different time instances.}
\label{fig:G1_phase_2}
\end{center}
\end{figure}

\begin{figure}
\begin{center}
\begin{tabular}{ccc}
\scalebox{0.38}{\includegraphics[angle=0, trim=0 0 0 0]{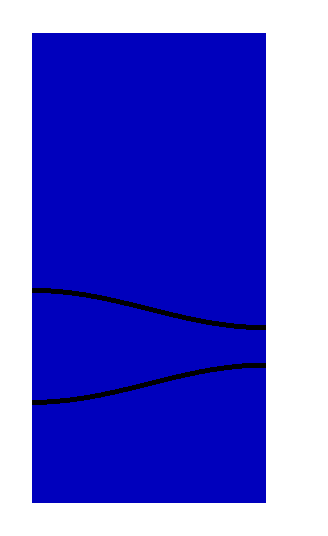}}  &
\scalebox{0.38}{\includegraphics[angle=0, trim=0 0 0 0]{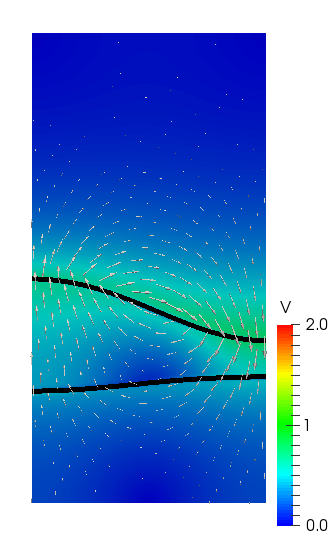}}  & 
\scalebox{0.38}{\includegraphics[angle=0, trim=0 0 0 0]{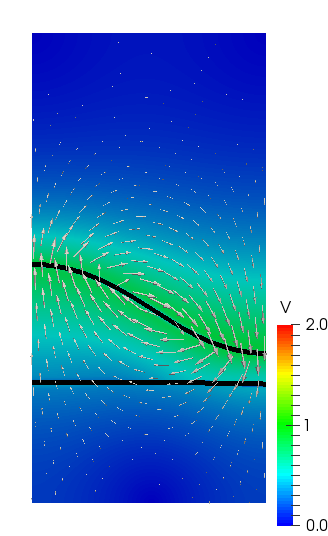}} \\[-0.3cm]
$t= 0.0$ & $t=5.000 \cdot 10^{-2}$ & $t=1.000 \cdot 10^{-1}$ \\ 
\scalebox{0.38}{\includegraphics[angle=0, trim=0 0 0 0]{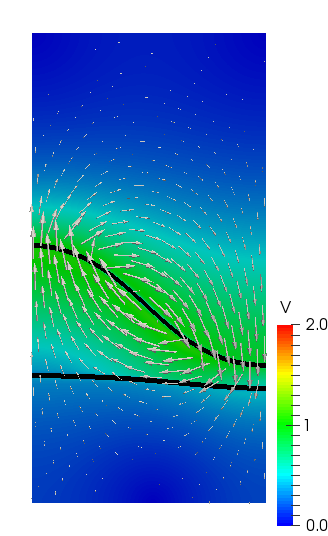}}  &
\scalebox{0.38}{\includegraphics[angle=0, trim=0 0 0 0]{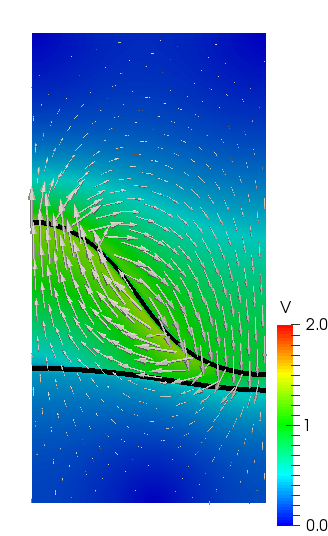}} &
\scalebox{0.38}{\includegraphics[angle=0, trim=0 0 0 0]{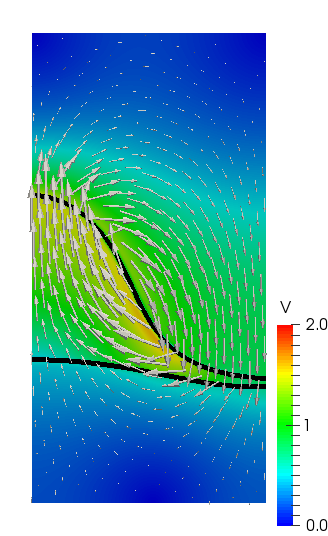}}  \\[-0.3cm]
$t=1.500 \cdot 10^{-1}$ & $t=2.000 \cdot 10^{-1}$ & $t=2.490 \cdot 10^{-1}$ \\ 
\scalebox{0.38}{\includegraphics[angle=0, trim=0 0 0 0]{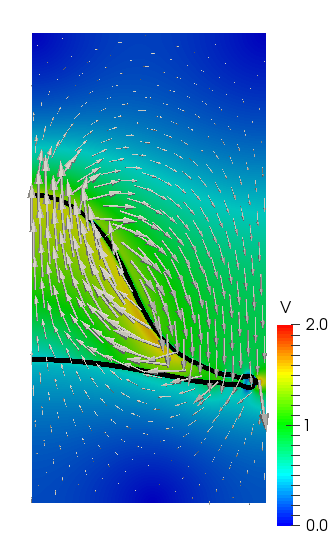}}  &
\scalebox{0.38}{\includegraphics[angle=0, trim=0 0 0 0]{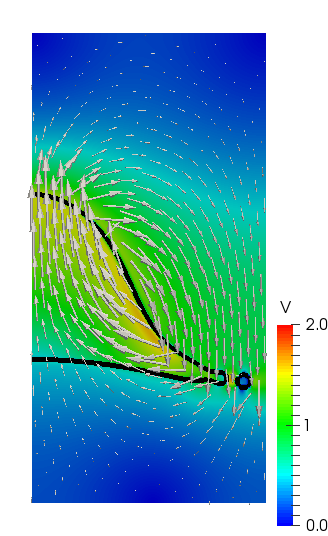}}  &
\scalebox{0.38}{\includegraphics[angle=0, trim=0 0 0 0]{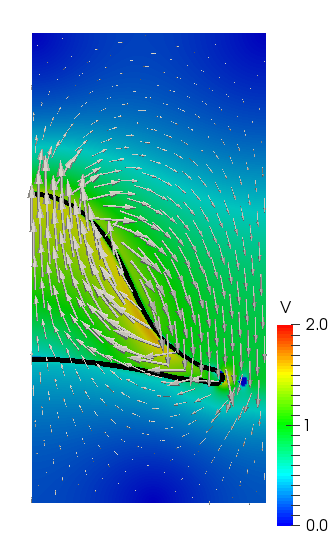}} \\ [-0.3cm]
$t = 2.500 \cdot 10^{-1}$ & $t = 2.510 \cdot 10^{-1}$ & $t = 2.515 \cdot 10^{-1}$ \\
\end{tabular}
\caption{Test~$1$. Rising bubble for $\rho_{1}=5$ and $\rho_{2}=1$. Velocity field at different time instances; the black contour lines highlight the interface among the phases.}
\label{fig:G1_vel_1}
\end{center}
\end{figure}

\begin{figure}
\begin{center}
\begin{tabular}{ccc}
\scalebox{0.38}{\includegraphics[angle=0, trim=0 0 0 0]{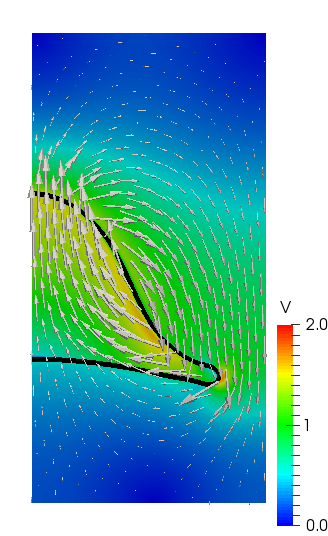}} &
\scalebox{0.38}{\includegraphics[angle=0, trim=0 0 0 0]{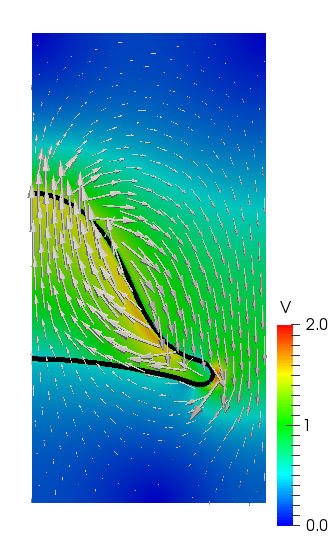}}  &
\scalebox{0.38}{\includegraphics[angle=0, trim=0 0 0 0]{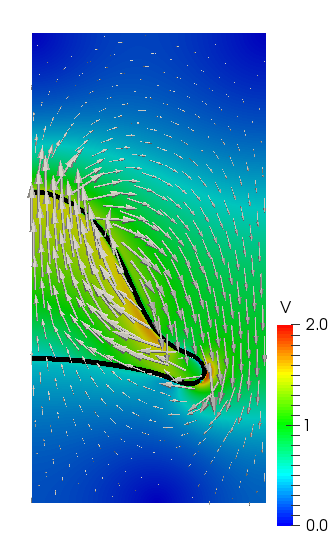}}  \\ [-0.3cm]
$t = 2.520 \cdot 10^{-1}$ & $t = 2.530 \cdot 10^{-1}$ & $t = 2.550 \cdot 10^{-1}$ \\
\scalebox{0.38}{\includegraphics[angle=0, trim=0 0 0 0]{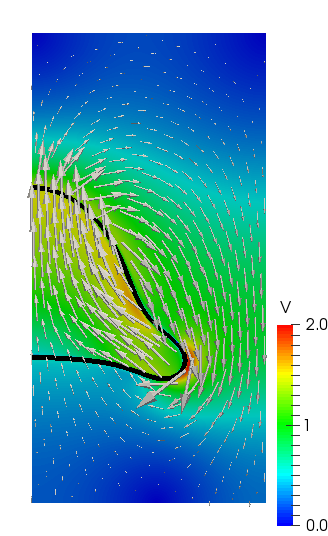}}  &
\scalebox{0.38}{\includegraphics[angle=0, trim=0 0 0 0]{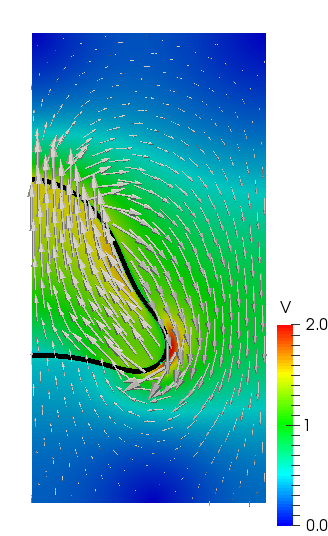}} &
\scalebox{0.38}{\includegraphics[angle=0, trim=0 0 0 0]{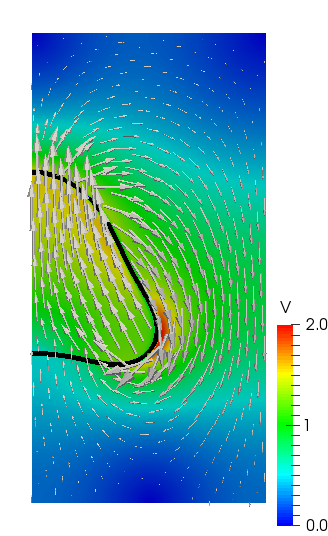}} \\[-0.3cm]
$t=2.625 \cdot 10^{-1}$ & $t=2.750 \cdot 10^{-1}$ & $t=2.875 \cdot 10^{-1}$ \\ 
\scalebox{0.38}{\includegraphics[angle=0, trim=0 0 0 0]{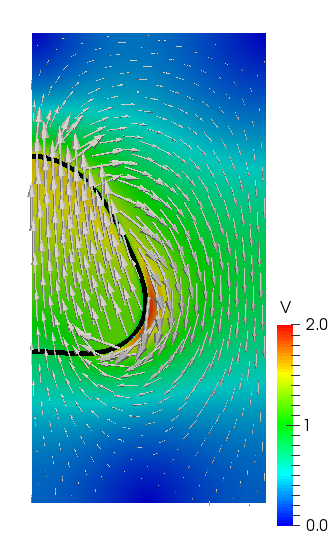}} &
\scalebox{0.38}{\includegraphics[angle=0, trim=0 0 0 0]{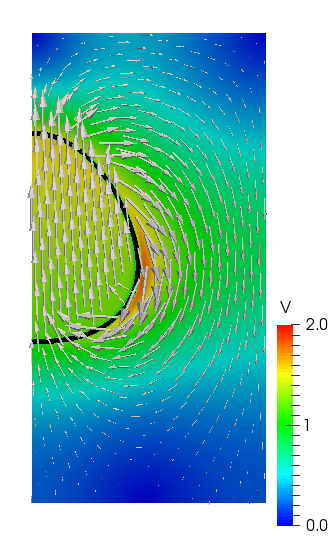}}  &
\scalebox{0.38}{\includegraphics[angle=0, trim=0 0 0 0]{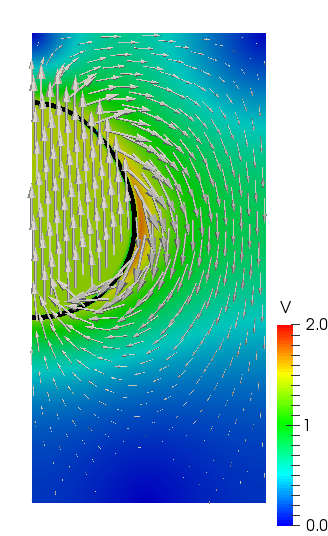}} \\[-0.3cm]
$t = 3.125 \cdot 10^{-1}$ & $t = 3.500 \cdot 10^{-1}$ & $t = 4.000 \cdot 10^{-1}$
\end{tabular}
\caption{Test~$1$. Rising bubble for $\rho_{1}=5$ and $\rho_{2}=1$. Velocity field at different time instances; the black contour lines highlight the interface among the phases.}
\label{fig:G1_vel_2}
\end{center}
\end{figure}

\begin{figure}
\begin{center}
\begin{tabular}{ccc}
\scalebox{0.38}{\includegraphics[angle=0, trim=0 0 0 0]{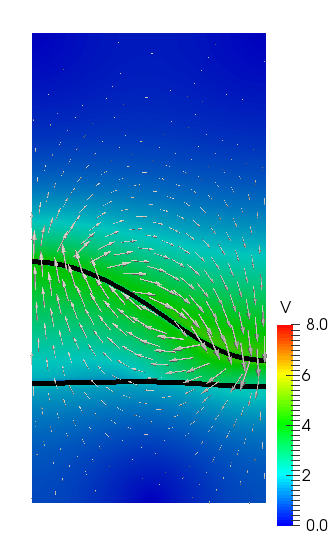}}  &
\scalebox{0.38}{\includegraphics[angle=0, trim=0 0 0 0]{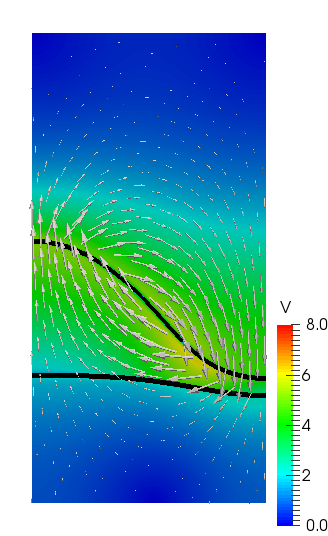}} &
\scalebox{0.38}{\includegraphics[angle=0, trim=0 0 0 0]{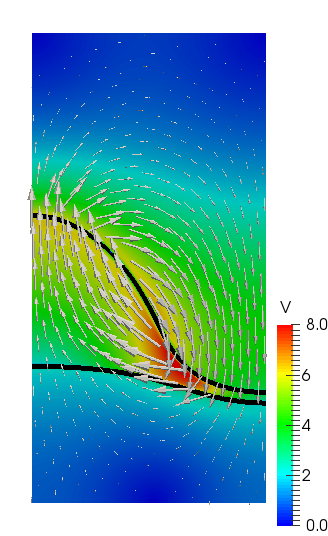}} \\[-0.3cm]
$t= 2.00 \cdot 10^{-2}$ & $t=3.00 \cdot 10^{-2}$ & $t=4.00 \cdot 10^{-2}$ \\ 
\scalebox{0.38}{\includegraphics[angle=0, trim=0 0 0 0]{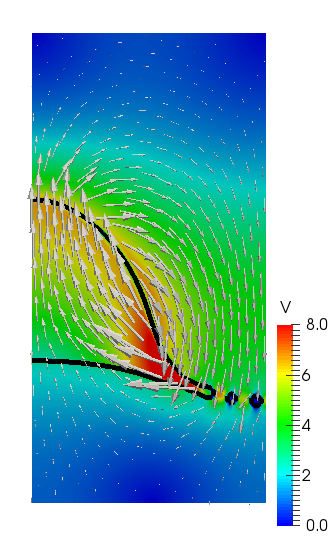}}  &
\scalebox{0.38}{\includegraphics[angle=0, trim=0 0 0 0]{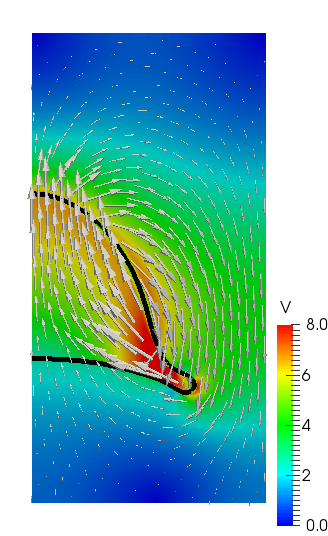}} &
\scalebox{0.38}{\includegraphics[angle=0, trim=0 0 0 0]{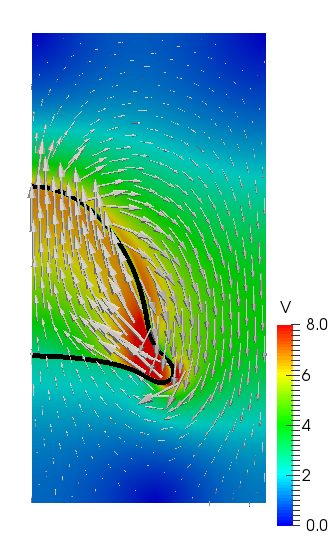}} \\[-0.3cm]
$t=4.55 \cdot 10^{-2}$ & $t=4.75 \cdot 10^{-2}$ & $t=5.00 \cdot 10^{-2}$ \\ 
\scalebox{0.38}{\includegraphics[angle=0, trim=0 0 0 0]{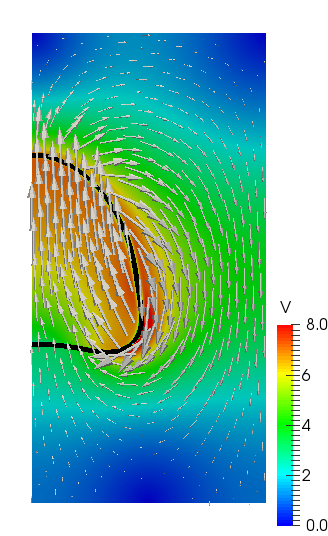}}  &
\scalebox{0.38}{\includegraphics[angle=0, trim=0 0 0 0]{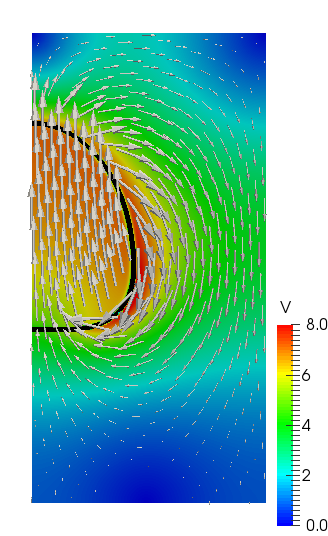}} &
\scalebox{0.38}{\includegraphics[angle=0, trim=0 0 0 0]{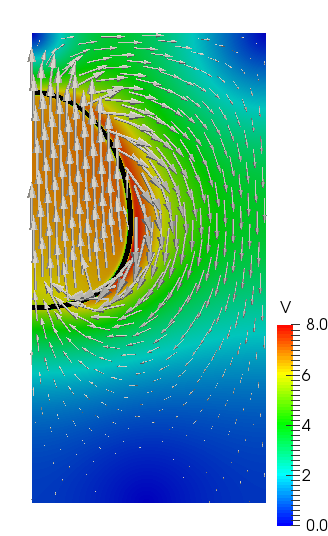}} \\[-0.3cm]
$t=6.00 \cdot 10^{-2}$ & $t=7.00 \cdot 10^{-2}$ & $t=8.00 \cdot 10^{-2}$ \\ 
\end{tabular}
\caption{Test~$1$. Rising bubble for $\rho_{1}=20$ and $\rho_{2}=1$. Velocity field at different time instances; the black contour lines highlight the interface among the phases.}
\label{fig:G5_vel_1}
\end{center}
\end{figure}


\subsection{Test~$1$: rising bubble}\label{sec:reso.bubble}
For this test, we set $\Gamma_{D} = \left\{ (\mathrm{x},\mathrm{y}) \in \Gamma \ : \ \mathrm{y} = 0 \right\}$ and $\Gamma_{N} \equiv \Gamma \backslash \Gamma_{D}$ by referring to the boundary conditions in \eqref{eq:HSCHstrong4th} with $f_{N} = 0$. Then, we set $\rho_{2}=1$, $\eta_{1}=\eta_{2}=0.1$, $\sigma=10^{-5}$, $g = 9.80665$, the characteristic length $L=1$, $\delta=0.1$, $V=0.1$, and $\eps = 2 \, h = \frac{1}{128}$.  In this manner, we have $\mathrm{Ca} = 10^6$ and $\mathrm{Bo} = 0.980665$.  For the time discretization, we set $\Delta t = 2.5 \cdot 10^{-5}$ for $T = 4.0 \cdot 10^{-1}$.  As initial condition for the order parameter, we choose $\varphi_0(\mathrm{x},\mathrm{y}) = \frac{1}{2} \zeta_{+}(\mathrm{x},\mathrm{y}) \zeta_{-}(\mathrm{x},\mathrm{y}) - 1$, with 
\begin{align*}
\zeta_{\pm} (\mathrm{x},\mathrm{y}) = 1 \pm \tanh \left( \frac{4}{3 \eps} \left ( \mathrm{y} - \frac{1}{3} \pm  \frac{1}{4 \pi} \left(1 + \frac{\cos( 2 \pi \mathrm{x})}{2} \right) \right )\right),
\end{align*}
which yields the set-up in Fig.~\ref{fig:G1_phase_1} (top-left), where the blue color is associated to $\varphi = -1$ -- the pure phase labeled ``1" corresponding to the ``heavy'' fluid -- while the red color is associated to $\varphi = +1$ -- the pure phase ``2" corresponding to the ``light" fluid.

We start by considering the case $\rho_{1}=5$, for which $\Theta_{1} = -2$ and $\Theta_{2} = 3$. We report in Figs.~\ref{fig:G1_phase_1}, and ~\ref{fig:G1_phase_2} the time evolution of the order parameter, which highlights the formation and rising of the bubble of light fluid, including topological changes.  Correspondingly, we report in Figs.~\ref{fig:G1_vel_1}, and ~\ref{fig:G1_vel_2} the evolution of the computed velocity field $\bm{v}(q_{h}^{n+1},\varphi_{h}^{n+1})$; as we can observe, relatively high magnitudes of the velocity occur at pinch-off and when the curvature of the interface is significant.

We also consider the case where the density of the heavier fluid is larger, say $\rho_{1}=20$ (for which $\Theta_{1} = -\frac{19}{2}$ and $\Theta_{2}=\frac{21}{2}$) yielding the result highlighted in Fig.~\ref{fig:G5_vel_1} with the velocity field.

\subsection{Test~$2$: viscous fingering}\label{sec:reso.finger}
\begin{figure}
\begin{center}
\begin{tabular}{ccc}
\scalebox{0.24}{\includegraphics[angle=0, trim=10 10 10 10]{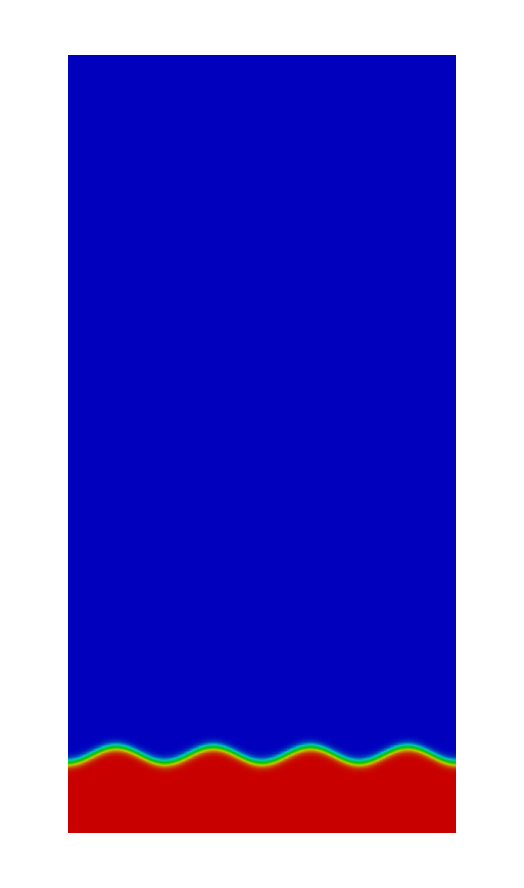}}  &
\scalebox{0.24}{\includegraphics[angle=0, trim=10 10 10 10]{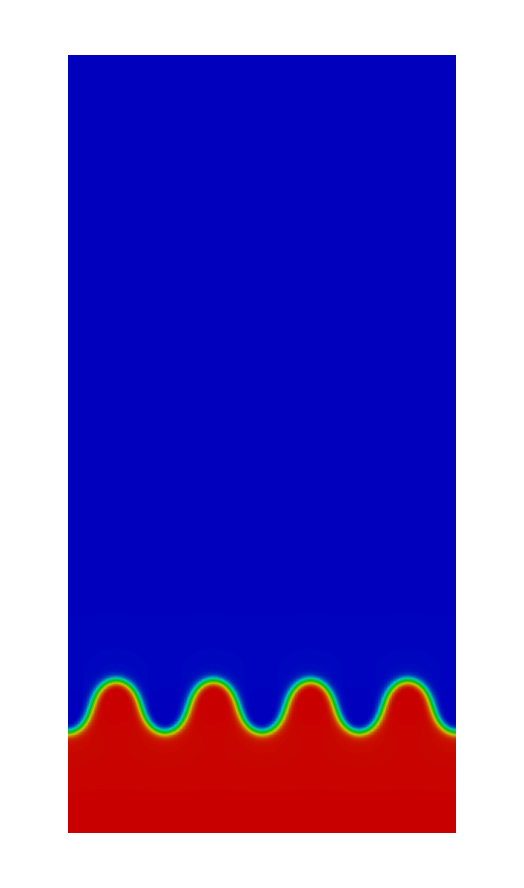}} &
\scalebox{0.24}{\includegraphics[angle=0, trim=10 10 10 10]{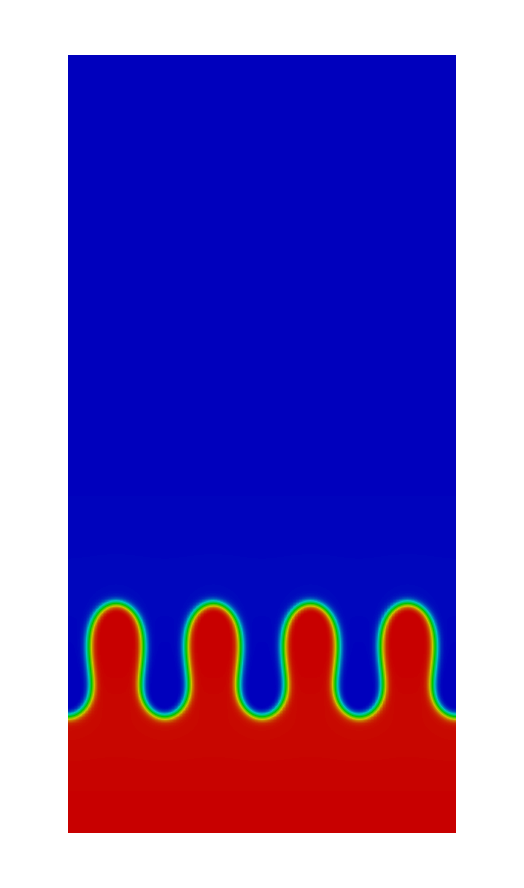}} \\[-0.3cm]
$t= 0.0$ & $t=1.25 \cdot 10^{-4}$ & $t=2.50 \cdot 10^{-4}$ \\ 
\scalebox{0.24}{\includegraphics[angle=0, trim=10 10 10 10]{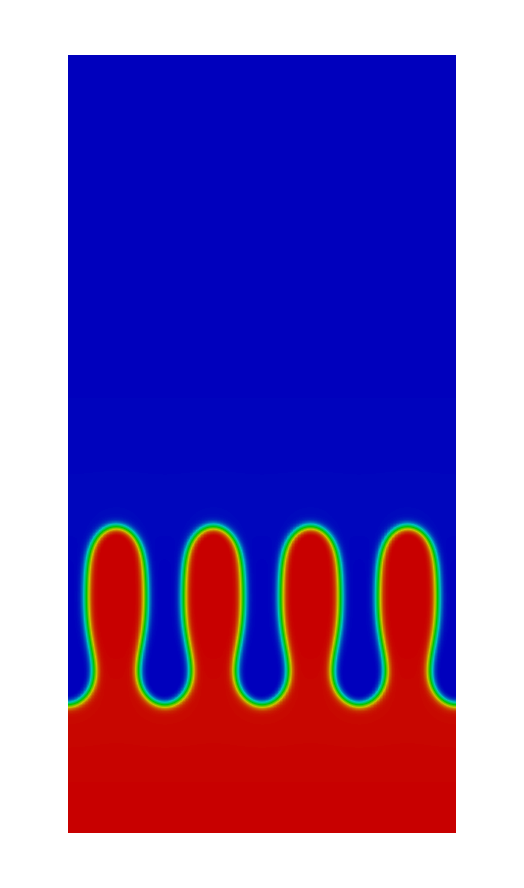}}  &
\scalebox{0.24}{\includegraphics[angle=0, trim=10 10 10 10]{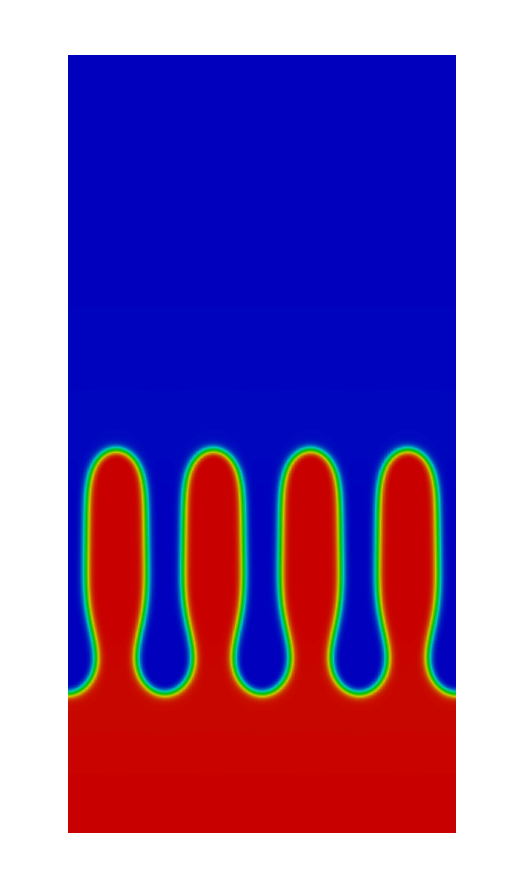}} &
\scalebox{0.24}{\includegraphics[angle=0, trim=10 10 10 10]{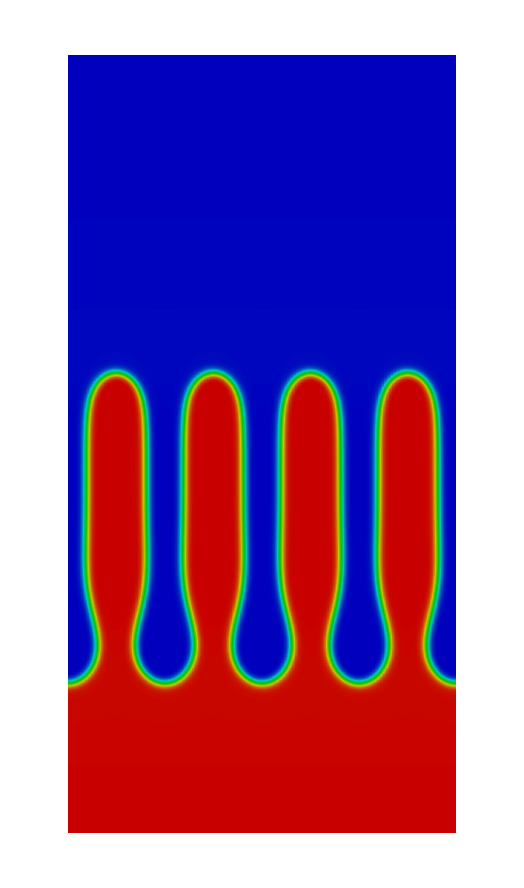}} \\[-0.3cm]
$t=3.75 \cdot 10^{-4}$ & $t=5.00 \cdot 10^{-4}$ & $t=6.25 \cdot 10^{-4}$ \\ 
\scalebox{0.24}{\includegraphics[angle=0, trim=10 10 10 10]{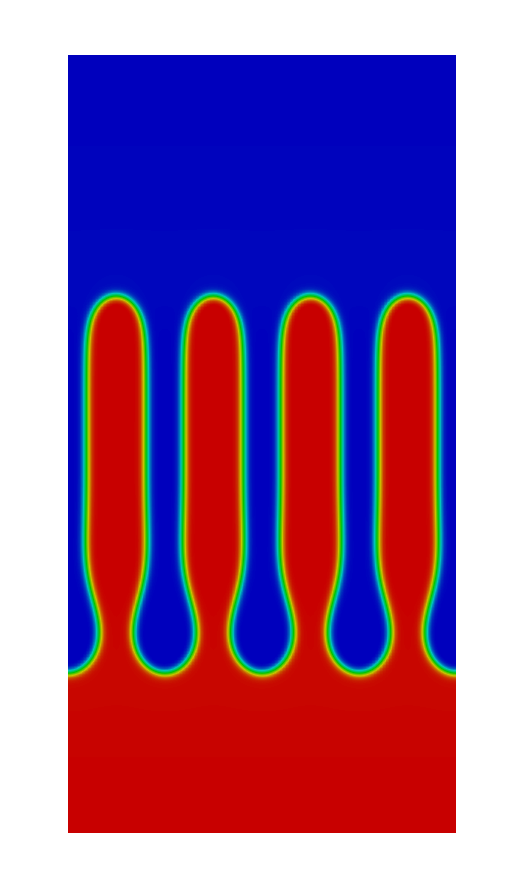}}  &
\scalebox{0.24}{\includegraphics[angle=0, trim=10 10 10 10]{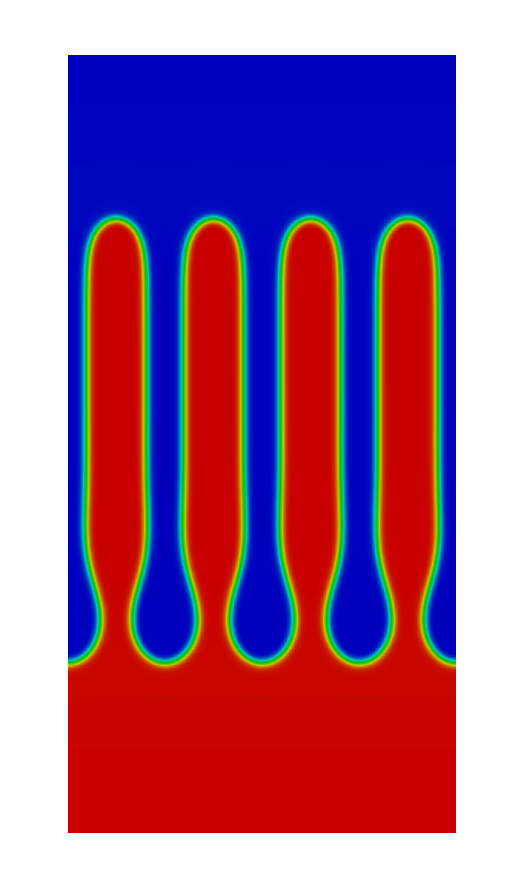}} &
\scalebox{0.24}{\includegraphics[angle=0, trim=10 10 10 10]{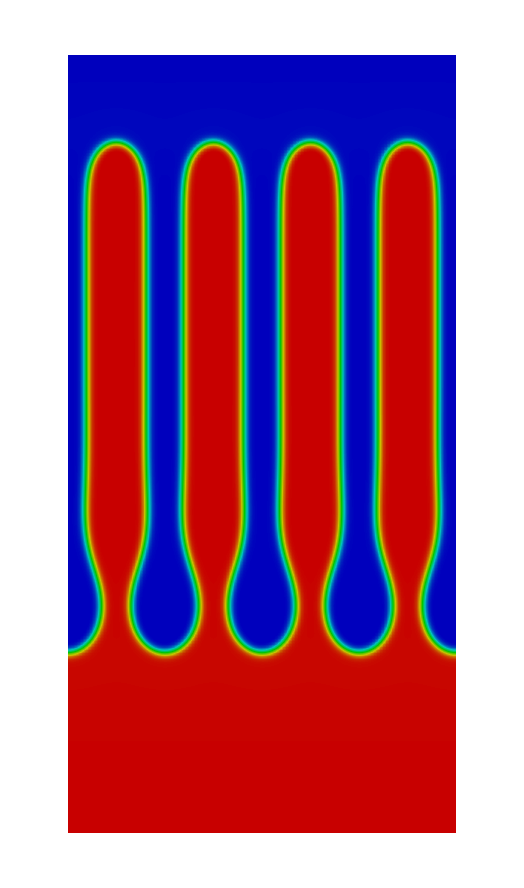}} \\[-0.3cm]
$t=7.50 \cdot 10^{-4}$ & $t=8.75 \cdot 10^{-4}$ & $t=1.00 \cdot 10^{-3}$ \\ 
\end{tabular}
\caption{Test~$2$. Viscous fingering for $\eta_{1}=50$, $\eta_{2}=1$, and $\sigma = 10^{-5}$. Phases evolution at different time instances.}
\label{fig:F1_phase}
\end{center}
\end{figure}

We set $\Gamma_{N} \equiv \Gamma$ with $\Gamma_{D} = \emptyset$ and, in order to enforce the injection of the fluid into the domain, $f_{N}= -V$ on $\Gamma_{N,b} = \left\{ (\mathrm{x},\mathrm{y}) \in \Gamma \ : \ \mathrm{y} = 0 \right\}$, $f_{N} = V$ on $\Gamma_{N,t} = \left\{ (\mathrm{x},\mathrm{y}) \in \Gamma \ : \ \mathrm{y} = 1 \right\}$, and $f_{N}=0$ on $\Gamma_{N} \backslash \left( \Gamma_{N,b} \bigcup \Gamma_{N,t}\right)$, for some injection velocity $V>0$. In this case, in order to obtain a well-posed problem, we prescribe the values of the control coefficients of the pressure field (approximated by IGA) $q_{32,319}= q_{32,320}=0$ for all $n \geq 1$.  Then, we choose $\rho_{1}=\rho_{2}=1$, $\eta_{2}=1$, $g = 0$, $L=1$, $\delta=0.1$, $V=50$, and $\eps = 2 \, h = \frac{1}{128}$, for which $\mathrm{Bo} = 0$.  For the time discretization, we set $\Delta t = 2.5 \cdot 10^{-6}$ for $T = 10^{-3}$.  The initial condition is 
\begin{align*}
\varphi_{0}(\mathrm{x},\mathrm{y}) = - \tanh \left( \frac{4}{3\eps} \left ( \mathrm{y} - \frac{1}{10} +  \frac{\cos( 16 \pi \mathrm{x})}{100} \right )\right),
\end{align*}
which yields the set-up in Fig.~\ref{fig:F1_phase} (top-left); we recall that the blue color is associated to $\varphi=-1$ -- the phase ``1" indicating the more viscous fluid -- while the red color is associated to $\varphi=+1$ -- the phase ``2" indicating the less viscous fluid.

\begin{figure}
\begin{center}
\begin{tabular}{ccc}
\scalebox{0.24}{\includegraphics[angle=0, trim=10 10 10 10]{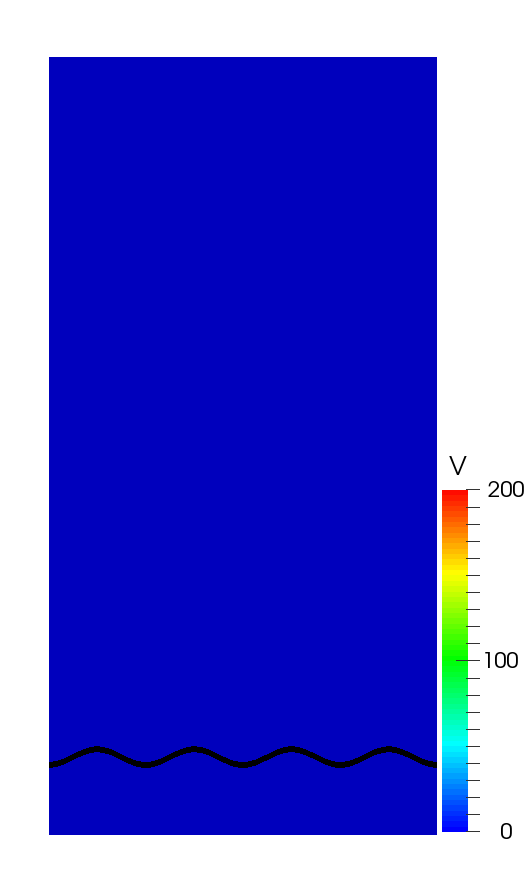}}  &
\scalebox{0.24}{\includegraphics[angle=0, trim=10 10 10 10]{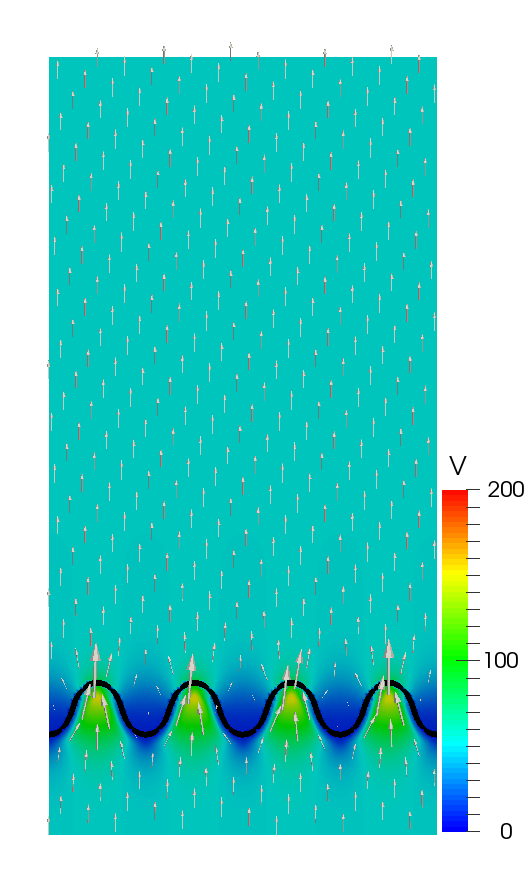}} &
\scalebox{0.24}{\includegraphics[angle=0, trim=10 10 10 10]{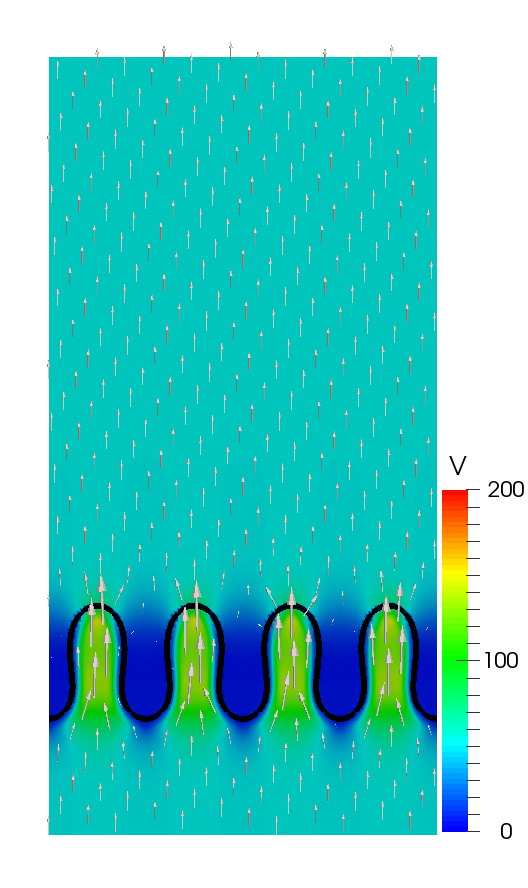}} \\[-0.3cm]
$t= 0.0$ & $t=1.25 \cdot 10^{-4}$ & $t=2.50 \cdot 10^{-4}$ \\ 
\scalebox{0.24}{\includegraphics[angle=0, trim=10 10 10 10]{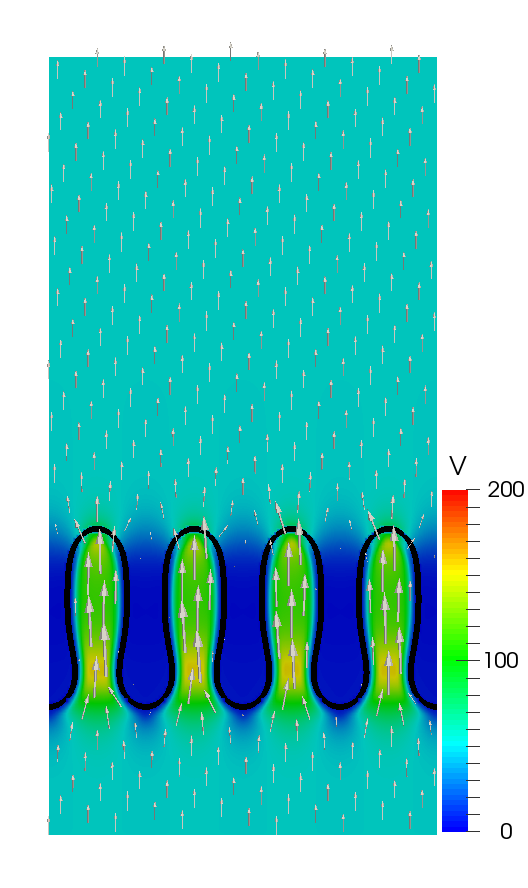}}  &
\scalebox{0.24}{\includegraphics[angle=0, trim=10 10 10 10]{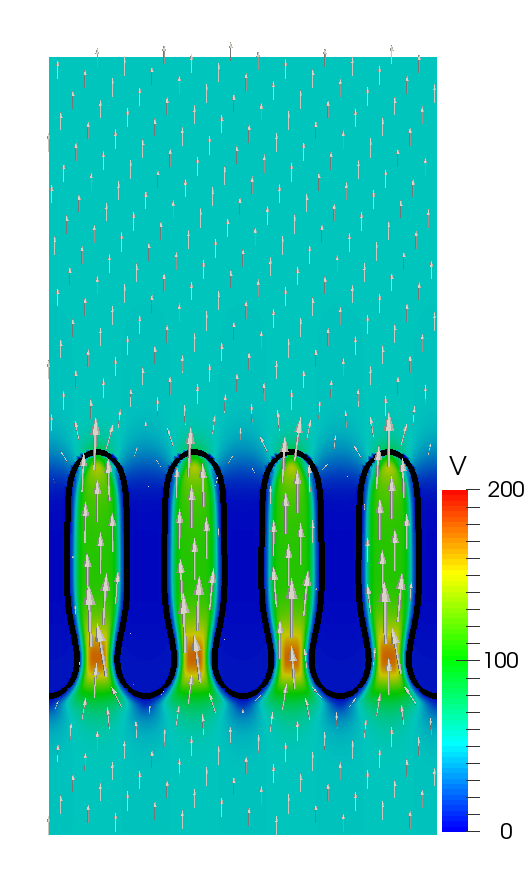}} &
\scalebox{0.24}{\includegraphics[angle=0, trim=10 10 10 10]{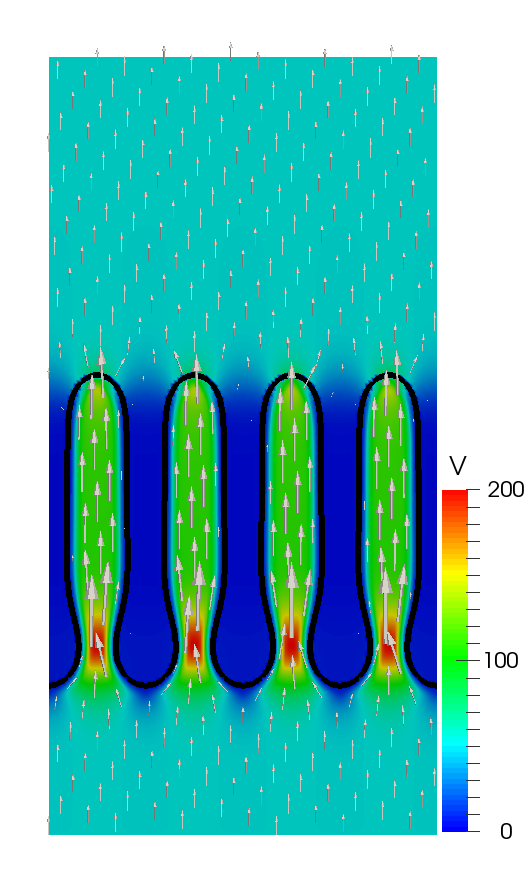}} \\[-0.3cm]
$t=3.75 \cdot 10^{-4}$ & $t=5.00 \cdot 10^{-4}$ & $t=6.25 \cdot 10^{-4}$ \\ 
\scalebox{0.24}{\includegraphics[angle=0, trim=10 10 10 10]{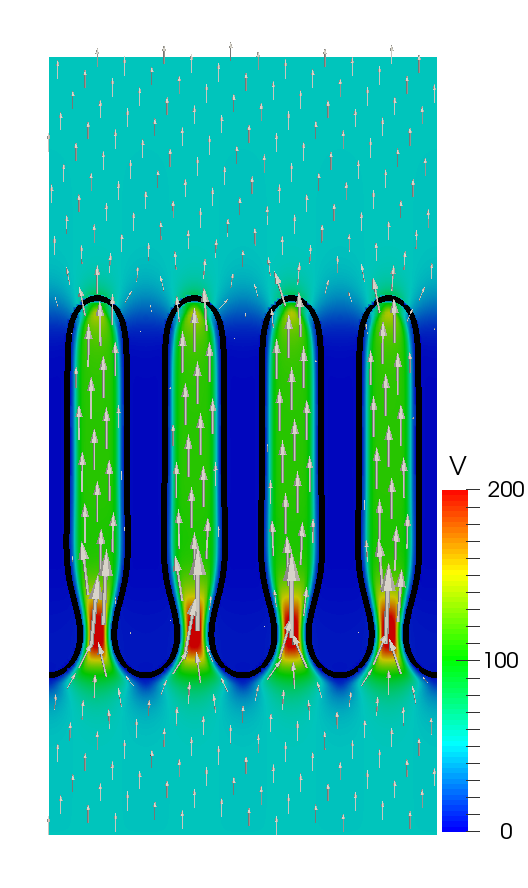}}  &
\scalebox{0.24}{\includegraphics[angle=0, trim=10 10 10 10]{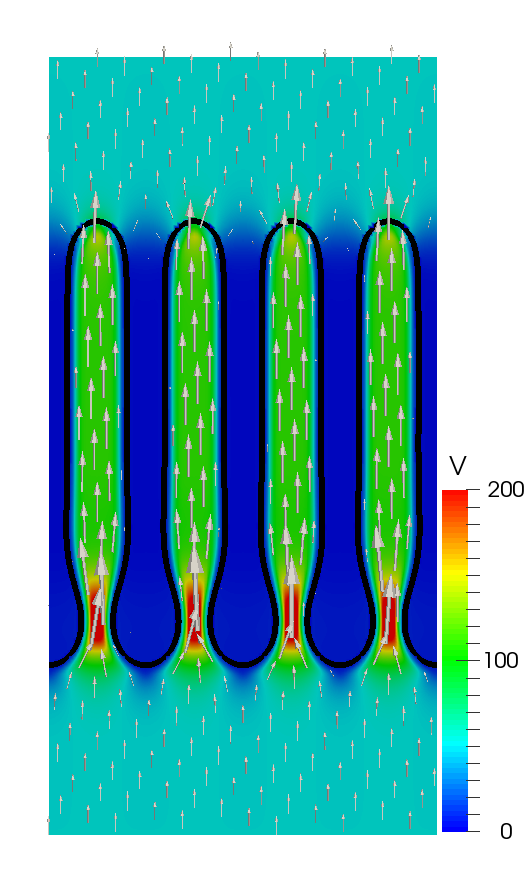}} &
\scalebox{0.24}{\includegraphics[angle=0, trim=10 10 10 10]{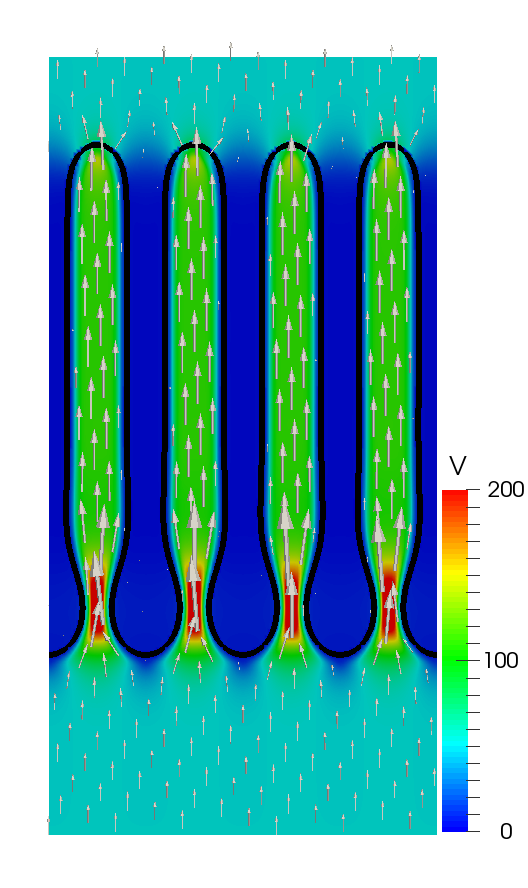}} \\[-0.3cm]
$t=7.50 \cdot 10^{-4}$ & $t=8.75 \cdot 10^{-4}$ & $t=1.00 \cdot 10^{-3}$ \\ 
\end{tabular}
\caption{Test~$2$. Viscous fingering for $\eta_{1}=50$, $\eta_{2}=1$, and $\sigma = 10^{-5}$. Velocity field at different time instances; the black contourline highlights the interface among the phases.}
\label{fig:F1_vel}
\end{center}
\end{figure}

We set $\eta_{1}=50$ and $\sigma=10^{-5}$, for which $\mathrm{Ca} = 5.0 \cdot 10^8$.  The time evolution of the computed order parameter and velocity are reported in Figs.~\ref{fig:F1_phase} and~\ref{fig:F1_vel}, respectively, which highlight the insurgency of the viscous fingering phenomenon.

\begin{figure}
\begin{center}
\begin{tabular}{ccc}
\scalebox{0.24}{\includegraphics[angle=0, trim=10 10 10 10]{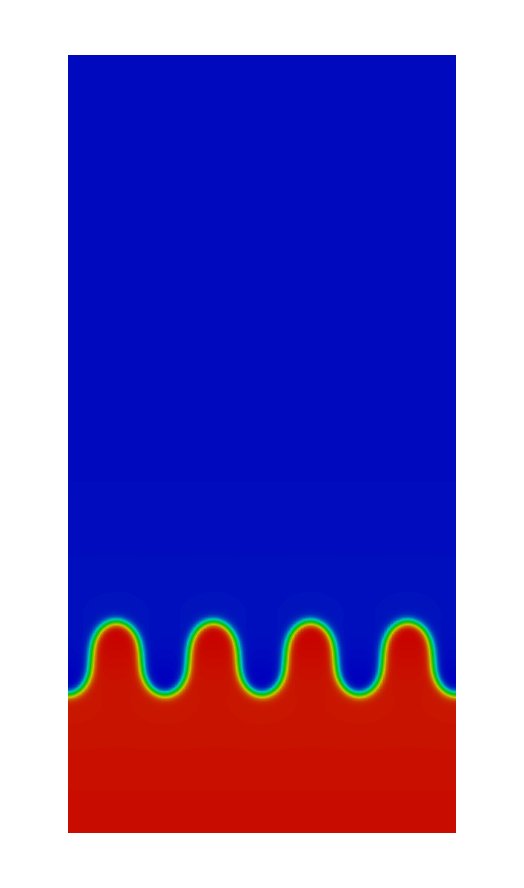}}  &
\scalebox{0.24}{\includegraphics[angle=0, trim=10 10 10 10]{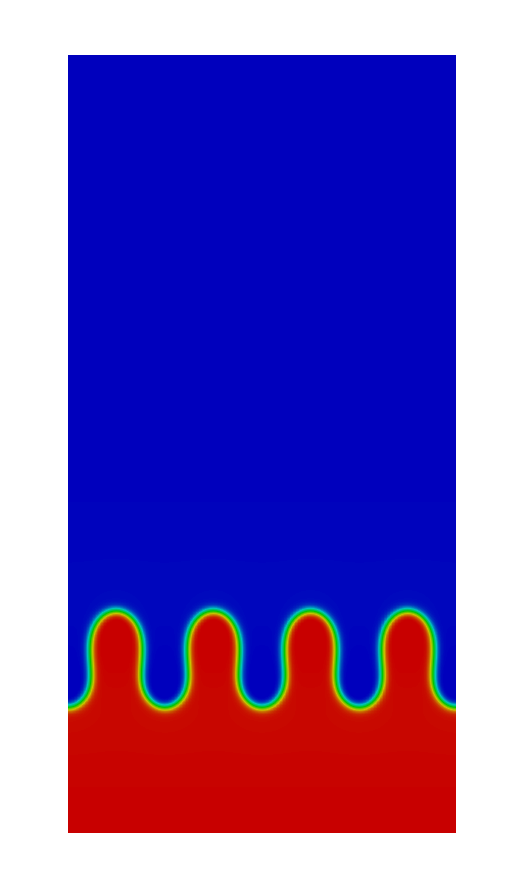}} &
\scalebox{0.24}{\includegraphics[angle=0, trim=10 10 10 10]{./F1phase_0100.png}} \\[-0.3cm]
$\eta_{1}=10$, $t=2.50 \cdot 10^{-4}$ & $\eta_{1}=20$, $t=2.50 \cdot 10^{-4}$ & $\eta_{1}=50$, $t=2.50 \cdot 10^{-4}$ \\ 
\scalebox{0.24}{\includegraphics[angle=0, trim=10 10 10 10]{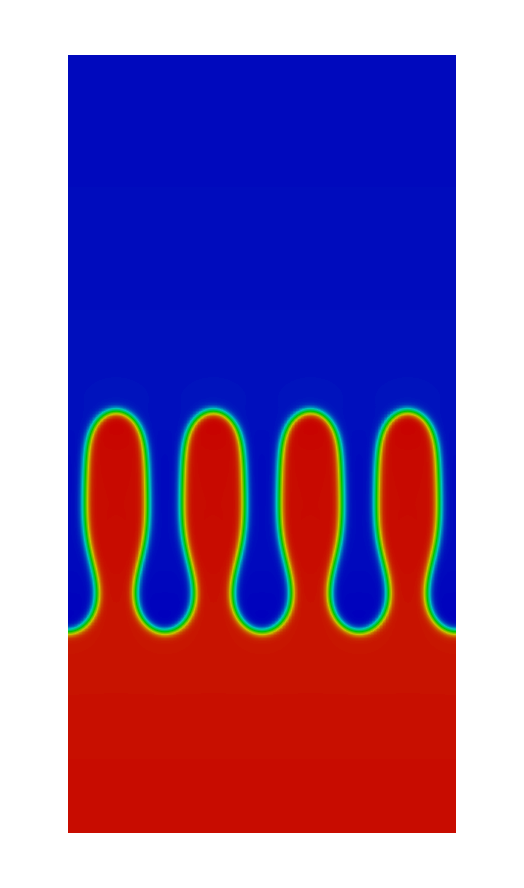}}  &
\scalebox{0.24}{\includegraphics[angle=0, trim=10 10 10 10]{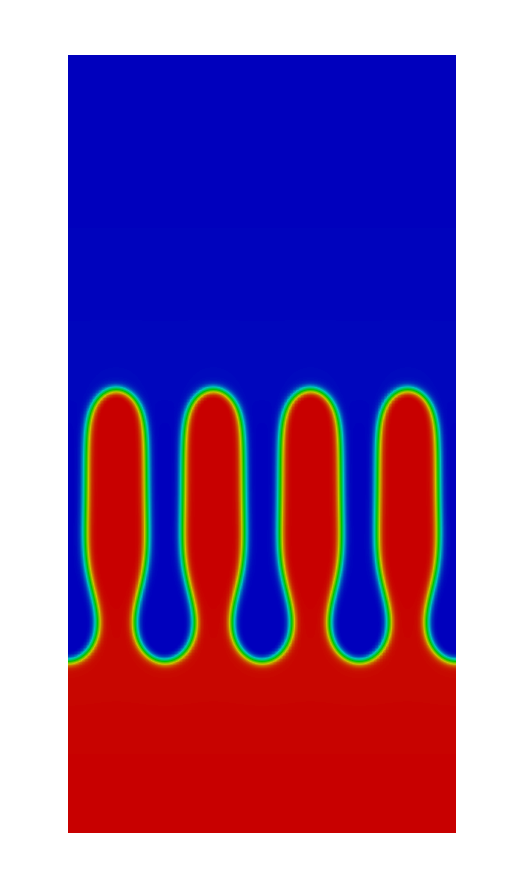}} &
\scalebox{0.24}{\includegraphics[angle=0, trim=10 10 10 10]{./F1phase_0250.png}} \\[-0.3cm]
$\eta_{1}=10$, $t=6.25 \cdot 10^{-4}$ & $\eta_{1}=20$, $t=6.25 \cdot 10^{-4}$ & $\eta_{1}=50$, $t=6.25 \cdot 10^{-4}$ \\ 
\scalebox{0.24}{\includegraphics[angle=0, trim=10 10 10 10]{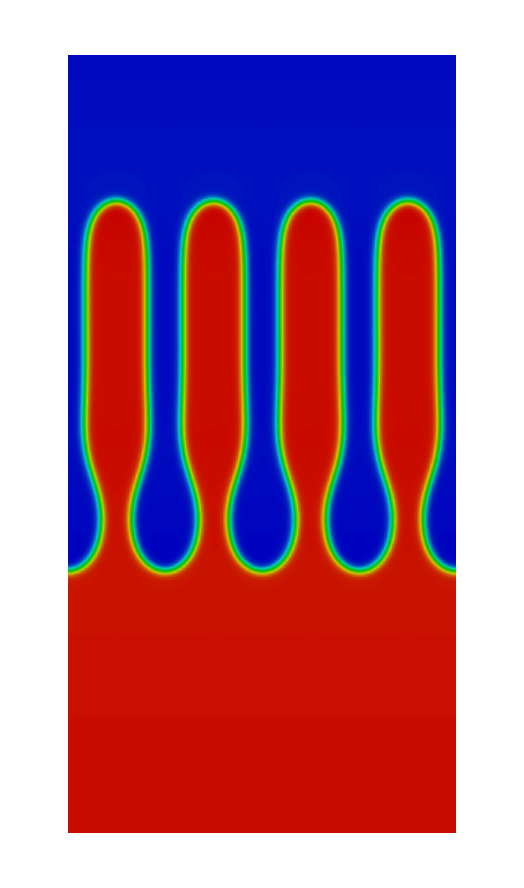}}  &
\scalebox{0.24}{\includegraphics[angle=0, trim=10 10 10 10]{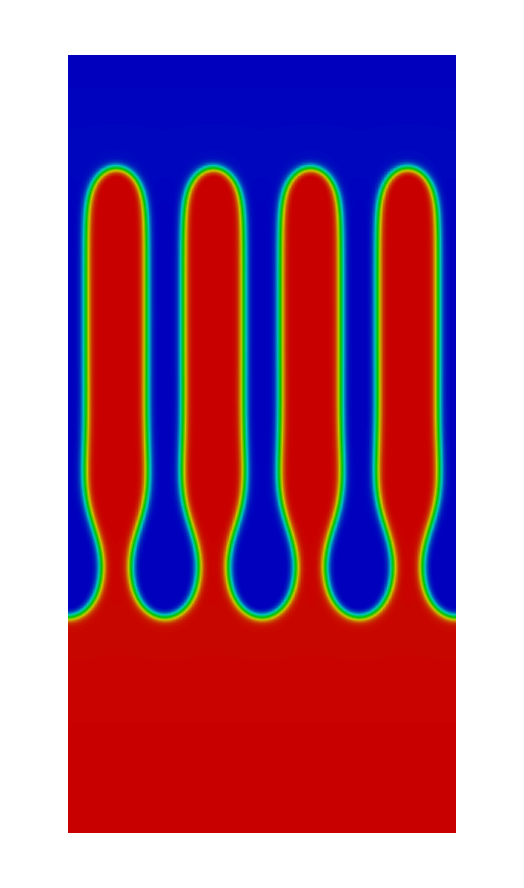}} &
\scalebox{0.24}{\includegraphics[angle=0, trim=10 10 10 10]{./F1phase_0400.png}} \\[-0.3cm]
$\eta_{1}=10$, $t=1.00 \cdot 10^{-3}$ & $\eta_{1}=20$, $t=1.00 \cdot 10^{-3}$ & $\eta_{1}=50$, $t=1.00 \cdot 10^{-3}$ \\ 
\end{tabular}
\caption{Test~$2$. Viscous fingering for $\eta_{2}=1$, $\sigma = 10^{-5}$, and different values of $\eta_{1}$. Phases at different time instances.}
\label{fig:F1F2F3_phase}
\end{center}
\end{figure}

In Fig.~\ref{fig:F1F2F3_phase} we compare the order parameters at different time instances obtained for the values of the viscosity $\eta_{1}=10$, $20$, and $50$ for $\sigma=10^{-5}$, thus yielding $\mathrm{Ca} = 10^{8}$, $2.0 \cdot 10^{8}$, and $5.0 \cdot 10^{8}$, respectively.  We remark that the more viscous the fluid ``1", the longer the fingers.

\begin{figure}
\begin{center}
\begin{tabular}{ccc}
\scalebox{0.24}{\includegraphics[angle=0, trim=10 10 10 10]{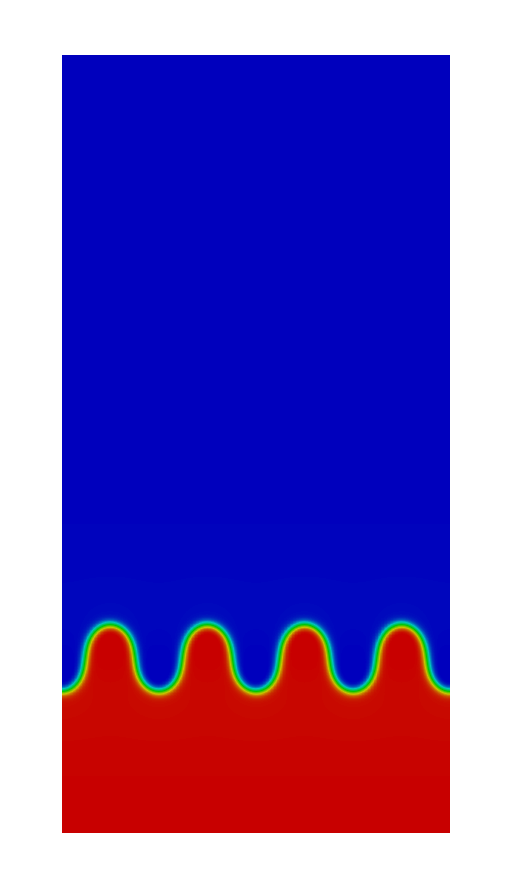}}  &
\scalebox{0.24}{\includegraphics[angle=0, trim=10 10 10 10]{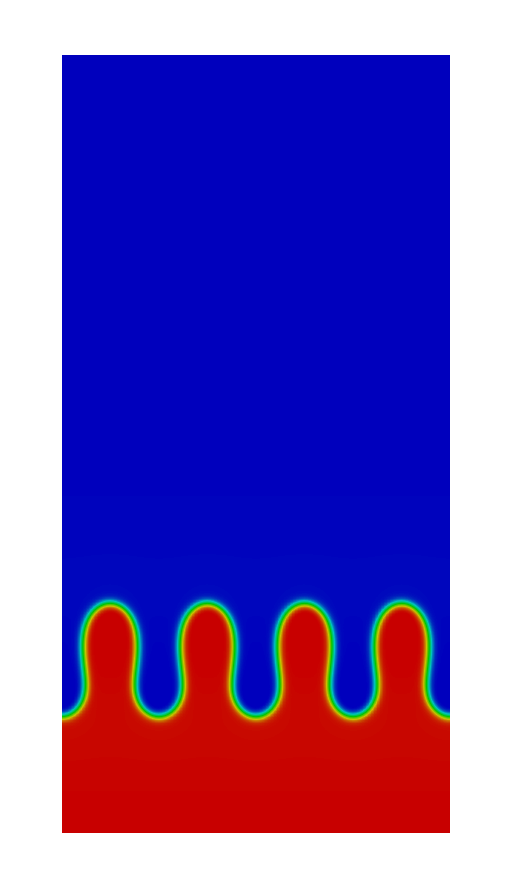}} &
\scalebox{0.24}{\includegraphics[angle=0, trim=10 10 10 10]{./F1phase_0100.png}} \\[-0.3cm]
$\sigma=10^4$, $t=2.50 \cdot 10^{-4}$ & $\sigma=10$, $t=2.50 \cdot 10^{-4}$ & $\sigma=10^{-5}$, $t=2.50 \cdot 10^{-4}$ \\ 
\scalebox{0.24}{\includegraphics[angle=0, trim=10 10 10 10]{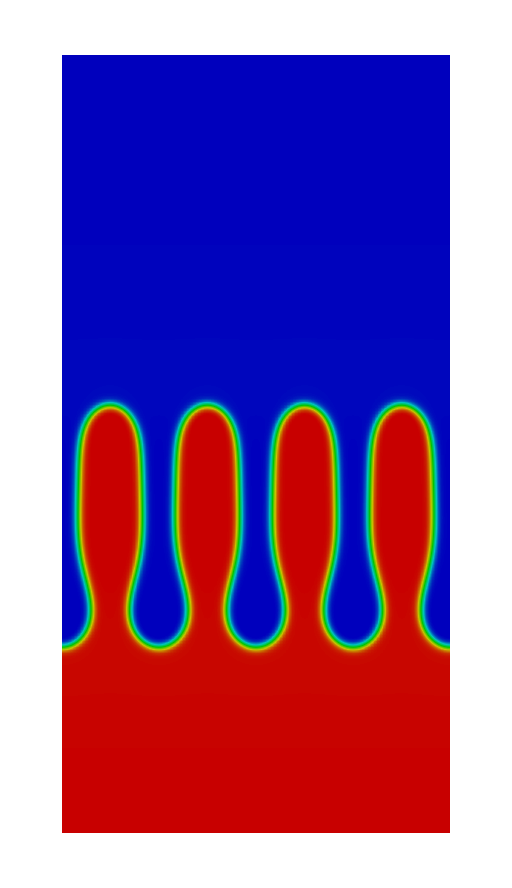}}  &
\scalebox{0.24}{\includegraphics[angle=0, trim=10 10 10 10]{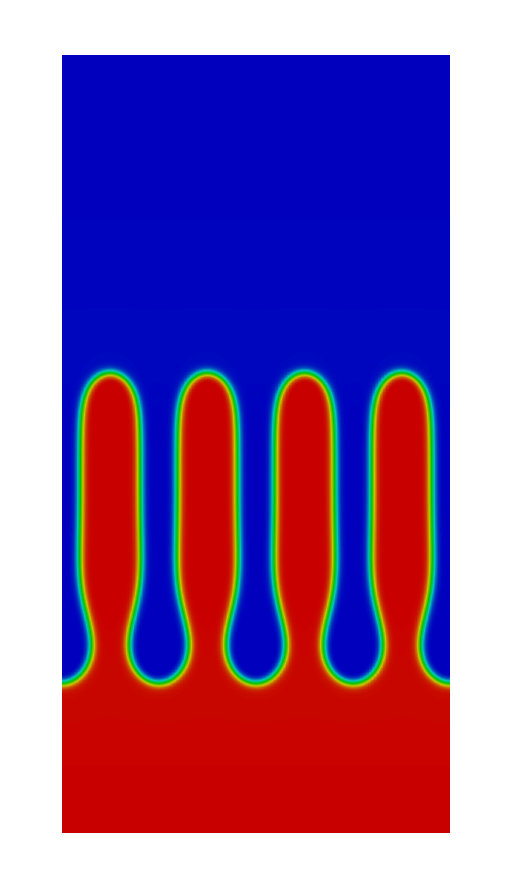}} &
\scalebox{0.24}{\includegraphics[angle=0, trim=10 10 10 10]{./F1phase_0250.png}} \\[-0.3cm]
$\sigma=10^4$, $t=6.25 \cdot 10^{-4}$ & $\sigma=10$, $t=6.25 \cdot 10^{-4}$ & $\sigma=10^{-5}$, $t=6.25 \cdot 10^{-4}$ \\ 
\scalebox{0.24}{\includegraphics[angle=0, trim=10 10 10 10]{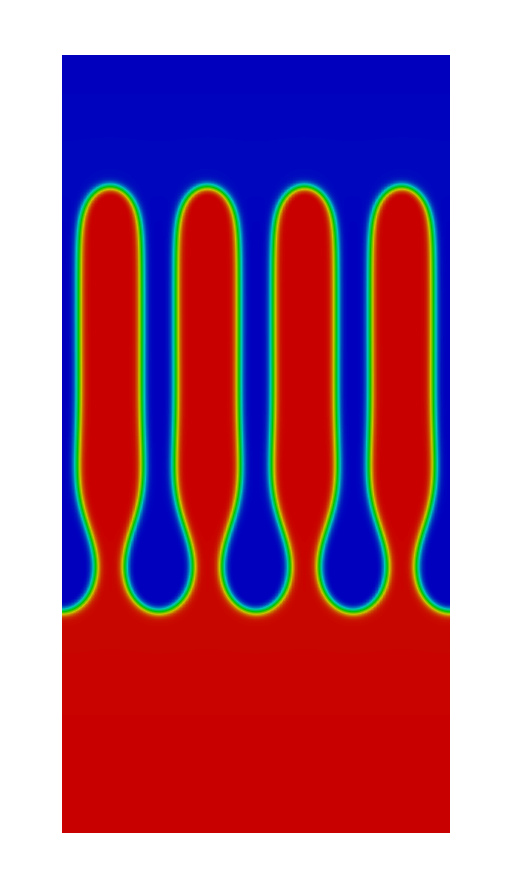}}  &
\scalebox{0.24}{\includegraphics[angle=0, trim=10 10 10 10]{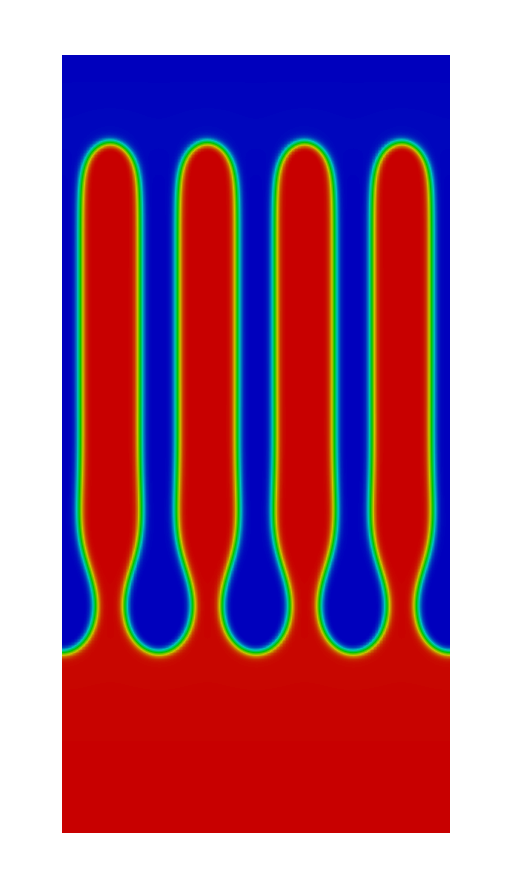}} &
\scalebox{0.24}{\includegraphics[angle=0, trim=10 10 10 10]{./F1phase_0400.png}} \\[-0.3cm]
$\sigma=10^4$, $t=1.00 \cdot 10^{-3}$ & $\sigma=10$, $t=1.00 \cdot 10^{-3}$ & $\sigma=10^{-5}$, $t=1.00 \cdot 10^{-3}$ \\ 
\end{tabular}
\caption{Test~$2$. Viscous fingering for $\eta_{1}=50$, $\eta_{2}=1$, and different values of $\sigma$. Phases at different time instances.}
\label{fig:F1F7F9_phase}
\end{center}
\end{figure}

Finally, in Fig.~\ref{fig:F1F7F9_phase} we show the order parameter at different time instances obtained for the viscosity $\eta_{1}=50$ and values of the surface tension $\sigma=10^{4}$, $10$, and $10^{-5}$ thus yielding $\mathrm{Ca} =  0.5$, $5.0 \cdot 10^{2}$, and $5.0 \cdot 10^{8}$, respectively.  We observe that, the smaller the surface tension, the longer the fingers.



\bibliographystyle{plain}
\bibliography{HSCHvol}
\end{document}